\documentclass[twocolumn]{aastex631}

\usepackage{CJK}
\usepackage{multirow}
\usepackage{
makecell}
\usepackage{graphicx}
\begin{document}

\title{JCMT 850 $\micron$ continuum observations
of density structures in the G35 molecular complex}

\shorttitle{Density structures in G35 complex}
\shortauthors{Shen et al.}

\author{Xianjin Shen}
\affiliation{School of Physics and Astronomy, Yunnan University, Kunming, 650091, People's Republic of China\\}

\author{Hong-Li Liu}
\affiliation{School of Physics and Astronomy, Yunnan University, Kunming, 650091, People's Republic of China\\}

\author{Zhiyuan Ren}
\affiliation{National Astronomical Observatories, Chinese Academy of Sciences, Datun Road A20, Beijing, People's Republic of China}
\affiliation{CAS Key Laboratory of FAST, NAOC, Chinese Academy of Sciences, Beijing, China}
\affiliation{University of Chinese Academy of Sciences, Beijing, People's Republic of China}

\author{Anandmayee Tej}
\affiliation{Indian Institute of Space Science and Technology, Thiruvananthapuram, Kerala 695 547, India}

\author{Di Li}
\affiliation{National Astronomical Observatories, Chinese Academy of Sciences, Datun Road A20, Beijing, People's Republic of China}
\affiliation{CAS Key Laboratory of FAST, NAOC, Chinese Academy of Sciences, Beijing, China}
\affiliation{University of Chinese Academy of Sciences, Beijing, People's Republic of China}
\affiliation{NAOC-UKZN Computational Astrophysics Centre (NUCAC), University of KwaZulu-Natal, Durban 4000, South Africa}

\author{Hauyu Baobab Liu}
\affiliation{Department of Physics, National Sun Yat-Sen University, No. 70, Lien-Hai Road, Kaohsiung City 80424, Taiwan, R.O.C.}
\affiliation{Center of Astronomy and Gravitation, National Taiwan Normal University, Taipei 116, Taiwan}

\author{Gary A. Fuller}
\affiliation{Jodrell Bank Centre for Astrophysics, School of Physics and Astronomy, University of Manchester, Oxford Road, Manchester, M13 9PL, UK}
\affiliation{Physikalisches Institut, University of Cologne, Zülpicher Str. 77, D-50937 Köln, Germany}

\author{Jinjin Xie}
\affiliation{National Astronomical Observatories, Chinese Academy of Sciences, Datun Road A20, Beijing, People's Republic of China}

\author{Sihan Jiao}
\affiliation{CAS Key Laboratory of FAST, NAOC, Chinese Academy of Sciences, Beijing, China}
\affiliation{University of Chinese Academy of Sciences, Beijing, People's Republic of China}

\author{Aiyuan Yang}
\affiliation{National Astronomical Observatories, Chinese Academy of Sciences, Datun Road A20, Beijing, People's Republic of China}
\affiliation{Key Laboratory of Radio Astronomy and Technology, Chinese Academy of Sciences, A20 Datun Road, Datun Road A20, Beijing, People's Republic of China}
\affiliation{Max-Plank-Institut für Radioastronomie, Auf dem Hügel 69, D-53121 Bonn, Germany}

\author{Patrick M. Koch}
\affiliation{Academia Sinica Institute of Astronomy and Astrophysics, No. 1, Sec. 4, Roosevelt Road, Taipei 10617, Taiwan}

\author{Fengwei Xu}
\affiliation{Kavli Institute for Astronomy and Astrophysics, Peking University, 5 Yiheyuan Road, Haidian District, Beijing 100871, China}
\affiliation{Department of Astronomy, School of Physics, Peking University, Beijing 100871, People's Republic of China}

\author{Patricio Sanhueza}
\affiliation{National Astronomical Observatory of Japan, National Institutes of Natural Sciences, 2-21-1 Osawa, Mitaka, Tokyo 181-8588, Japan}
\affiliation{Astronomical Science Program, The Graduate University for Advanced Studies, SOKENDAI, 2-21-1 Osawa, Mitaka, Tokyo 181-8588, Japan}

\author{Pham Ngoc Diep}
\affiliation{ Vietnam National Space Center, Vietnam Academy of Science and Technology, 18 Hoang Quoc Viet, Hanoi, Vietnam} 

\author{Nicolas Peretto}
\affiliation{Cardiff Hub for Astrophysics Research \& Technology, School of Physics \& Astronomy, Cardiff University, Queens Buildings, The Parade, Cardiff CF24 3AA, UK}

\author{R.K. Yadav}
\affiliation{National Astronomical Research Institute of Thailand (NARIT), Sirindhorn AstroPark, 260 Moo 4, T. Donkaew, A. Maerim, Chiangmai 50180, Thailand}

\author{Busaba H. Kramer}
\affiliation{Max-Plank-Institut für Radioastronomie, Auf dem Hügel 69, D-53121 Bonn, Germany}
\affiliation{National Astronomical Research Institute of Thailand (NARIT), Sirindhorn AstroPark, 260 Moo 4, T. Donkaew, A. Maerim, Chiangmai 50180, Thailand}

\author{Koichiro Sugiyama}
\affiliation{National Astronomical Research Institute of Thailand (NARIT), Sirindhorn AstroPark, 260 Moo 4, T. Donkaew, A. Maerim, Chiangmai 50180, Thailand}

\author{Mark G. Rawlings}
\affiliation{Gemini Observatory/NSF's NOIRLab, 670 N. A`ohōkū Place, Hilo, HI 96720, USA}

\author{Chang Won Lee}
\affiliation{ Korea Astronomy and Space Science Institute, 776 Daedeokdae-ro, Yuseong-gu, Daejeon 34055, Republic of Korea}
\affiliation{University of Science and Technology, Korea (UST), 217 Gajeong-ro, Yuseong-gu, Daejeon 34113, Republic of Korea}

\author{Ken'ichi Tatematsu }
\affiliation{National Astronomical Observatory of Japan, National Institutes of Natural Sciences, 2-21-1 Osawa, Mitaka, Tokyo 181-8588, Japan}
\affiliation{Astronomical Science Program, The Graduate University for Advanced Studies, SOKENDAI, 2-21-1 Osawa, Mitaka, Tokyo 181-8588, Japan}

\author{Daniel Harsono}
\affiliation{Institute of Astronomy, Department of Physics, National Tsing Hua University, Hsinchu 30013, Taiwan}

\author{David Eden}
\affiliation{ Armagh Observatory and Planetarium, College Hill, Armagh, BT61 9DB, United Kingdom}

\author{Woojin Kwon}
\affiliation{Department of Earth Science Education, Seoul National University, 1 Gwanak-ro, Gwanak-gu, Seoul 08826, Republic of Korea}
\affiliation{SNU Astronomy Research Center, Seoul National University, 1 Gwanak-ro, Gwanak-gu, Seoul 08826, Republic of Korea}

\author{Chao-Wei Tsai}
\affiliation{National Astronomical Observatories, Chinese Academy of Sciences, Datun Road A20, Beijing, People's Republic of China}
\affiliation{Institute for Frontiers in Astronomy and Astrophysics, Beijing Normal University, Beijing 102206, People's Republic of China}
\affiliation{University of Chinese Academy of Sciences, Beijing, People's Republic of China}

\author{Glenn J. White}
\affiliation{School of Physical Sciences, The Open University, Walton Hall, milton Keynes, MK7 6AA, UK}
\affiliation{RAL Space, STFC Rutherford Appleton Laboratory, Chilton, Didcot, Oxfordshire, OX11 0QX, UK}

\author{Kee-Tae Kim}
\affiliation{Korea Astronomy and Space Science Institute, 776 Daedeokdae-ro, Yuseong-gu, Daejeon 34055, Republic of Korea}
\affiliation{University of Science and Technology, Korea (UST), 217 Gajeong-ro, Yuseong-gu, Daejeon 34113, Republic of Korea}

\author{Tie Liu}
\affiliation{Shanghai Astronomical Observatory, Chinese Academy of Sciences, 80 Nandan Road, Shanghai 200030, People's Republic of China}

\author{Ke Wang}
\affiliation{Kavli Institute for Astronomy and Astrophysics, Peking University, 5 Yiheyuan Road, Haidian District, Beijing 100871, China}

\author{Siju Zhang}
\affiliation{Kavli Institute for Astronomy and Astrophysics, Peking University, 5 Yiheyuan Road, Haidian District, Beijing 100871, China}

\author{Wenyu Jiao}
\affiliation{Kavli Institute for Astronomy and Astrophysics, Peking University, 5 Yiheyuan Road, Haidian District, Beijing 100871, China}
\affiliation{Department of Astronomy, School of Physics, Peking University, Beijing 100871, People's Republic of China}

\author{Dongting Yang}
\affiliation{School of Physics and Astronomy, Yunnan University, Kunming, 650091, People's Republic of China}

\author{Swagat R. Das}
\affiliation{Departamento de Astronomia, Universidad de Chile, Las Condes, 7591245 Santiago, Chile}
\affiliation{Indian Institute of Science Education and Research (IISER) Tirupati, Rami Reddy Nagar, Karakambadi Road, Mangalam (P.O.), Tirupati 517507, India}

\author{Jingwen Wu}
\affiliation{National Astronomical Observatories, Chinese Academy of Sciences, Datun Road A20, Beijing, People's Republic of China}
\affiliation{University of Chinese Academy of Sciences, Beijing, People's Republic of China}

\author{Chen Wang}
\affiliation{National Astronomical Observatories, Chinese Academy of Sciences, Datun Road A20, Beijing, People's Republic of China}

\correspondingauthor{Hong-Li Liu, Zhiyuan Ren, Anandmayee Tej}
\email{hongliliu2012@gmail.com, jeremyrzy@gmail.com, tej@iist.ac.in}

\begin{abstract}
Filaments are believed to play a key role in high-mass star formation. We present a systematic study of the filaments and their hosting clumps in the G35 molecular complex using JCMT SCUBA-2 850 $\micron$ continuum data. We identified five clouds in the complex and 91 filaments within them, some of which form 10 hub-filament systems (HFSs), each with at least 3 hub-composing filaments. We also compiled a catalogue of 350 dense clumps, 183 of which are associated with the filaments. We investigated the physical properties of the filaments and clumps, such as mass, density, and size, and their relation to star formation. We find that the global mass-length trend of the filaments is consistent with a turbulent origin, while the hub-composing filaments of high line masses ($m_{\rm l}\,>$\,230\,$\mathrm{M_{\odot}~pc^{-1}}$) in HFSs deviate from this relation, possibly due to feedback from massive star formation. We also find that the most massive and densest clumps (R\,$>$\,0.2\,pc, M\,$>35\,\mathrm{M_{\odot}}$, $\mathrm{\Sigma}>\,0.05\,\mathrm{g~cm^{-2}}$)  are located in the filaments and in the hubs of HFS with the latter bearing a higher probability of occurrence of high-mass star-forming signatures, highlighting the preferential sites of HFSs for high-mass star formation. We do not find significant variation in the clump mass surface density across different evolutionary environments of the clouds, which may reflect the balance between mass accretion and stellar feedback.
\end{abstract}

\keywords{Star formation; Star forming regions; Molecular Clouds; Interstellar filaments; Molecular gas; Dust continuum emission}

\section{Introduction} 
\label{sec:introduction}

High-mass stars (M $>$ 8\,$ \mathrm{M_{\sun}}$) 
significantly contribute to the energy and momentum of galaxies through their radiation, stellar winds, massive outflows, expanding HII regions, and supernova explosions \citep{Zinnecker&Yorke2007,Motte2018}. Despite their importance, the formation process of these stars remains less understood than that of their low-mass counterparts, primarily due to their short lifetime scales, far distances, intense accretion processes and significant feedback effects. 

Observational studies have revealed the ubiquity of filamentary structures in the Milky Way  \citep[e.g.,][and references therein]{Andre2010,HiGAL2010,Kumar2020,Schisano2020,Annie2023}. These studies have also highlighted a strong correlation between star formation and filamentary molecular clouds \citep[e.g.,][]{Andro2014,Lu2014,Lu2018,LHL2020,LHL2022}. 
Intersecting filaments create distinct web structures, known as hub-filament systems (HFSs), representing a unique class of filaments in star formation, particularly for high-mass stars \citep[e.g.,][]{Myers2009,Liu2012,Wang2016,Kumar2020,Zhou2022, LiuHL2023}. In the context of HFSs, the central node is referred to as the hub, while the individual filaments associated with it are termed hub-composing filaments. 

Recent studies of HFSs \citep[e.g.,][]{LiuH2012,Peretto2013,Peretto2014,Yuan2018,LiuHL2019,Ren2021,Sanhueza2021,Saha22,Xu2023,He23,Yang23,Pan24} have invoked the latest high-mass star formation models, such as Global Hierarchical Collapse (GHC, \citealt{GHC2019}) and Inertial Inflow (I2, \citealt{Pad20}) to 
understand star formation in these systems.
In such models, filamentary structures are predicted to undergo self-growth through radial accretion from their surrounding environment and supply the formed clumps or cores through longitudinal accretion. Therefore, investigating how the physical proprieties (e.g., mass, density, size) of these hierarchical density structures evolve with time is helpful to understanding the complete picture of star formation, especially for high-mass star formation.  To this end, conducting 
observational studies of large samples of filaments and embedded dense structures at different stages of evolution become the focus of this study.

The structure of the paper is as follows: Sections\,\ref{sec:presentation} and \ref{sec:observations} describe the targeted region, and observations, respectively. 
Section\,\ref{sec:results} presents the results and analysis, focusing on distance measurement, identification of structures, and derivation of relevant physical parameters. Section\,\ref{sec:discussion} discusses  the star formation scenario in the complex investigated. Specifically, Section\,\ref{sec:5.1} delves into the fragmentation of filaments; Section\,\ref{sec:5.2} discusses the role of HFS clouds in massive star formation; Section\,\ref{sec:5.3} explores the evolution of clump density over time. Section\,\ref{sec:conclusions} gives the summary and conclusion.

\section{Presentation of the targeted region} \label{sec:presentation}

\begin{figure*}
    \centering
    \includegraphics[scale=0.5]{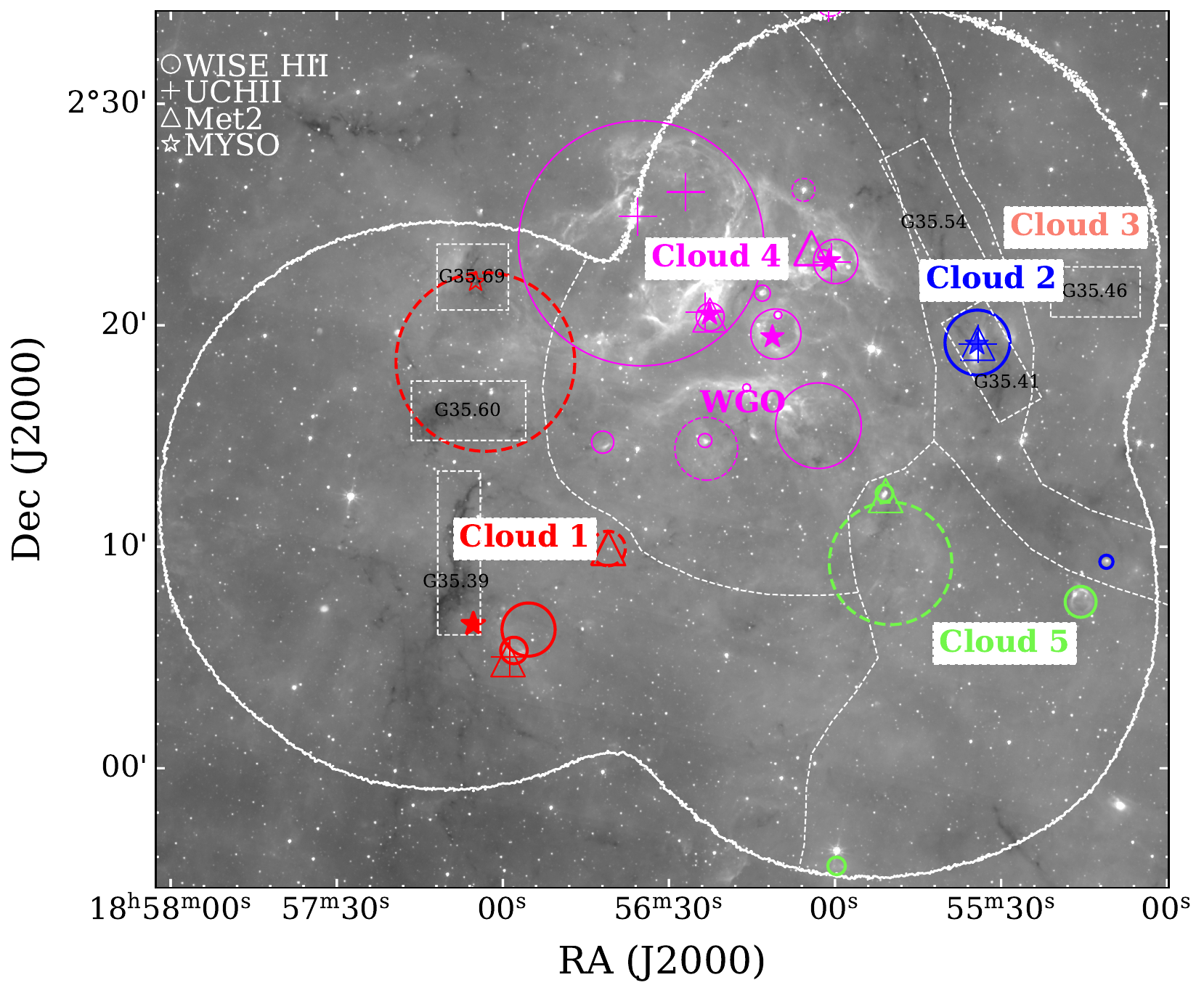}
    \caption{ Overview of 
    the G35 complex in  {\it Spitzer} 8\,\micron\ emission. The solid and dashed circles correspond to the HII regions and  candidates from the WISE catalogue by \citet{HIIcat2014}, respectively, with the symbol size representing the actual size of those objects. The plus symbols identify the positions of ultracompact HII regions from different literature catalogs \citep{Bronfman1996,Hou2014,Hu2016,Kalcheva2018,Ouyang2019}. The class II methanol masers retrieved from \citet{Ladeyschikov2019, Nguyen2022} are marked by open triangles. The solid stars identify the Red MSX MYSOs 
    \citep{Urquhart2011} while the open stars identify  massive protostars \citep{LiQ2018}. Six dark clouds mentioned in Sect.\,\ref{sec:presentation} with their names are marked using white dashed rectangles. The white dashed lines approximately delineate boundaries of different clouds. The region of 850\,\micron\ continuum for final analysis is  limited within a white solid contour of the  20\,mJy~beam$^{-1}$ {\it rms} level.}             
    \label{fig:cross}
\end{figure*}

The G35 molecular complex is selected as the target region investigated here due to the presence of many filamentary structures at different evolutionary stages.
The target region, centered at $\alpha_{2000}=18^{\mathrm{h}}56^{\mathrm{m}}28^{\mathrm{s}}.16,\delta_{2000}=2\degr14\arcmin25.\arcsec71$ ({\it l}\,=\,35.499\degr, {\it b}\,=\,-0.097\degr),  with an approximate radius of 0.41\degr, covers five individual clouds which correspond to different systemic velocities (see Section\,\ref{sec:4.1}) and thus are designated as Clouds 1, 2, 3, 4, and 5 (see Fig.\,\ref{fig:cross}). 

Cloud\,1, Cloud\,2 and Cloud\,3  harbour
infrared dark clouds (IRDCs). Six IRDCs (i.e., G35.39, G35.60 and G35.69  in Cloud\,1, G35.41 and G35.54 in Cloud\,2, G35.46 in Cloud\,3 ),  as candidate sites for high-mass star formation, were cataloged by \citet{IRDC2009}.
One of them, G35.39, located at $\sim 2.9$\,kpc \citep{Simon2006}, is a well-known star-forming ridge \citep{ministarburst2011,Motte2018} that has been extensively 
studied for its fragmentation, kinematics, dust polarization, and/or chemical properties \citep[e.g.,][]{Sanhueza2012,Jim2014,Barnes2016,G35Td2017,Liu2018,Barnes2021,Xie2021}. 
In comparison, IRDC\,G35.41, exhibiting a 
bipolar bubble feature in the 8\,\micron\ image, is at a more evolved stage. 
The elongated IRDC\,G35.46 in Cloud\,3, together with IRDC\,G35.41 and G35.54 in Cloud\,2, and IRDC\,G35.60 and G35.69 in Cloud\,1, are less well studied.
In addition, the entire targeted region contains
37 ATLASGAL clumps \citep{csc2014,Urquhart2018}, about a quarter of which are located within the IRDCs. 

The  other molecular cloud complex in the northern region is Cloud\,4 centred at $\alpha_{2000}=18^{\mathrm{h}}56^{\mathrm{m}}09^{\mathrm{s}}.55,\delta_{2000}=2\degr22\arcmin58.\arcsec99$. This cloud, approximately 0.$2{\degr}$ in radius  
(corresponding to a physcial scale of $\sim$10\,pc), exhibits  extended morphology and is bright in the IR.
Its eastern side is connected to a large IR bubble (i.e., G356521-0002439; \citealt{MWP2019}), denoted as the largest magenta circle. The entire cloud encompasses several smaller IR bubbles surrounding HII regions \citep{HIIcat2014}, also marked with
colored circles in Fig.\,\ref{fig:cross}. Three massive young stellar objects (MYSOs) and a WISE Green Object (WGO) G35.417-0.285 denoted in Fig.\,\ref{fig:cross} can be found within Cloud\,4 \citep{MYSO2008,WEGO2023}. Furthermore, the Spitzer/IRAC Candidate young stellar object (YSO) Catalog for the Inner Galactic Midplane project (SPICY; \citealt{SPICY2021}) has revealed a group of over 300 young YSO candidates, which are not shown here.  Fourteen HII regions and six methanol masers at 6.7 GHz, both suggestive of high-mass star formation, have been observed in this large area containing the four primary clouds by the Global View on Star Formation (GLOSTAR) survey with high-resolution VLA observations \citep{glostar2021A&A,Nguyen2022,Dzib2023,YangAY2023}. 

Cloud\,5 appears to harbour four bubbles identified as (candidate) HII regions \citep{HIIcat2014}, one of which is associated with a methanol maser at 6.7 GHz. In this cloud along with Cloud\,3, most of the region show relatively weak cold dust emission as traced by 850\,\micron\ radiation (see Sect.\,4.2, and Fig.\,\ref{fig:850micron}), which is similar to those low intensity contrast observed in 8\,\micron\ emission.

In summary, the entire target complex encompasses active high-mass star-forming clouds at various evolutionary phases, from quiescent IRDCs to evolved bubbles or HII regions.

\section{Observations} \label{sec:observations}
\subsection{SCUBA-2 850\,\micron\ Continuum Data}

The G35 molecular complex was observed as part of the 850\,\micron\  dust continuum survey, called "A Lei of the Habitat and Assembly of Infrared Dark Clouds" (ALOHA IRDCs \footnote{\url{https://www.eaobservatory.org/jcmt/science/large-programs/aloha-irdcs/}}) under a JCMT large program  (Proj. ID. M20AL021; Ren et al. 2024, in prep.).  
The survey was performed using the SCUBA-2 rotating Pong900 pattern. The entire observing coverage of the target region, which is delineated as the outermost white contour in Fig.\,1,  consists of three fields of view (FoV), each observed with a radius of approximately 15\arcmin\ at a scan spacing of 30\arcsec.
The entire target was observed with an integration time of 22 hours. Raw data were reduced using the \textit{Starlink}
\footnote{\url{https://starlink.eao.hawaii.edu/starlink}} software, yielding a final reduced 850\,\micron\ image and a rms noise map. 
The noise level varies between 5 to 75\,mJy~beam$^{-1}$ from the inner region to the edge, and the global {\it rms} distribution appears, except for the regions approaching the edge, reasonably uniform at an approximate 10\,mJy~beam$^{-1}$ (see Fig.\,\ref{fig:rms}). 
The beam size of the reduced data is 14.1\arcsec\ and the adopted pixel size for mapping is 4\arcsec. The flux calibration uncertainty of SCUBA-2 at 850\,\micron\ is $\sim$\,6$\%$ \citep{jcmt2021}.
Note that ground-based (sub)millimeter continuum observations usually suffer from  missing flux at large scales \citep[e.g.,][]{Jiao2022} in the data-processing stage, since a filter has to be applied to remove large-scale atmospheric noises. Following the SCUBA-2 Data Reduction Cookbook\footnote{\url{http://starlink.eao.hawaii.edu/devdocs/sc21.htx/sc21.html}}, a default filter scale of 480\arcsec\  was adopted.
This translates to scales $>4.5$\,pc at the distance of the G35 complex and thus is assumed not to significantly affect the smaller-scale dense structures of filaments (i.e., $\sim 0.5$\,pc in typical width) and clumps investigated here \citep[see also][]{ATLASGAL2009,Li2016, Urquhart2018}.

\subsection{Archival Data}
We used multi-wavelength infrared images from the {\it Spitzer}
and {\it Herschel} telescopes to study the dust properties of the cloud complex.
These include  {\it Spitzer} 8\,$\mu$m data from the GLIMPSE \citep{GLIMPSE2003} survey, and {\it Herschel} data (160\,$\mu$m and 250\,$\mu$m) from the Hi-GAL survey \citep{HiGAL2010}. 
The angular resolutions at 8\,$\mu$m, 160\,$\mu$m and 250\,$\mu$m are approximately 2\arcsec, 12\arcsec, and 18\arcsec, respectively.

Molecular line observations of $^{13}\rm{CO}$ (\textit{J}=1-0) were obtained as part of Boston University–Five College Radio Astronomy Observatory Galactic Ring Survey (GRS; \citealt{GRS2006}) using the FCRAO 14 m telescope. The data has a {\it rms} sensitivity $<$\,0.4\,K, a spectral resolution of 0.21\,km~s$^{-1}$, and an angular resolution of 46\farcs\ 

\section{Results}\label{sec:results}
\subsection{Distance determination}
\label{sec:4.1}
Except for the well-studied IRDC G35.39-0.33 located at 2.9 kpc \citep{Simon2006,Liu2018}, the distances for the remaining clouds in the region targeted here have not yet been determined. 
Fig.\,\ref{fig:sp} displays the average spectra of $^{13}$CO over the entire target region. Five strong velocity components are seen in the $^{13}$CO line with velocity ranges [5, 20], [20, 40], [40, 50], [50, 65], and [70, 85] km~s$^{-1}$. With the assumption that cold dust emission comes from the the same region as molecular gas,
the velocity-integrated intensity of $^{13}$CO of these velocity components can be used to investigate their spatial association with the 850\,$\mu$m dust continuum. Based on this, we exclude the second velocity component from further analysis. That component is most likely a non-associated component, as the line and dust emission do not spatially match  well. The integrated intensity  of $^{13}$CO of four velocity components is presented in Fig.\,\ref{fig:int}, showing a good association between gas and dust emission, excluding the second component. Note that there are overlapping distributions of relatively weak gas emission from different velocity components, such as the overlap between Cloud\,5 and Cloud\,2 (Fig.\,\ref{fig:int}).
This indicates that the observed dust emission is likely contaminated by the line of sight contamination of  clouds at different distances.

Based on the above-mentioned spatial association in 850\,\micron\ dust continuum emission, five primary clouds shown in Fig.\,\ref{fig:cross} have been approximately delineated. The separation between Cloud\,1 and Cloud\,4 is roughly based on the appearance of relatively dark and bright 8\micron\ features and hence, is to be considered with caution for guidance purpose solely.
The associated four velocity components are considered to derive kinematic distances using the latest distance calculation tool, the "Parallax-Based Distance Calculator V2" \citep{bessel2016,bessel2019}, leading to distance estimates 
of 2.1\,kpc (i.e., Cloud\,5), 2.5\,kpc (i.e., Cloud\,1, Cloud\,3 and Cloud \,4), and 5.1\,kpc (i.e., Cloud\,2), all with an uncertainty of $\sim$10\%.
The distance estimates of Clouds\,1, 3, and 4 are very close to that of IRDC 035.39-0.33.
Therefore, they are assumed to be in the same complex, and are treated for consistency to have the same distance as that of the well-studied IRDC G35.39 (i.e., 2.9 kpc).

\subsection{Dust Continuum Emission}
\begin{figure*}
    \centering
    \includegraphics[scale=0.5]{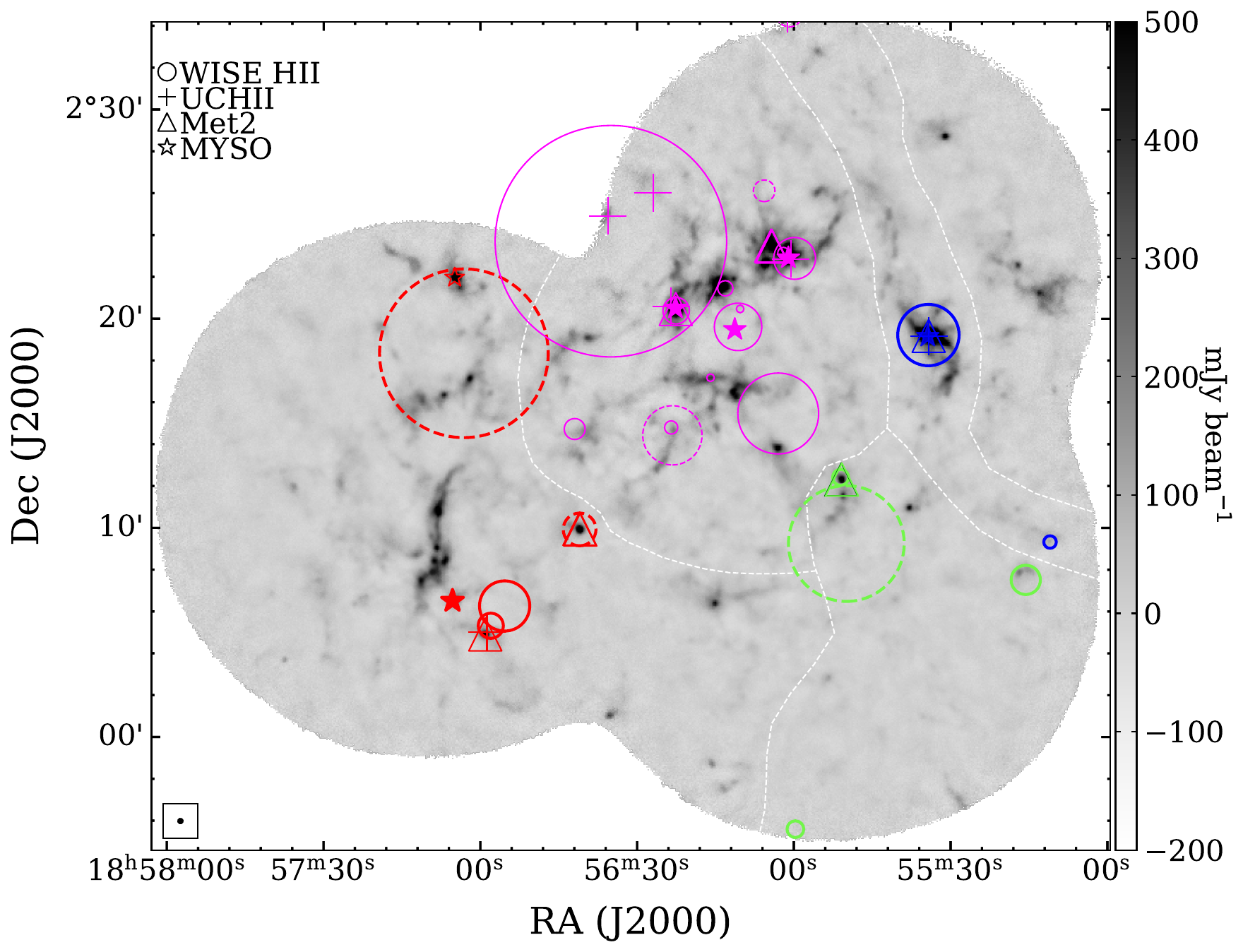}
    \caption{Dust emission maps as traced by 850\,\micron\ continuum for the G35 complex. All symbols are the same as those in Fig.\,\ref{fig:cross}. The beam size of 850\,\micron\ observations is displayed at the bottom left corner.}
    \label{fig:850micron}
\end{figure*}

Fig.\,\ref{fig:850micron} shows the 850\,\micron\ dust emisson image. As an excellent tracer of column density, the continuum peak emission matches the 8\,$\mu$m dark lanes. Overall, Cloud\,1, 2, and 4 account for the vast majority of the continuum emission, while the continuum emission from Cloud\,3 and 5 are relatively weak.

\subsection{Identification of Filaments and Clumps}
\begin{figure}
    \centering
    \includegraphics[scale=0.35]{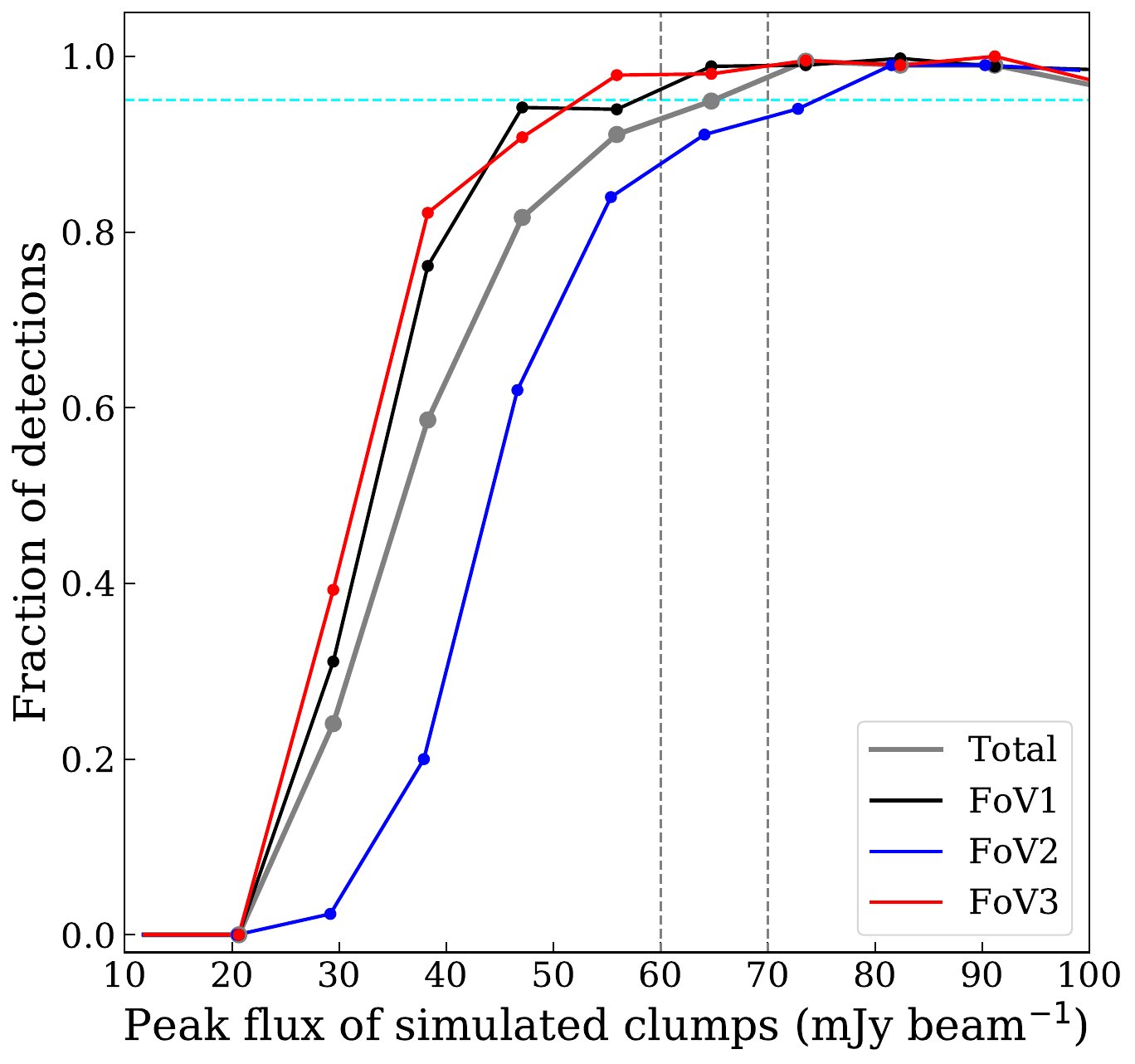}
    \caption{Detection fraction of simulated sources as a function of peak flux density. Different curves correspond to different FoV analysis, including FoVs\,1, 2 and 3 and their sum as the total FoV. The dashed cyan line indicates the 95$\%$ detection fraction/flux completeness level. Two dashed gray lines (from left to right) show the 6$rms$ and 7\,$rms$ levels ($1\,rms=$10mJy~beam$^{-1}$), respectively.}
    \label{fig:completeness}
\end{figure}
\begin{figure*}
    \centering
    \includegraphics[scale=0.41]{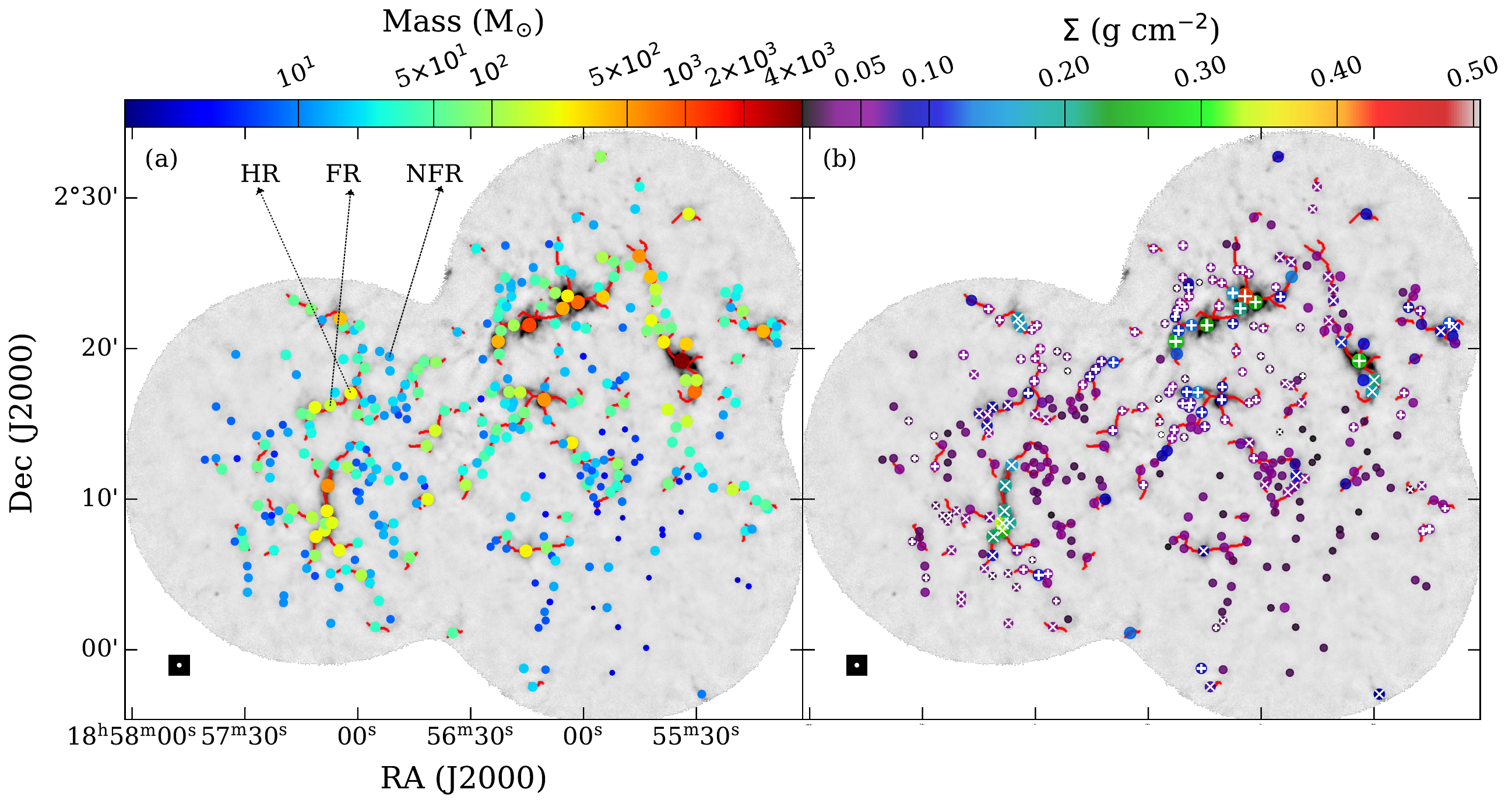}
    \caption{Spatial distributions of filaments (red curves) and clumps (colored dots). In panel\,(a), the color of the dot corresponds to the value of the clump mass while in panel\,(b), it corresponds to the value of the mass surface density. In both panels, the white plus and cross symbols indicate IR-bright, and IR-dark clumps inferred from {\it Spitzer} 8\,\micron\ image, respectively. The examples of HR clumps, FR clumps, and NFR clumps mentioned in Section\,\ref{sec:5.1} are marked with arrows.}
    \label{fig:clump_filament}
\end{figure*}
\subsubsection{Filament extraction}
Filamentary structures were extracted from the 850\,\micron\ continuum image, using the algorithm \texttt{FILFINDER} \citep{FilFinder2015}. We set the minimum threshold (\texttt{glob$\_$thresh}) to five times the {\it rms} ($\sim$\,50\,mJy~beam$^{-1}$), the minimum structure area (\texttt{size$\_$thresh}) to 18\,pixels (twice the beam area), and the hole-filling size (\texttt{fill$\_$hole$\_$size}) to 150\,pixels. Here, the first two parameters were determined to improve the robustness of the identified skeletons, while the third one was to avoid the presence of holes within a skeleton. During our post selection, we discarded structures at the image edges and those in the areas with {\it rms} greater than 20\,mJy~beam$^{-1}$. We also removed structures with a major axis (length) shorter than 3 beam sizes, following \cite{Suri2019}. As a result, we obtained 91 skeletons representing individual filaments, some of which form HFSs (see Sect.\,1 for their definition). 

The identified filaments are shown in Fig.\,\ref{fig:clump_filament} and \ref{fig:Tg}. Overall, Cloud\,1 is found to contain 17 individual filaments and 3 HFSs (i.e., each HFSs corresponding to a web of at least three individual filaments\footnote{The definition of the HFS structure is required to have at least three hub-composing filaments. This would allow to get rule out of the possibility of a single filament appearing as an HFS structure due to the kink by external pressures instead of the merge process by several filaments.}). 
Cloud\,2 has 7 individual filaments and 2 HFSs. Cloud\,3 hosts 4 individual filaments and 1 HFSs. Cloud\,4 consists of 15 individual filaments and 4 HFSs. Cloud\,5 shows 3 individual filaments.  Note that there could be a few filaments that were not identified due to our adopted strict criteria (e.g., 5 rms as the minimum threshold) for a robustness. This incompleteness caused by a few unidentified filaments, however, does not affect our following analysis. In addition, we find that all HFSs identified here are included in the catalogue of filaments by \cite{Li2016} using ATLASGAL 870\,\micron\  dust emission observed at a similar angular resolution but with a lower sensitivity (i.e., 50-70\,mJy~beam$^{-1}$). In all, except for more detection of fainter structures in our observations, the dense filament detection has a good correspondence between the two different observations.

\subsubsection{Clump extraction}
We created the clump catalog using the source extraction algorithm \texttt{SExtractor} \citep{SExtractor1996}, an aperture photometry tool widely used to detect compact objects \citep[e.g.,][]{csc2013,csc2014}.  The 850\,\micron\ map and the noise {\it rms} map serve as input to the \texttt{SExtractor}, with a detection threshold of five times the local {\it rms}. The total flux of each object is measured from the 850\,\micron\ map  adopting the ''Bijection" paradigm  \citep[see Fig.\,4 of][]{Rosolowsky2008} using ellipse aperture photometry. The final uncertainty 
in the measured flux arises from the combination in quadrature of the local noise  with the systematic flux uncertainty \citep[e.g.,][]{Thompson2006,csc2013}. As a result, 350 compact density structures are extracted, which are hereafter called clumps. Note that a few clumps appear not be extracted, for example in the northernmost of the observing region, which could be either due to the low intensity contrast (i.e., $<5$ signal-to-nosie ratio) with the local rms level or due to their sizes smaller than the beam size accessible to 850\,um observations.
The latter possibility could be due to potential artefacts of continuum map, and thus those associated sources are not considered in our clump catalogue analysis. 

It is necessary to investigate the flux completeness level of our catalogued clumps given the limitation of the observing sensitivity. To this end,  we randomly inserted simulated clumps into the 850\,\micron\ noise map \citep[e.g.,][]{csc2013,YangAY2023}. We simulated 5000 clumps for each FoV analysis, with the input flux densities ranging from 1 to 100\,mJy~beam$^{-1}$ (for reference, 1\,{\it rms}=10\,mJy~beam$^{-1}$ overall). Here, three individual FoVs (see Fig.\,\ref{fig:rms}), each representing different observations, and their sum as the entire FoV were considered, each having slightly different rms distributions. Subsequently, the same Sextractor algorithm was applied to the simulated map. Fig.\,\ref{fig:completeness} illustrates the detection fraction of the simulated clumps as a function of peak flux density. As a result, the 95$\%$ completeness levels are $\sim 6\,rms$ for FoVs\,1 and 3, $\sim 7\,rms$ for FoV\,2, and approximately $\sim 6.5\,rms$ ($N_{\rm H_{2}}\approx4.3\times10^{21}~$cm$^{-2}$) for the entire FoV. This suggests that the slight difference in different FoVs does not significantly affect the detection of clumps. In addition, the 5\,rms level adopted in practice in our clump extraction corresponds to a global flux completeness level of $\sim 90\%$ for the entire FoV.

The measured parameters of the identified 350 clumps are listed in Table\,\ref{table:clump properties}, including the position, peak flux and position angle. 
Fig.\,\ref{fig:clump_filament} shows the identified clumps overlaid on the 850\,\micron\ continuum. 
The majority of the detected clumps are located in Cloud\,1 and Cloud\,4, i.e., 141 and 123 sources, respectively. The number of extracted clumps is similar in Cloud\,2 and Cloud\,5, with 33 and 36, respectively. The remaining 18 clumps are distributed in Cloud\,3. 
The identified clumps include all 37 ATLASGAL clumps identified by \cite{csc2014} using the same \texttt{SExtractor} algorithm. From the present analysis, we infer that both high sensitivity and angular resolution of the ALOHA dataset allows us to detect additional hundreds of clumps located in extended diffuse background. Furthermore, the measured fluxes of the ATLASGAL clumps from previous 870\,\micron\ and our 850\,\micron\ observations are comparable within quoted uncertainties (see Appendix\,A.4 and Fig.\,\ref{fig:fp}).

\subsection{Physical Properties of Filaments and Clumps.}
\subsubsection{Width and flux of filaments}
\begin{figure}
    \centering
    \includegraphics[scale=0.36]{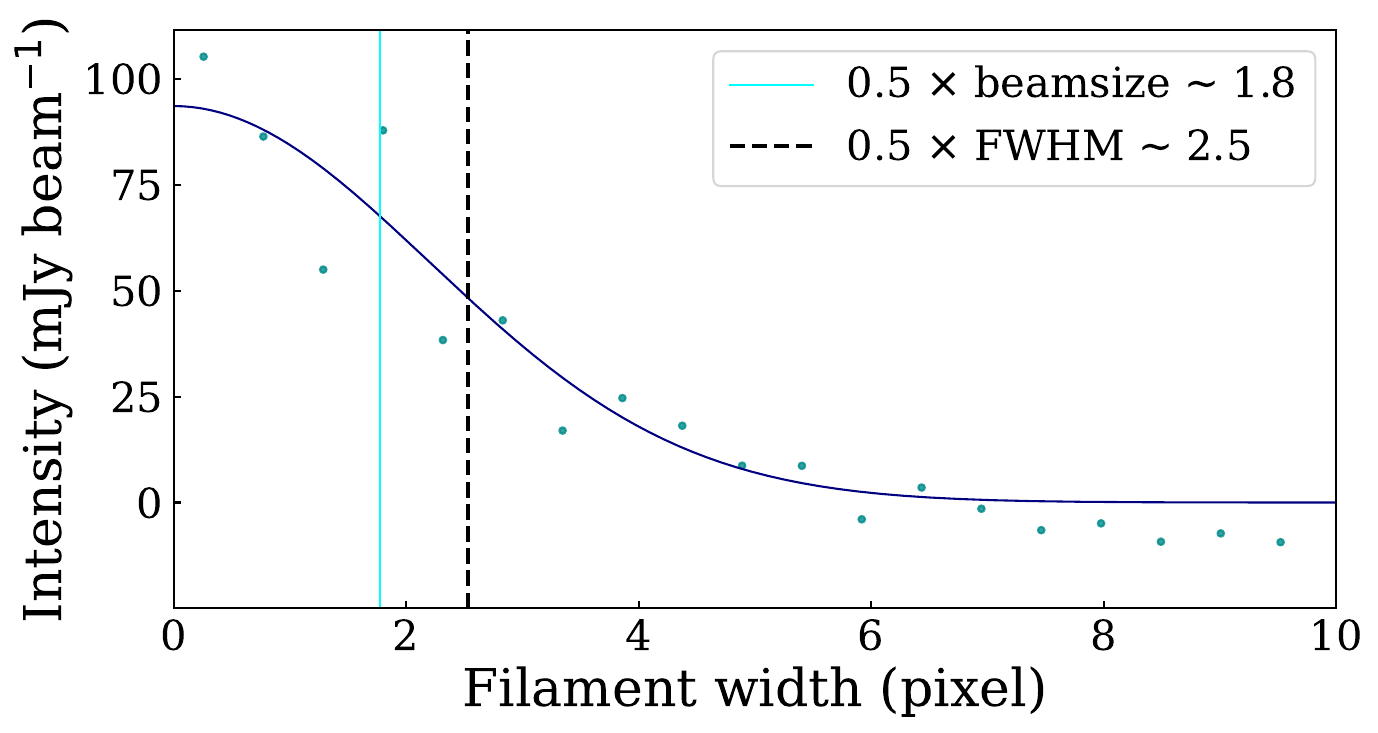}
    \caption{An example of the averaged radial profile from the \texttt{FILFINDER}. The solid dark blue line shows the profile fitted with a Gaussian curve. Both 0.5\,$\times$\,beam size and 0.5\,$\times$\,FWHM are presented in the plot.}
    \label{fig:width}
\end{figure}

Although the \texttt{FILFINDER} algorithm  gives  the positions and lengths of filament skeletons, the other resulting parameters (e.g., filament width) may be biased.
This is because the averaged radial intensity profiles across the filamentary skeletons are not always characterized by a Gaussian/Plummer-like function, which is usually applied to derive the filament width. This situation becomes severe for the hub-composing filaments in HFSs. For the individual filaments, we can derive the averaged radial intensity profile of each skeleton using \texttt{ FilFINDER} and an example of the Gaussian profile is shown in Fig.\,\ref{fig:width}. However, in HFSs, some hub-composing filaments display significantly different irregular profiles.
In some cases, limitations in the number of sample points (i.e., short filaments or those between multiple connections such as in HFSs) may also prevent obtaining accurate profiles, leading to a poor determination of the filament width. To account for the above mentioned inaccuracies, for all hub-composing filaments in HFSs, we adopt a common filament average width. 

For the total flux at 850\,\micron\ within each filament, we integrate all 850\,\micron\ intensity within a curved rectangle defined by the filament width and its length, which was assumed to represent a 2D projection of each filament.  During the measurement of total flux, we did not perform large-scale background subtraction. This is because such subtraction have been carried out during our data-processing stage (as detailed in Sect.\,3.1). To address flux uncertainties arising from a fixed filament width, we measured the total flux by varying each filament's width by $\pm 1/2$ times, which will be discussed in Sect.\,4.4.2. Note that we did not partition the hub flux into specific filaments within the HFSs. Instead, we incorporated the hub flux into the overall flux measurement for each filament composing the hub.
The total integrated flux at 850\,\micron\ for each filament are listed in Col.\,5 of Table \ref{table:filament properties1}.

\subsubsection{Dust temperature, mass, and related properties of both filaments and clumps}
The mass of clumps and filaments was calculated following the gray-body radiation model (e.g., \citealt{mass1983,LiuH21}):
\begin{equation}
M=\frac{D^{2}S_{\nu}R_{gd}}{B_{\nu}(T_{\mathrm{dust}})\kappa_{\nu}},    
\end{equation}
where $D$ is the source distance, $S_{\nu}$ is the integrated flux of 850\,$\mu$m emission, $R_{\rm gd}$ is the gas-to-dust mass ratio taken as $R$=100, $B_{\nu}$ is the Planck function for a dust temperature $T_{\mathrm{dust}}$, and $\kappa_{\nu}$ is the dust absorption coefficient taken as $\kappa_{_{850\, \mu m}}=1.85$ $\mathrm{cm^2~g^{-1}}$.  This value was derived from the interpolation of the values listed in Col.\,5 of Table 1 in \cite{kv1994}, called the OH5 model \citep{kv2011}. Here, $T_{\mathrm{dust}}$ was estimated from the 18\arcsec-resolution temperature map (see Fig.\,\ref{fig:Tg} in Appendix), which presents a range of 14\,K to 48\,K with a median value of 18\,K. It was created {\bf using} 
the 160\,$\mu$m and 250\,$\mu$m images from the Hi-GAL survey, where the ratio of the two images serves as a temperature probe (see details in \citealt{Tdust2016, Pan23}). 

We estimated the average H$_2$ column density ($N_{\mathrm{H_2}}$) as follows:
\begin{equation}
N_{\mathrm{H_2}}=\frac{M}{A\mu_gm_H},    
\end{equation}
where $A$ is the measured area of density structures (clumps or filaments), $\mu_g$ is the molecular weight per hydrogen molecule taken as 2.8 \citep{mug2008}, and $m_H$ is the mass of an atomic hydrogen.

For clumps only, their mass surface densities ($\Sigma_{\mathrm{clump}}$) were computed using the following equation:
\begin{equation}
\Sigma_{\mathrm{clump}}=\frac{M_{\mathrm{clump}}}{\pi R_{\mathrm{clump}}^{2}},
\end{equation}
where $M_{\mathrm{clump}}$ and $R_{\mathrm{clump}}$ represent the mass and radius of clumps, respectively.
The derived parameters for the filaments and clumps are tabulated in Table\,\ref{table:filament properties1} and Table\,\ref{table:clump properties}, respectively.

The uncertainties of these properties come from various aspects. We first consider a kinematic distance uncertainty of 10\% which is directly linked to the uncertainties of length and width of the filaments or the radius of clumps. The dust opacity ($\kappa_{\nu}$) at the same frequency can exhibit variations across different clouds, ranging from diffuse to dense environments. This variability is observed in studies such as those by Martin (2012), Roy (2013), and Webb (2017).
Similarly, the gas-to-dust ratio ($R_{gd}$) is also seen to vary.
According to \citet{Sanhueza2017G28,Xu2024ASSEMBLE}, $\kappa_{\nu}$ may vary by approximately 28\%, conservatively estimated from the OH5 model. Additionally, $R_{gd}$ can deviate by 23\% relative to a reference value of $R_{gd} = 100$. To account for uncertainties stemming from these sources, we assumed a uniform distribution between extreme values. Employing the Monte Carlo technique with 10,000 random samples, we calculated the standard deviation as the final uncertainty in related parameter calculations.

For clumps, mass uncertainties range from 43$\%$ to 47$\%$, with a mean value of approximately 44$\%$. The column/surface density uncertainties span 38\% to 42\%, with a mean value around 39\%. In the case of filaments, the related uncertainties remain the same as for clumps. However, when accounting for potential variations in filament width (up to ±1/2 times, see Sect.\,4.4.1), total flux uncertainties for each filament range from 38\% to 77\%, resulting in higher mass uncertainties of 58\% to 91\% with a mean value of approximately 78\%.

\subsubsection{Statistical results of physical properties}
\begin{figure*}
    \centering
    \includegraphics[scale=0.38]{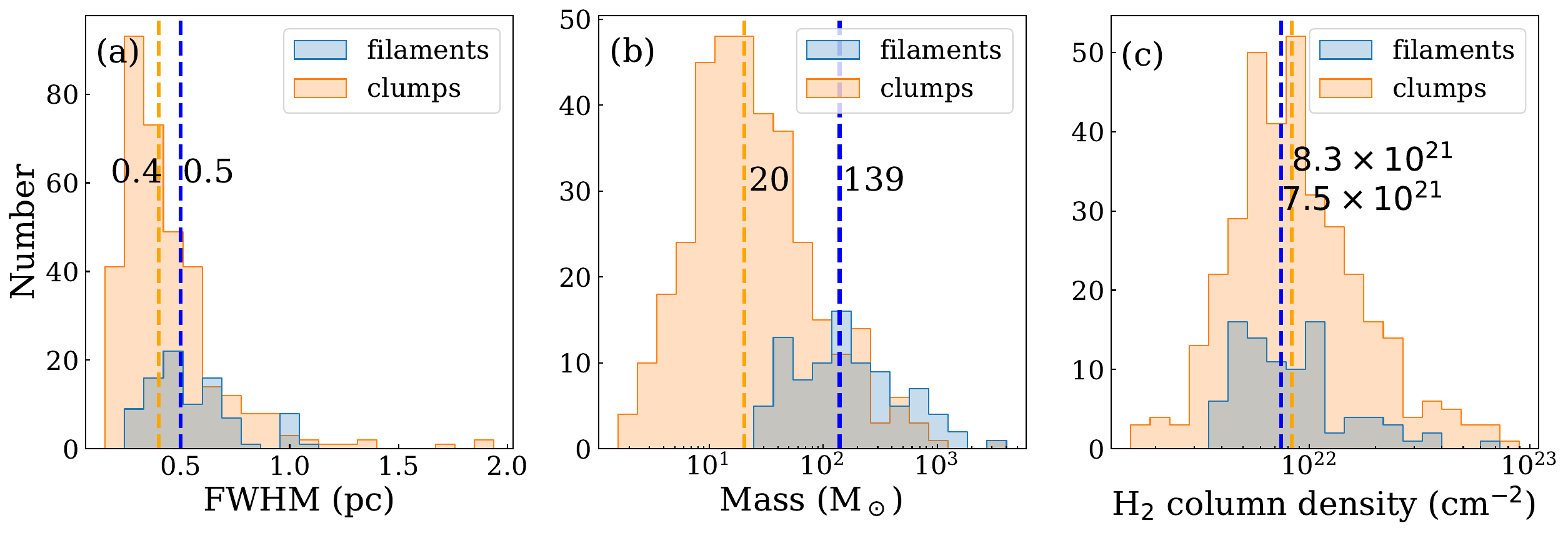}
    \caption{Distribution of parameters (i.e., FWHM, mass, column density) of filaments (blue) and clumps (orange). The corresponding median value is indicated with a vertical dashed line.} 
    \label{fig:combined_histograms1}
\end{figure*}
\begin{figure*}
    \centering
    \includegraphics[scale=0.45]{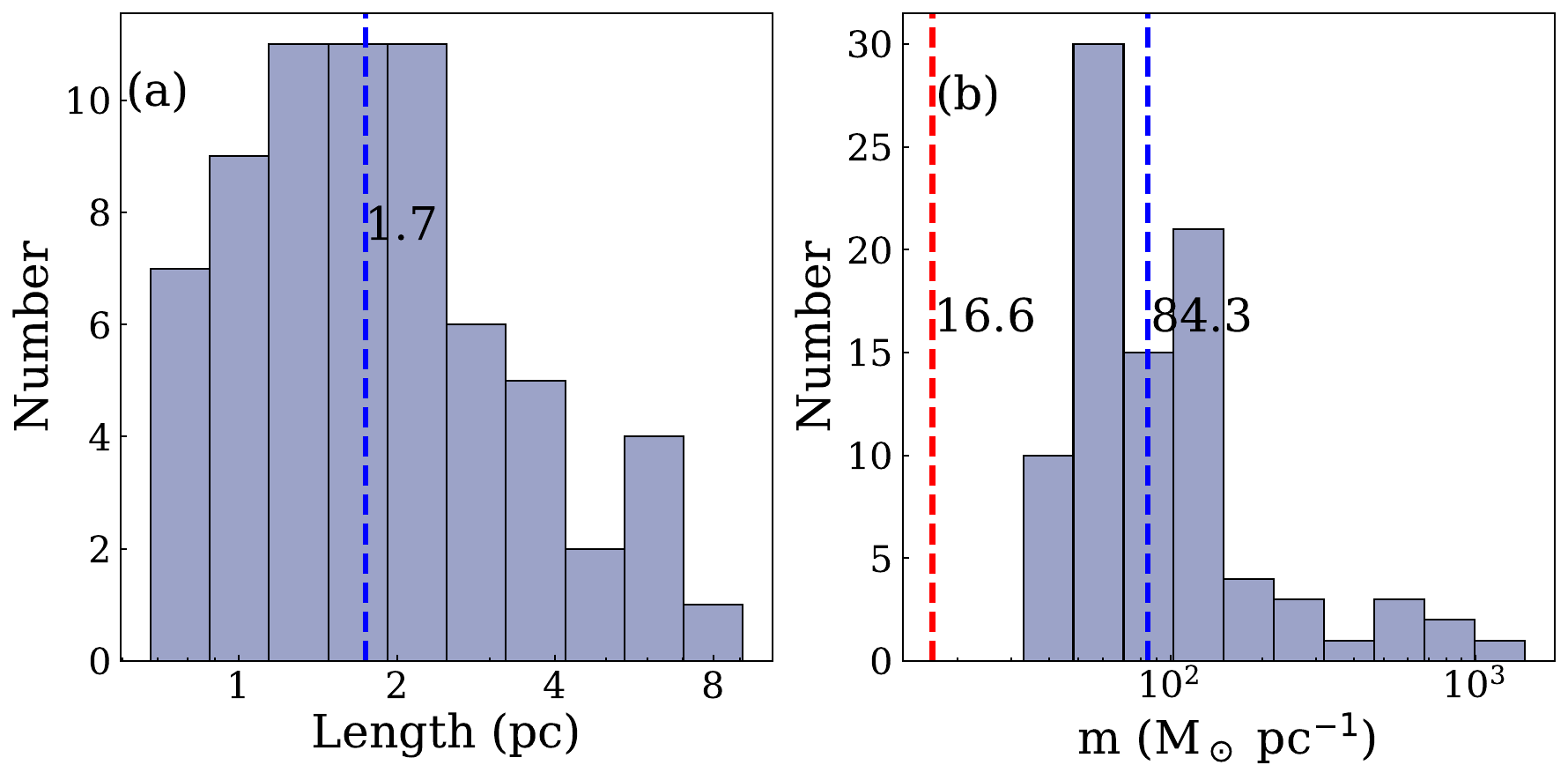}    
    \caption{Length and line mass distributions of filaments. The vertical blue line in both panels presents the median value, while the red line in panel\,b gives a critical line mass of 16.6\,$\mathrm{M_{\sun}~pc^{-1}}$.}
    \label{fig:histograms2}
\end{figure*}
Fig.\,\ref{fig:combined_histograms1} shows the summary statistics of the physical properties of filaments and clumps. Fig.\,\ref{fig:combined_histograms1}\,(a)  shows that widths of filaments have a median value of  $\sim 0.51$\,pc. 
The width of filament, considered as one of their fundamental properties, is subject to debate \citep[see reviews by][]{Andro2014, Hacar2023,Pineda2023}.
In contrast, the size distribution of clumps is wider, ranging from  $\sim 0.25$ to 2\,pc, with a median value of 0.37\,pc slightly smaller than the filament width. 

Fig.\,\ref{fig:combined_histograms1}\,(b) shows an evident difference between the distributions of the mass for filaments and clumps, with the peaks of the two types of samples clearly offset from each other. The filaments have masses of 26 to 2962\,$\mathrm{M_{\sun}}$ with a median value of 139\,$\mathrm{M_{\sun}}$, while the clump masses range from 2 to 4023 \,$\mathrm{M_{\sun}}$ with a median value of 20\,$\mathrm{M_{\sun}}$.

Fig.\,\ref{fig:combined_histograms1}\,(c) displays the distributions of the averaged column density for filaments and clumps. The two distributions have a similar dynamical range, with a similar median value of $\sim 8\,\times$~10$^{21}$\,cm$^{-2}$. 

Fig.\,\ref{fig:histograms2} provides the distributions of the length and line mass $m$ of filaments. The filament length ranges from 0.7 to 9.1\,pc, with a median value of 1.7\,pc. The line mass $m$ is defined as the ratio of the mass to the length of a filament in a unit of $\mathrm{M_{\sun}~pc^{-1}}$. Overall, the observed line masses fall within the range of the statistical results reported in previous Galactic surveys, as summarized in Table\,2 of \cite{Hacar2023} and references therein.

The observed line mass, in conjunction with the critical one $m_{crit}$, is a standard criterion for assessing the stability of the filament. 
$m_{crit}$ is 
16.6\,$\mathrm{M_{\sun}~pc^{-1}}$ for an ideal isothermal cylindrical model \citep{Ostriker1964}  at a kinematic temperature of 10\,K. As a result, the  observed line masses are greater than the theoretical value, indicating a dynamical state of being gravitationally bound. Furthermore, we investigated the turbulent contributions for the dense filaments using the NH$_3$(1-1) data with a beam size $\sim$\,32$\arcsec$ from the RAMPS survey (\citealt{RAMPS2018ApJS}, and see Sect.\,\ref{sec:NH3}). Due to the low detection rates of the RAMPS survey across the entire G35 complex (see Fig.\,\ref{fig:lw}), we only report the typical values in the following.  We find that IR-bright Cloud\,4 exhibits larger linewidths ($\sim$\,0.87\,km~s$^{-1}$) compared to the remaining relatively quiescent clouds ($\sim$\,0.49\,km~s$^{-1}$ for Clouds\,1, 2, 3, and 5). The critical line mass $m_{\mathrm{vir,crit}}$ including the contributions of the non-thermal motions \citep{Fiege2000}, also called virial line mass, was calculated as follows:
\begin{equation}
m_{\mathrm{vir,crit}}=\frac{2\sigma_v^2}G
\end{equation}
where $\sigma_v$ is the total velocity dispersion of the gas and {\it G} is the gravitational constant. 
Using the two typical linewidths estimated above in Eq. (4), we derived two critical line masses of 352\,$\mathrm{M_{\sun}~pc^{-1}}$ for the IR-bright cloud and 112\,$\mathrm{M_{\sun}~pc^{-1}}$ for the IR-quiet counterparts. The latter is comparable to the observed median line mass in Fig.\,\ref{fig:histograms2}, implying considerable turbulent contribution to the dynamical stability in early stages of filament evolution. However, in more evolved IR-bright environments, the turbulent contribution is observed to be higher by a factor of three, implying the potential role of external pressures in maintaining the dynamical stability of filaments. 

\subsubsection{Mass-length diagram }
\begin{figure}
    \centering
    \includegraphics[scale=0.246]{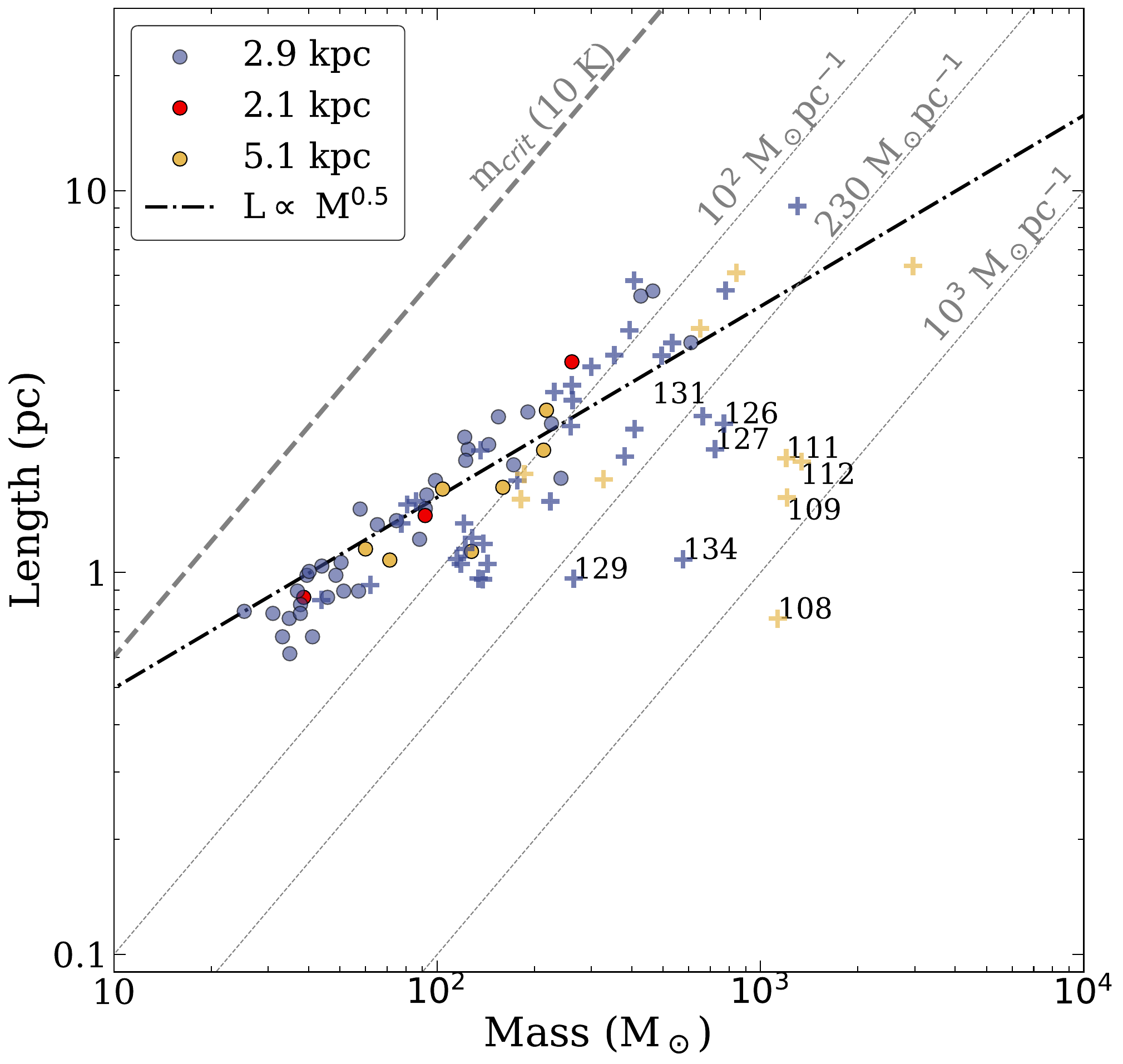}
    \caption{Mass-length diagram of individual filaments and hub-composing filaments within HFSs. The solid circles denote the individual filaments located at different distances, while the plus symbols mark the hub-composing filaments within HFSs. 
    Numbers indicate hub-composing filaments characterized  by notable deviations from the global trend (see text). The bold dash-dotted line presents the relation $\mathrm{L\propto~{M^{0.5}}}$ described by the simple model of \citet{Hacar2023}.
    The thick dashed line at the top of the diagram represents a typical critical line mass of 16.6\,$\mathrm{M_{\sun}~pc^{-1}}$, while the thick dashed-dotted line represents the global trend between mass and length. The light dashed lines represent different reference line masses. 
    }
    \label{fig:ML}
\end{figure}
\begin{table}
    \centering
    \caption{ Summary of correlation coefficients}
    \begin{tabular}{ccc}
      \hline
      \hline
       & entire samples & hub-composing filaments \\
      \hline
      $\rho_{12}$ & 0.837 & 0.637\\
      $\rho_{13}$ & 0.303 & 0.468\\
      $\rho_{23}$ & 0.135 & 0.115\\
      $\rho_{12,3}$ & \textbf{0.844} & \textbf{0.664}\\
      \hline
    \end{tabular}
    \vspace{10pt}
    \noindent\textbf{Notes.} \raggedright{“1”, “2”, “3” correspond to mass, length and distance, respectively. The notation $\rho_{xy}$ indicates the Spearman's rank correlation coefficient of each pairs, while $\rho_{12,3}$ indicates the partial correlation coefficient (highlighted in bold).}
    \label{tab:pct}
\end{table}
Fig.\,\ref{fig:ML} shows the mass-length diagram of filaments, including individual filaments (i.e., small solid circles) and hub-composing ones (i.e., plus symbols) within HFSs.
We find the populations of individual filaments and hub-composing filaments to be differently distributed, with the latter forming a wider spread in length and mass. 

In the above analysis, both parameters have distance dependence. To remove mutual dependence of mass and length on the distance, the first-order partial correlation test \citep[e.g.,][]{Urquhart2018,YangAY2018,Taniguchi2023} was adopted. The partial correlation coefficient, $\rho_{12,3}$, is as follows:
\begin{equation}
\rho_{12,3}=\frac{\rho_{12}-\rho_{13}\rho_{23}}{\sqrt{(1-\rho_{13}^{2})(1-\rho_{23}^{2})}}    
\end{equation}
Here, “1” and “2” denote the two dependent parameters (mass and length, respectively) that we prioritize, while “3” signifies the independent distance that may influence the aforementioned ones. $\rho_{xy}$ (i.e., $\rho_{12}$, $\rho_{13}$, $\rho_{13}$) represents the Spearman's rank correlation coefficient between pairs. Table\,\ref{tab:pct}  summarizes all of the correlation coefficients for the entire samples and hub-composing filaments alone.
As shown in Table\,\ref{tab:pct}, the correlation between mass and length is already strong in the entire sample ($\rho_{12,3}\,>$\,0.84 with $p-$value $\ll0.001$), while the hub-composing ones show relatively weaker correlations ($\rho_{12,3}\sim$\,0.66 with $p-$value $\ll0.001$).
These results suggest that the dispersion of the mass-length distribution can mostly be attributed to the presence of hub-composing filaments. Meanwhile, 
the consistency between the Spearman coefficient and the partial correlation coefficient suggests that the effect of distance on the distribution of mass and length is negligible. 

\section{Discussion}
\label{sec:discussion}

\subsection{Implication of mass-length scaling of filaments}
\label{sec:5.1}
The mass-length diagram presented in Fig.\,\ref{fig:ML} shows interesting trends. All the individual filaments (small solid circles) and most of the hub-composing filaments (plus symbols not numbered) are seen to approach a global trend (but with a somehow steeper slope), proposed by \citet{Hacar2023} as a power-law scaling relation of $\mathrm{L\propto~{M^{\alpha}}}~(\alpha=0.5)$. This scaling relation is similar to Larson's M\,$\propto~\mathrm{R^2}$ \citep{Larson1981,Solomon1987,Heyer2009} and could be attributed to turbulent fragmentation from clouds to filaments \citep{Tafalla2015,Smith2016,Hacar2023}.  In such a turbulent fragmentation scenario, filaments increase in mass and length as they undergo turbulence-dominated either longitudinal or radial accretion from their natal clouds. As the radial accretion progresses, the length could remain nearly steady, while the mass continues to increase, likely causing a slight deviation from the global scaling relation \citep[e.g.,][]{Feng2024}. This could account for the somewhat steeper trend observed in the global distribution of most filaments, compared to the proposed relation $L\propto~{M^{0.5}}$.

Moreover, a  number of the hub-composing filaments (numbered plus symbol) are seen to significantly deviate from the aforementioned power-law scaling. This deviation is reliable since even when considering the uncertainty in the mass of the filament, which could be up to $\sim 80\%$ (see Sect.\,4.4.2), we do not anticipate a significant deviation in mass, on the order of magnitude, when the length of the filaments is held constant. In addition,
similar deviation has been reported in the hub-composing filaments of nearby molecular clouds, such as Musca, L1495-B213 (Taurus), and Orion A (see \citealt{Hacar2023}). It is worth noting that these hub-composing filaments investigated here have high line masses greater than a threshold line mass of 230\,M$_{\sun}{~}\mathrm{pc}^{-1}$. 
This threshold indicates high-mass star formation as inferred by \cite{Li2016}, following Eq.\,12 of \citet{Fiege2000}, using a lower limit (0.7\,km s$^{-1}$) of line widths obtained from ammonia observations of massive star-forming clumps  \citep{Dunham2011,Urquhart2011,Wienen2012}. 

Furthermore, these numbered hub-composing filaments (see Fig.\,\ref{fig:ML}) are found {to be} associated with HII regions (see Col.\,12 of Table\,\ref{table:clump properties}). The dust temperatures of these filaments could be higher than those reported in Table\,B1. If the temperature were higher by a factor of two, the mass of the filament would be underestimated by the same factor. However, this would not significantly impact the evident deviation of these HFS filaments from the global trend observed in the mass-length distribution of most filaments. In light of this, these  hub-composing filaments could therefore be influenced by feedback from massive star formation, such as the expansion of HII regions and bubbles. 
This feedback, in conjunction with ambient turbulence, acts as external forces that can condense and break up the gas. Generally, this leads to formation of more thermally supercritical filaments that compose the hub, as discussed by \cite{Shimajiri2019}. Consequently, this causes the filaments to make a nearly vertical downward shift in the M-L plot.

It is important to note that the discussions we made above on the the M-L plot may be affected by observational biases. 
\cite{Schisano2020} found that the filament length measured from ground-based observations is generally 2-4 times shorter than those from space observations given a comparable filament mass. This difference, resulting from observational biases, could potentially explain why the HFS sample provided by space observations  does not deviate significantly from non-hub filaments in the M-L plot in \cite{Kumar2020}, as opposed to the result shown here in Fig.\,\ref{fig:ML}.

Additionally, we do not find a scaling relation (and hence not shown here) between filament density ($\langle n \rangle$) and L, similar to the third Larson’s relation.  \cite{Hacar2023} suggested that such a relation is equivalent to the assumption of constant column density for filaments, which does not hold for our samples. This result could be due to the fact that filament densities depend on the asymmetric structure of the cylinders, and thus the density estimate under the assumption of a spherical morphology could affect the resulting $\langle n \rangle$--L relation.  
\subsection{HFS clouds as a preferential site of HMSF}
\label{sec:5.2}
\begin{figure*}
    \centering
    \includegraphics[scale=0.287]{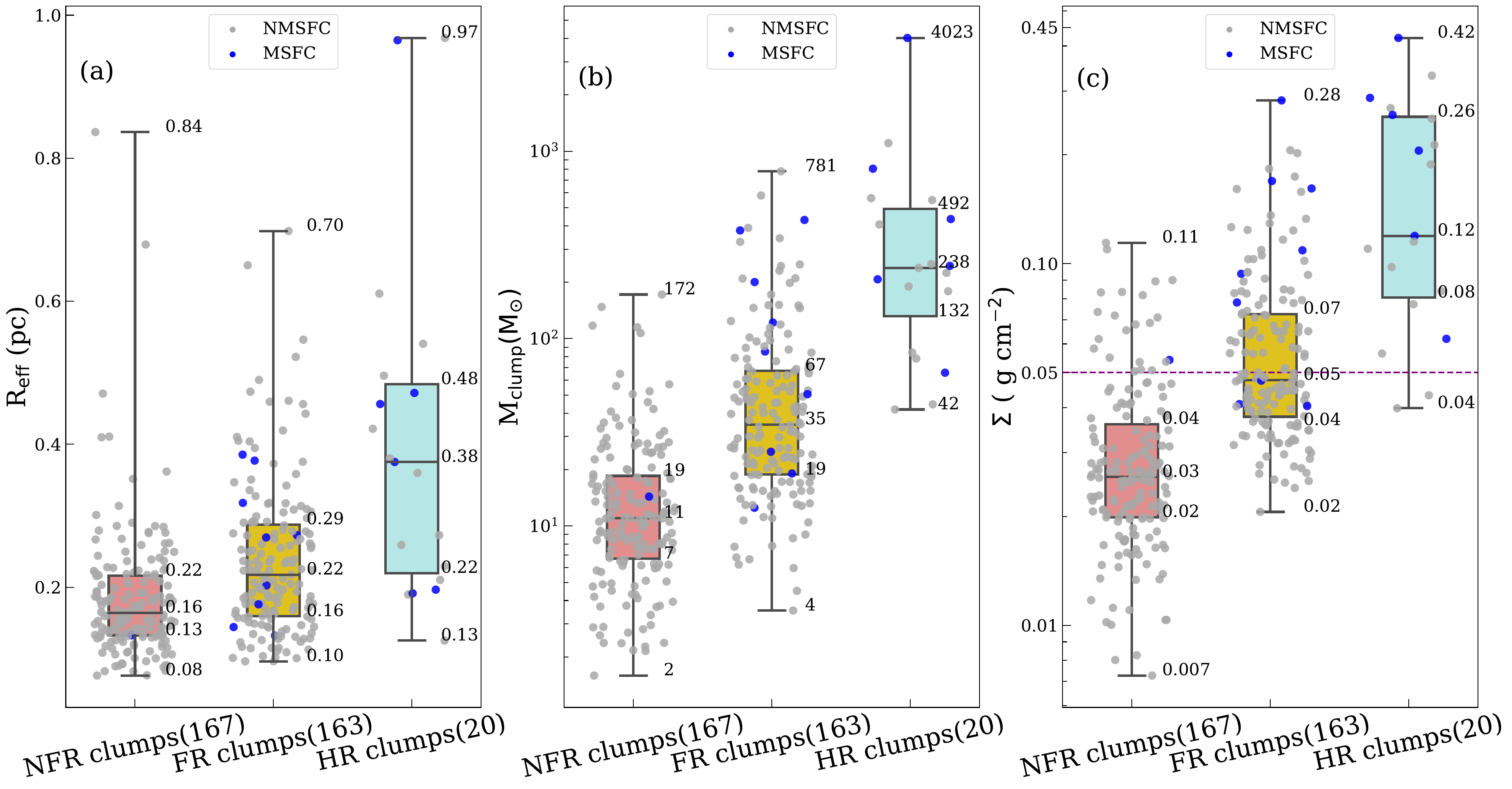}
    \caption{Box plots of the physical properties including effective radius (a), mass (b), and mass surface density (c) for three classes of clumps (i.e., NFR, FR, and HR clumps, see the text). The numbers within the parentheses in each classification indicate the quantity of clumps belonging to each class. The boxes represent the quartile range of the data, from the bottom to the top of each box, including the lower quartile (i.e., the 25th percentile), the median inside the box, and the upper quartile (i.e., the 75th percentile). The "whiskers" present the full extent of the data (i.e., minimum value and maximum value). The purple dashed line in panel (c) represents a mass surface density of 0.05\,g~cm$^{-2}$. Scatter plots show the concentration of data, while whiskers represent the extent of data distribution. Different colored scatters mark the MSF clumps (MSFC) and Non-MSF clumps (NMSFC). }
    \label{fig:box1}
\end{figure*}

\begin{table}
    \centering
    \caption{KWH test results of clump properties for three groups including NFR, FR and HR clumps}
    \begin{tabular}{cccc}
      \hline
      \hline
       Property group &  R & M & $\Sigma$\\
       & (pc) & (M$_{\sun}$) & (g~cm$^{-2})$\\
      \hline
     $p$-value  & 1.8 $\times 10^{-11}$ & 2.6 $\times 10^{-30}$ & 2.3 $\times 10^{-31}$\\
     $\chi^2$ & 49.5 & 136.3 & 141.1\\
      \hline
    \end{tabular}
    \noindent\textbf{Notes.} \raggedright{if $p-$value is smaller than the significance level, the null hypothesis will be rejected. A larger $\chi^2$ indicates a more significant difference between different groups.}
    \label{tab:table1}
\end{table}
Fig.\,\ref{fig:clump_filament} shows the spatial distribution of clumps along with their masses, and mass surface densities. In Fig.\,\ref{fig:clump_filament}\,(a), 75\% of massive clumps   (M\,$\geq20\,\mathrm{M_{\sun}}$) are located in the filaments, whereas low-mass clumps are distributed more widely.
The HFSs have many massive clumps at the hubs. For example, the longest one (i.e., HFS8 in Fig.\,\ref{fig:hfs}) contains massive clumps associated with MYSOs typical of high-mass protostars, as reported by the RMS survey \citep{MYSO2008}, and include a nascent stellar cluster G35.60-0.00 \citep{SPICY2021}. 

The mass surface density is another key parameter in star formation, especially for high-mass star formation. 
Thresholds for massive star formation, advocated by several authors \citep[e.g.,][]{Nature2008,Urquhart2014} lie in the range
from 0.05\,g $\mathrm{cm^{-2}}$ to 1.0\,g $\mathrm{cm^{-2}}$. As shown in Fig.\,\ref{fig:clump_filament}\,(b), the mass surface densities of clumps in hubs or ridges (i.e., IRDC G35.39 and 8\,\micron-bright ridge G35.6-0.0 ) are higher than the ATLASGAL clumps' lower limit of 0.05\,g~cm$^{-2}$\citep{Urquhart2014}. These observed trends are in good agreement with the previous observations which have characterised the essential roles of HFSs in massive star formation \citep[e.g.,][]{Schneider2012,Peretto2013,Kumar2020,Kumar2022,Arzoumanian2023}. 

Our observations of massive and dense clumps lying in filamentary structures or located at their junctions are qualitatively very similar to the previous results of MHD simulations (e.g., \citealt{Inoue2013,Gilberto2014,Chen2015,Gong2015,Inoue2018}). In these simulations, converging turbulent flows, colliding flows or cloud–cloud collisions are considered as the vital factors to produce the dense filaments and HFSs naturally. Meanwhile, the growth pattern of central clumps follows the "clump-fed" model \citep{clump-fed2009} that central massive clumps act as reservoirs of high-mass star formation in the HFSs, gaining mass through gas accretion from their large-scale environments such as filaments. Theoretically, the global hierarchical collapse scenario \citep{GHC2019} further emphasizes the pivotal role of filaments in molecular cloud evolution, accumulating material through radial accretion from ambient gas and feeding dense clumps through longitudinal contraction. 

For further analysis, we divide 350 clumps into three groups based on their location in filaments, in hubs or neither in both (see Fig.\,3 for example): 167 non filament-rooted (NFR), 163 filament-rooted (FR), and 20 hub-rooted (HR) clumps.
The numbers in each category could be influenced by the level of background subtraction achieved. However, this effect will not be further examined as such background subtraction was not specifically performed, except for the one incorporated during the data-processing stage (see Sect.\,3.1). Fig.\,\ref{fig:box1} reveals an increasing trend of the median value for all three physical parameters (i.e., clump radius, mass, mass surface density) from NFR, FR to HR clumps. Particularly, the median value of mass surface density increases by a factor of 4 from NFR, FR, to HR clumps, while the same for mass increases by a factor of 17.  
The observed trends persist even if we account for the uncertainty around 10\% for the radius of clumps, around 44\% for the mass, and around 39\% for the mass surface density (see Sect.\,4.4.2). In addition, similar increasing trends of clump radius/mass from the NFR clumps and FR clumps were also found in the study of cores in Orion \citep{Polychroni2013}. 
To evaluate such increasing trends further, we applied Kruskal-Wallis H test, which can infer the difference among the distributions of multiple independent samples from different populations.  
The null hypothesis usually assumes that the samples come from the same parent population. Table\,\ref{tab:table1} shows the $p$-values and $\chi^2$ of the Kruskal-Wallis H test for physical parameters of one group against others. Since the $p$-values of all cases are much lower than 0.05, we reject the null hypothesis, and conclude that all three classification groups of clumps have different distributions, in terms of the radius, mass, and mass surface density parameters. 
Therefore, the observed increasing trends of these three parameters from NFR, FR to HR clumps could be significant. Accordingly, these trends underscore the tendency for massive dense clumps to associate with filaments and HFSs. Particularly, the densest HR clumps offer conducive conditions for high-mass star formation, indicating HFSs as preferential sites for such star formation.

We also used different tracers from previous catalogs, such as class II methanol masers, MYSOs, and ultracompact (UC) HII regions, to match our clumps with signposts of massive star formation. As a result, we identified 16 such massive star-forming (MSF) clumps. Among them, 15  preferentially reside in filaments or in the hubs within HFSs (see Fig.\,\ref{fig:box1}).
Although the counts of high-mass star-forming clumps are similar between the HFS's hubs and the isolated filaments, the former has a much higher fraction of the observed population than the latter does. This quantitative analysis favors the importance of filaments in high-mass star formation, and even the HFS clouds as a preferential site of HMSF.

\subsection{Evolution of clump density}
\label{sec:5.3}
\begin{figure*}
    \centering
    \includegraphics[scale=0.4]{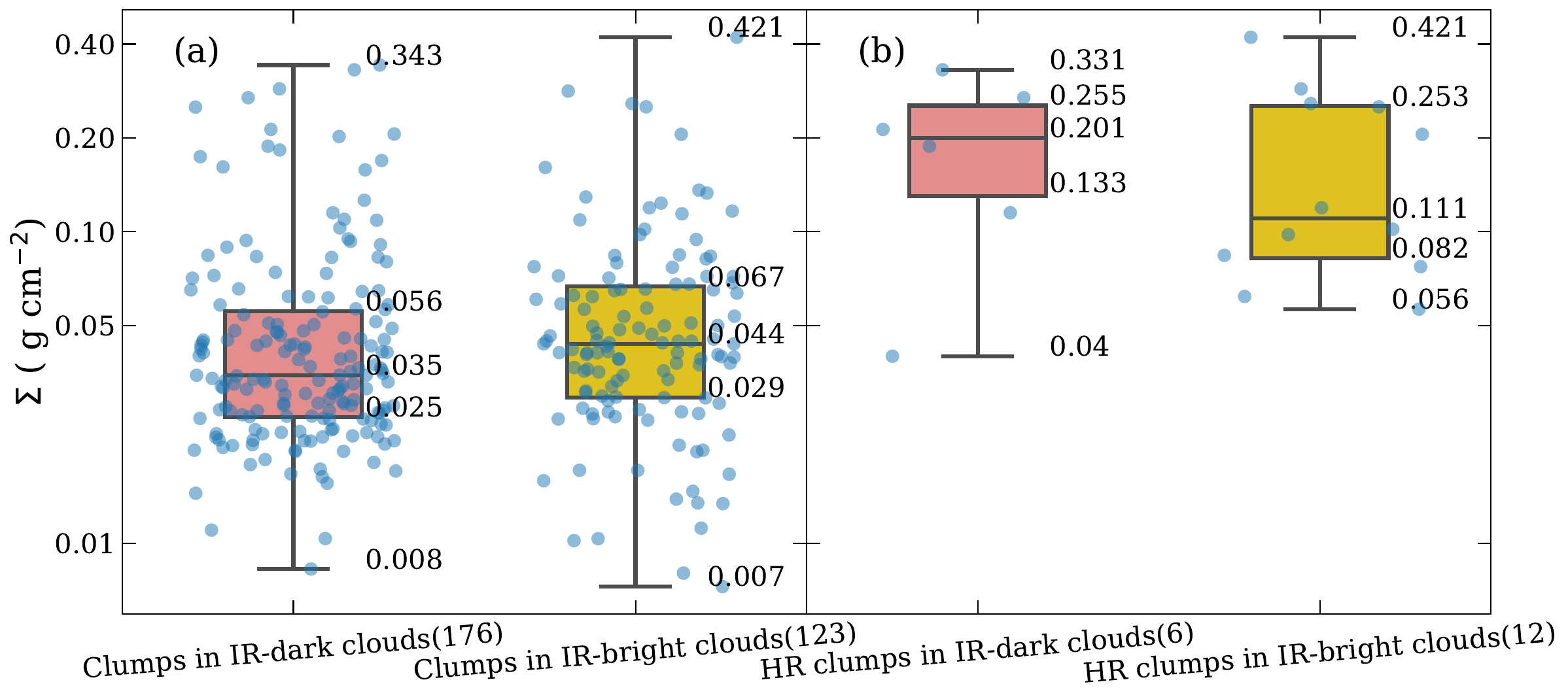}
    \caption{ Box plots for comparison of  mass surface density between clumps in IR-dark environment (i.e., in Cloud\,1 and/or  Cloud\,2) and those in IR-bright environment (i.e., in Cloud\,4). (a) Comparison between clumps in IR-dark  Cloud\,1 \& 2 and those in IR-bright Cloud\,4. 
    (b) Distribution of mass surface density of HR clumps (i.e., located at hubs) in both IR-dark  (i.e., in Cloud\,1 \& 2) and IR-bright (i.e., in Cloud\,4) environments. The numbers in parentheses of the horizontal axis label represent the sample counts in each group.
    }
    \label{fig:DBboxplot}
\end{figure*}

We select the distance-independent parameter of clump mass surface density to examine how critical parameters of HMSF evolve in different star-forming environments. For this analysis, we focus on clouds with evident infrared-dark (IR-dark) and infrared-bright (IR-bright) signatures. The presence of IRDCs (i.e., G35.39 and G35.41) suggest quiescent star formation in Clouds\,1 and 2. In comparison, the detection of several HII regions \citep{HIIcat2014} in Cloud\,4, which are seen as IR-bright signatures, indicates active star formation activity, and a relatively later evolutionary phase (see Sect.\,\ref{sec:presentation}). Given the weak 850\,\micron\ emission in Cloud\,3 and Cloud\,5 relative to other clouds, these have been excluded from the discussion here.

Fig.\,\ref{fig:DBboxplot}\,(a) shows the comparison of the densities of the clumps located at two different evolutionary stages of clouds at different distances (IR-dark Cloud\,1 \& 4 v.s. IR-bright Cloud\,4). 
As seen in Fig.\,\ref{fig:DBboxplot}\,(a),  
the mass surface densities of clumps are slightly higher in the IR-bright clouds than in the IR-dark counterparts with a median ratio of 1.3. We conduct the two-sample Kolmogorov-Smirnov (KS) test to investigate if they are statistically  identical. The statistical results, with yielding (a) in a $p-$value $\sim$\,0.005, 
suggest that the samples are unlikely to be drawn from the same parent population.  The weak trend implies a possible increasing trend for the mass surface density of clumps from the IR-dark to IR-bright stage. 
In their statistical study of a sample of 17 HFS clouds using ALMA data, \citet{LiuHL2023} have observed a similar increasing trend (by a factor of around 3) in the central massive clumps from IR-dark to IR-bright HFSs. Several other systematic studies \citep[e.g,][]{Giannetti2013,Urquhart2014,Rigby2021GASTON,Xu2024ASSEMBLE} also conjecture a similar positive correlation of density as a function of evolutionary sequence. It is also worth noting that there are other studies \citep[e.g.,][]{Lopez2010,Rathborne2010,Sanchez2013}, where no significant differences in column densities (equivalent to surface densities) were observed for clumps.
Furthermore, clumps in the HII region phase show an even greater difference in surface densities compared to the other stages \citep[e.g.,][]{Urquhart2014,He2015}. On the contrary, \cite{Guzman2015} presented results of 3000 massive MALT90 clumps in disagreement with the previous enhanced densities in the group of HII regions. Their study shows that the surface density of clumps tends to decrease during the development of HII regions. 
Note that the observed trend should be treated with caution, given the estimated uncertainties (see Sect.\,4.4.2 and 5.2).

If we consider the HR clumps depicted in Fig.\,\ref{fig:DBboxplot}\,(b), these massive structures, rooted in the central hub of HFS clouds, exhibit a decreasing surface density trend from an IR-dark to an IR-bright stage, contrasting with the previously mentioned weak increasing trend. A two-sample KS test also conducted on the two groups yielded a $p$-value of 0.76, suggesting a common origin. Consequently, the observed decreasing trend should be interpreted with caution. 
Under the framework of the latest theoretical models like GHC and I2, high-mass star formation in HFSs involves a multiscale mass accretion/transfer process from the cloud to filaments and then onto the central hub and finally to the star-forming cores. Thus, clumps located in the central hubs would be in a state of active accretion, increasing in mass and surface mass density.
However, for HR clumps in IR-bright clouds, 11 out of the 12 clumps are associated with HII regions. In this phase, ionizing feedback and radiation pressure from MYSOs would play an important role in 
influencing the temporal trend of mass surface density in these clumps. Given the critical role of this trend in understanding mass accretion and 
high-mass star formation, further dedicated observational studies are warranted.

\section{Conclusions}\label{sec:conclusions}
In this paper, we have carried out a systematic analysis on the filaments and their hierarchically-fragmented clumps in the G35 molecular complex, centered at $\alpha_{2000}=18^{\mathrm{h}}56^{\mathrm{m}}28^{s}.16$, $\delta_{2000}=2^{\circ}14^{\prime}25.^{\prime\prime}71$, with an approximate radius of $0.41^{\circ}$ from SCUBA-2 850 $\micron$ continuum observations. We focus on the basic physical properties of the filaments and clumps, including the mass, density, and size. The major results are as follows: 

(1) Five clouds, namely Cloud\,1--5, were identified. Their respective kinematic distances, estimated from the $^{13}$CO~(1--0) line data, are 2.9\,kpc, 5.1\,kpc, 2.9\,kpc, 2.9\,kpc, and 2.1\,kpc.

(2) We extracted a sample of 91 filaments, some of which can be divided into 10 HFSs, each composed of  at least 3 hub-composing filaments. A catalogue of 350 dense clumps was compiled, 183 out of which are associated with the filaments.

(3) The identified filaments have a median width of 0.51\,pc, lengths ranging from 0.7\,pc to 9.1\,pc, and masses between $\sim$ 26-2962 M$_{\sun}$. All exceed the thermally critical line mass of 16.6\,$\mathrm{M_{\sun}~pc^{-1}}$, with a median value of 84\,$\mathrm{M_{\sun}~pc^{-1}}$. The clumps, with a median size of $\sim$ 0.4\,pc and mass of $\sim$ 20 M$_{\sun}$, have a similar average column density to the filaments, suggesting a density inheritance from larger-scale filaments.

(4) The global mass-length trend of filaments, close to $\mathrm{L\propto~{M^{0.5}}}$, suggests that their physical origin is linked to turbulence. For those hub-composing filaments within HFSs, they deviate from the global mass-length trend, which could be due to feedback from massive star formation therein, particularly HII regions.

(5) Massive clumps, which most likely form high-mass stars,
are the densest in filaments and  in the hubs of HFSs with the latter bearing a higher probability of occurrence of high-mass star-forming signatures, favoring the HFSs as a preferential site of high-mass star formation.

(6) We examined the variation in clumps' mass surface density relative to their host cloud environment, from the IR-dark to IR-bright stage. No significant variation was observed. This could be attributed to the regulation of clump properties, such as mass surface density, by the interplay between mass accretion and feedback from HII regions.

Overall, we have provided filament and dense clump samples in the G35 molecular complex to explore their connection with star formation. The significance of filaments, especially hub-composing filaments in HFSs, for high-mass star formation is highlighted. Further observational studies particularly on kinematics and dynamics are needed to understand the mass accretion process in high-mass star formation through these density structures.   

\begin{acknowledgments}

This work has been supported by the National Key R\&D Program of China (No.\,2022YFA1603101). X.-J. Shen is supported by the 15th Graduate Student Research and Innovation Project of Yunnan University (Project ID:KC-23233964). H.-L. Liu is supported by National Natural Science Foundation of China (NSFC) through the grant No.\,12103045, by Yunnan Fundamental Research Project (grant No.\,202301AT070118, 202401AS070121), and by Xingdian Talent Support Plan--Youth Project. 
H.B.L. is supported by the National Science and Technology Council (NSTC) of Taiwan (Grant Nos.\,111-2112-M-110-022-MY3). PS was partially supported by a Grant-in-Aid for Scientific Research (KAKENHI Number JP22H01271 and JP23H01221) of JSPS. The work of M.G.R. is supported by NOIRLab, which is managed by the Association of Universities for Research in Astronomy (AURA) under a cooperative agreement with the National Science Foundation. K.T. was supported by JSPS KAKENHI (Grant Number JP20H05645).
\end{acknowledgments}

\software{Starlink \citep{starlink2014},
          Astropy \citep{astropy2013},  
          FILFINDER \citep{FilFinder2015},
          Aplpy \citep{aplpy2017},
          Source Extractor \citep{SExtractor1996,Barbary2016}
          }

\vspace{5mm}

\appendix
\section{Complementary figures}

\subsection{Noise map analysis of 850\,\micron\ dust observations}
Fig.\,\ref{fig:rms} displays the spatial distribution of the noise map (left panel) and the associated statistics (right panel).
\setcounter{figure}{0}
\renewcommand{\thefigure}{A\arabic{figure}}
\begin{figure}
    \centering
    \includegraphics[scale=0.35]{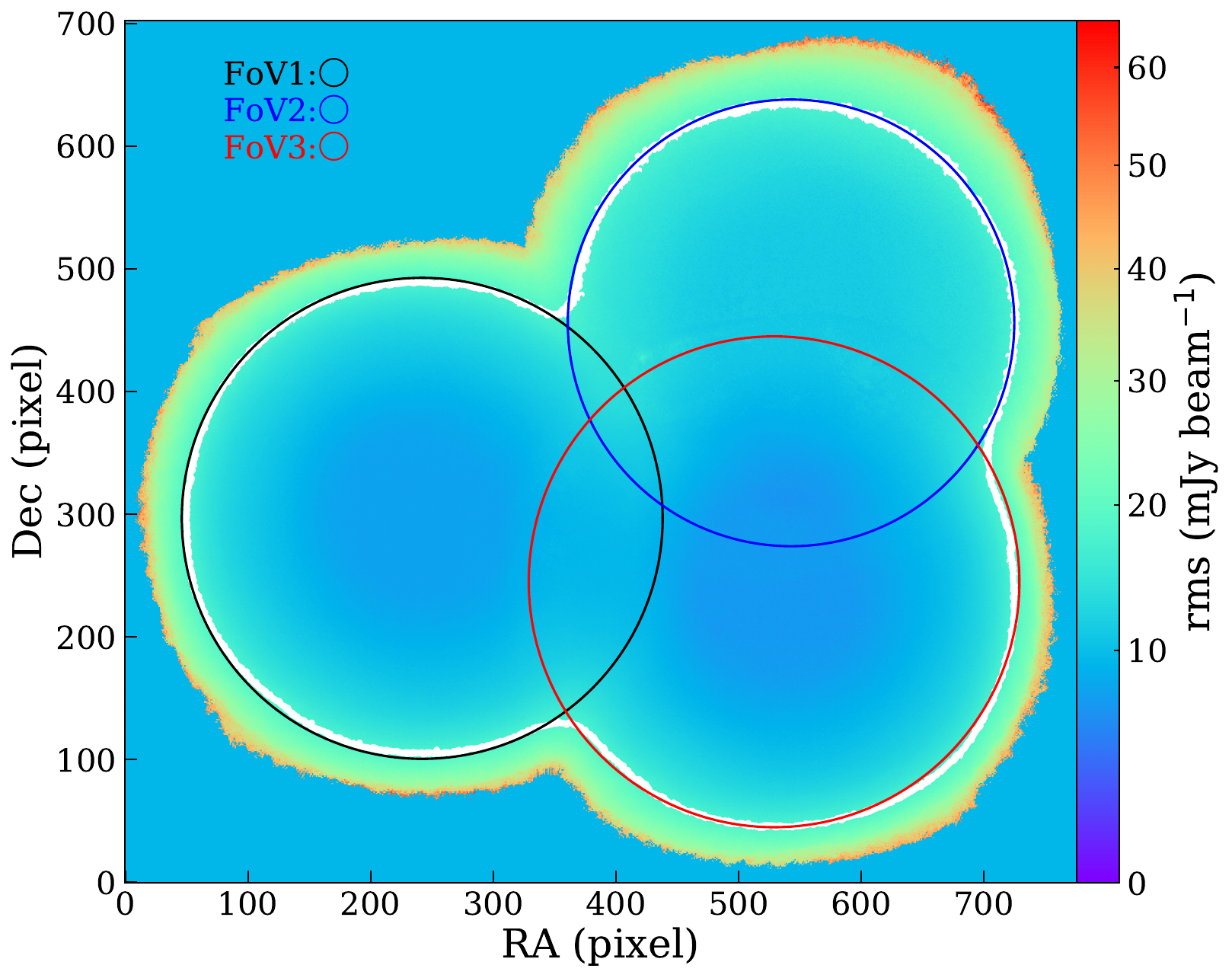}
    \includegraphics[scale=0.35]{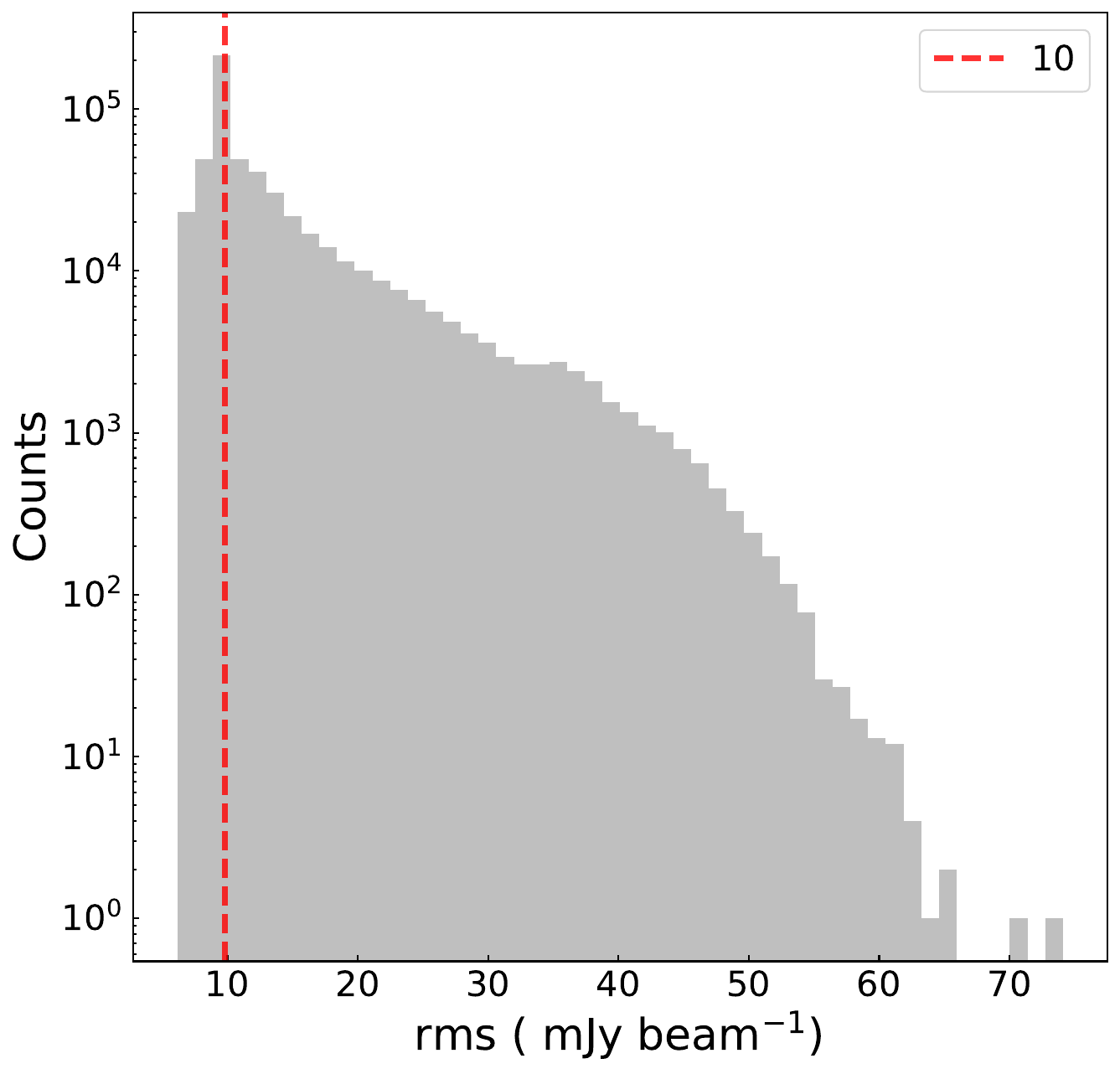}
    \caption{Noise map and statistical histogram of the {\it rms}. {\it Left:} the color map present the {\it rms} varying from 5 to 75\,mJy~beam$^{-1}$ and the white contour indicates {\it rms} value of 20\,mJy~beam$^{-1}$. The colored circles pinpoint three FoVs considered for analysis.  {\it Right:} the histogram shows the distribution of the {\it rms} with a median value of 10\,mJy~beam$^{-1}$ (red dashed line).}
    \label{fig:rms}
\end{figure}

\subsection{Molecular gas distribution in the G35 complex}
To approximately identify each cloud in the G35 molecular complex, we provide in Fig.\,\ref{fig:sp} the average spectrum of $^{13}$CO ({\it J=1--0}) over the entire complex, and in Fig.\,\ref{fig:int} the JCMT 850\,\micron\ continuum image overlaid with the intensity contours of $^{13}$CO for each individual cloud.

\setcounter{figure}{1}
\renewcommand{\thefigure}{A\arabic{figure}}
\begin{figure}
    \centering
    \includegraphics[scale=0.33]{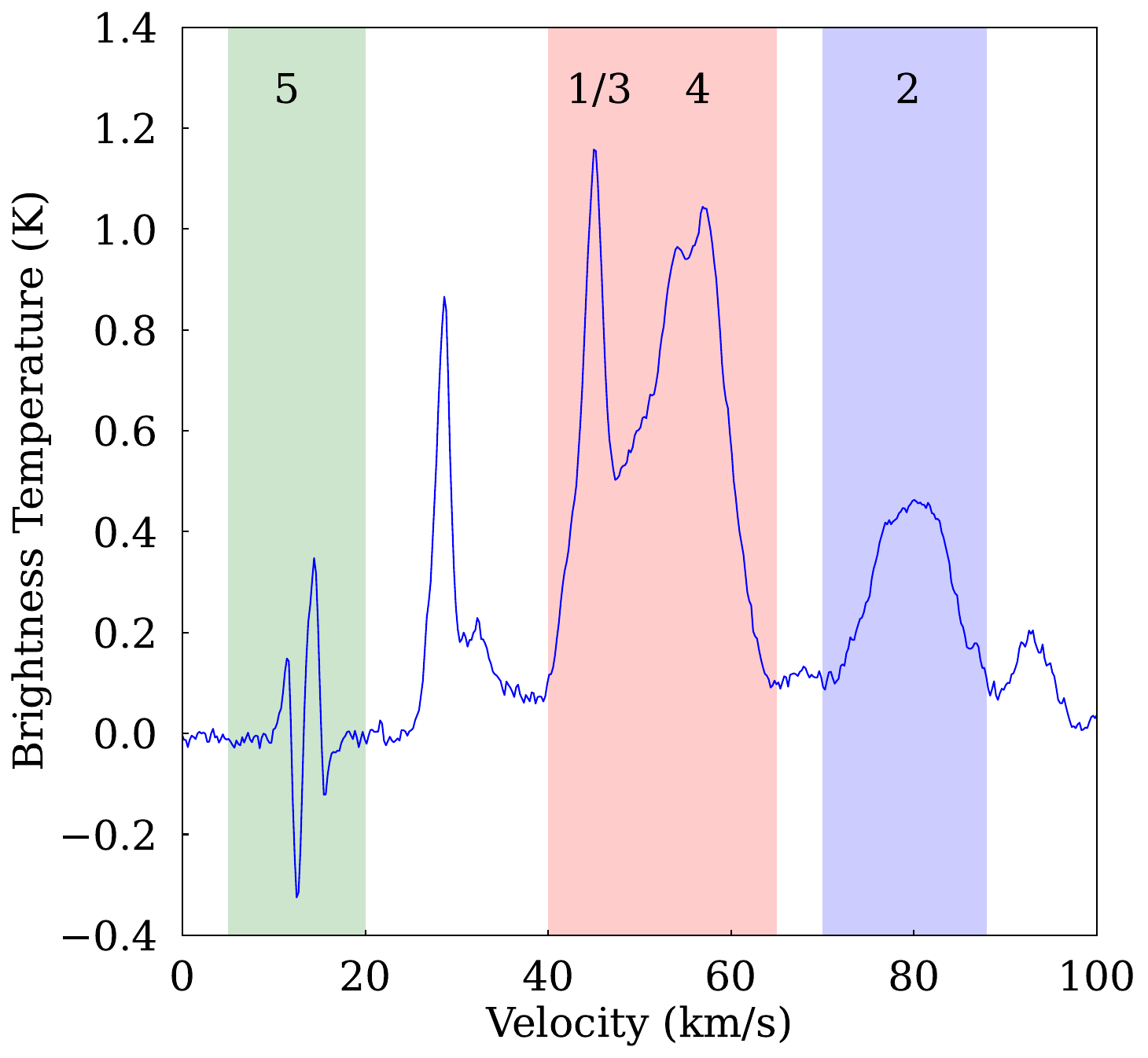}
    \caption{Average spectrum of $^{13}$CO~(1--0) emission over the G35 molecular complex. The color bands highlight three major velocity components, corresponding to [5, 20], [40, 65], and [70, 85] km~s${^{-1}}$ in green, red, and blue, respectively. The numbers represent the cloud IDs.}
    \label{fig:sp}
\end{figure}

\setcounter{figure}{2}
\renewcommand{\thefigure}{A\arabic{figure}}
\begin{figure*}[htbp]
    \centering
    \includegraphics[scale=0.35]{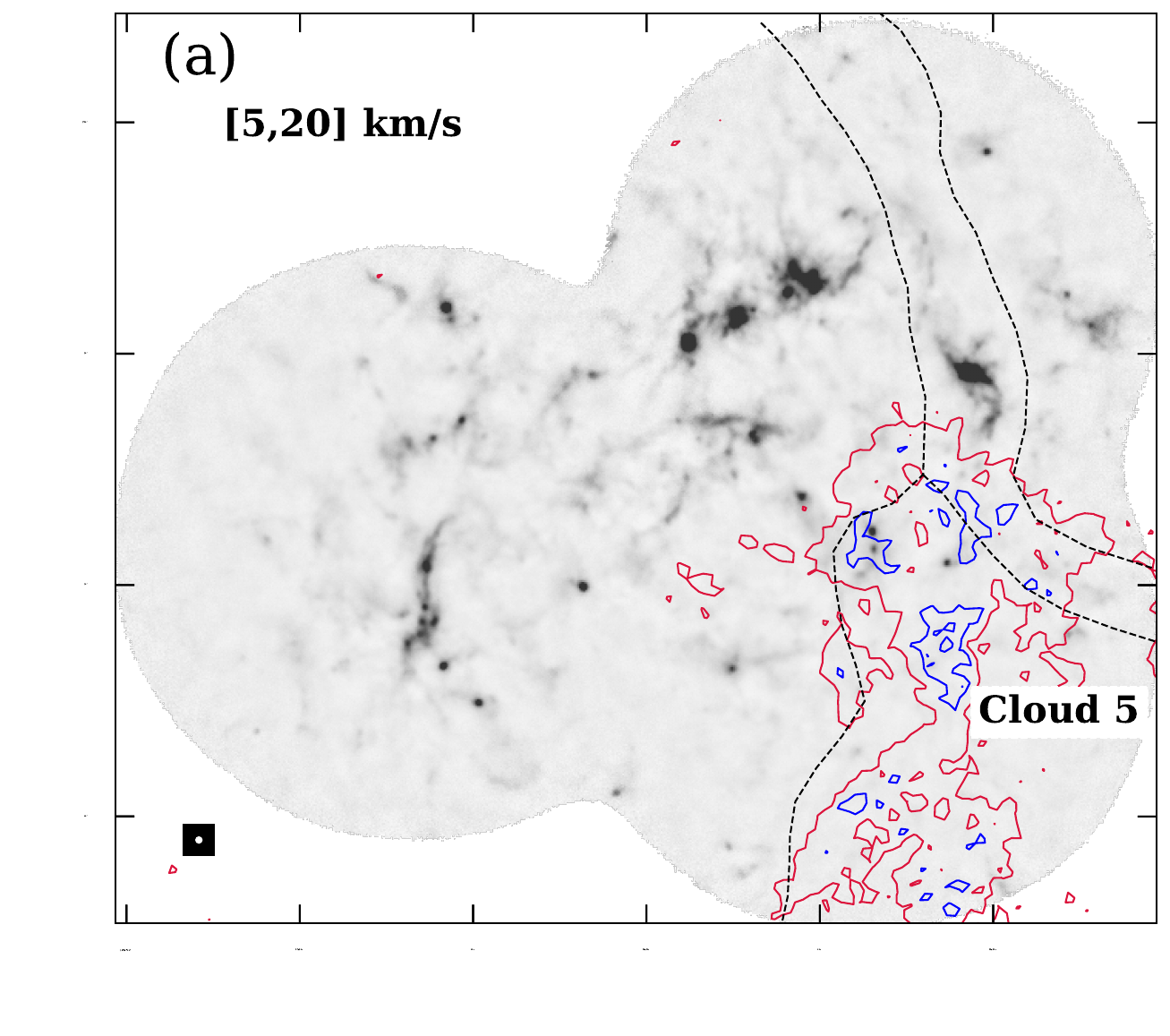}
    \includegraphics[scale=0.35]{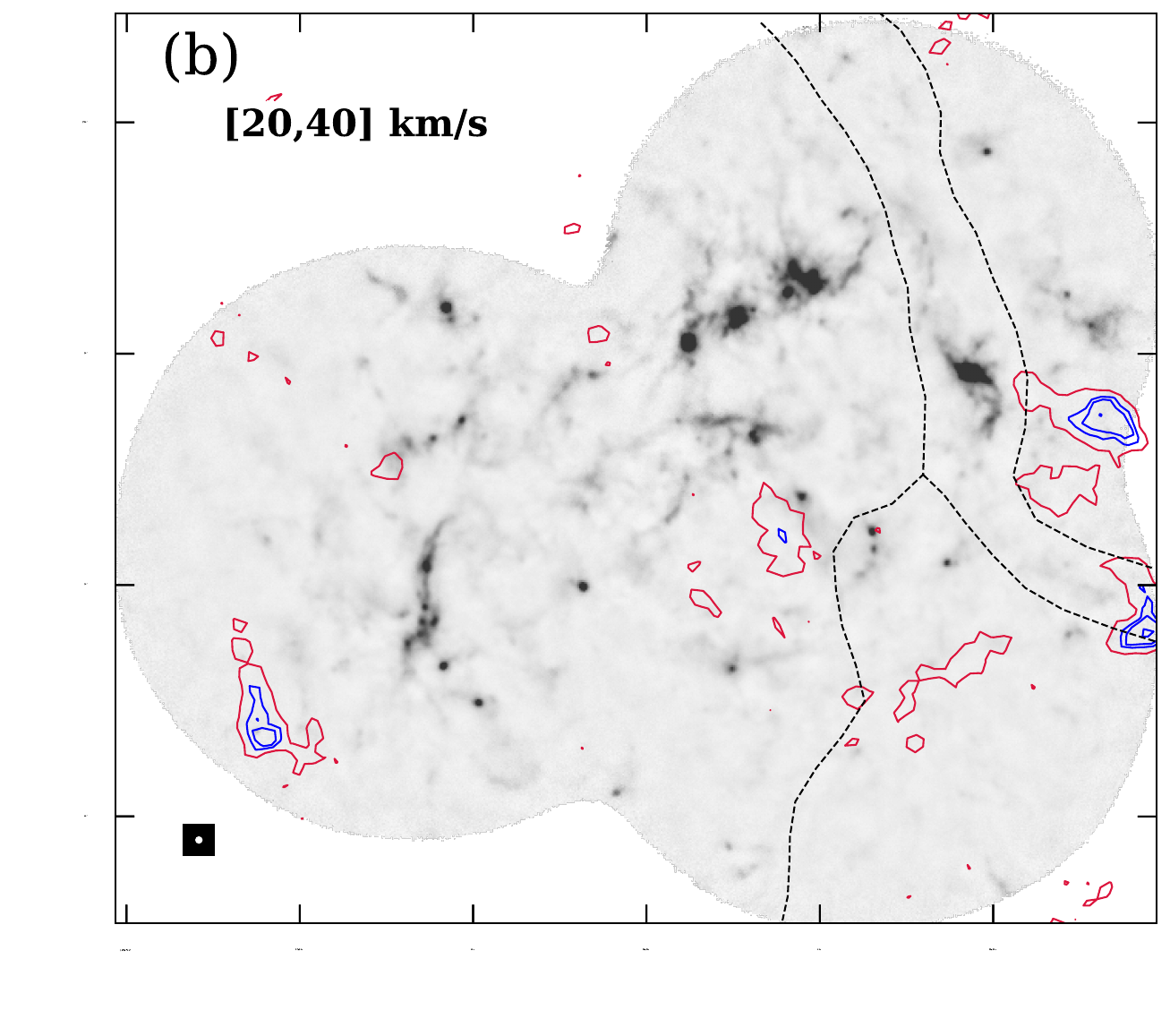}
    \includegraphics[scale=0.35]{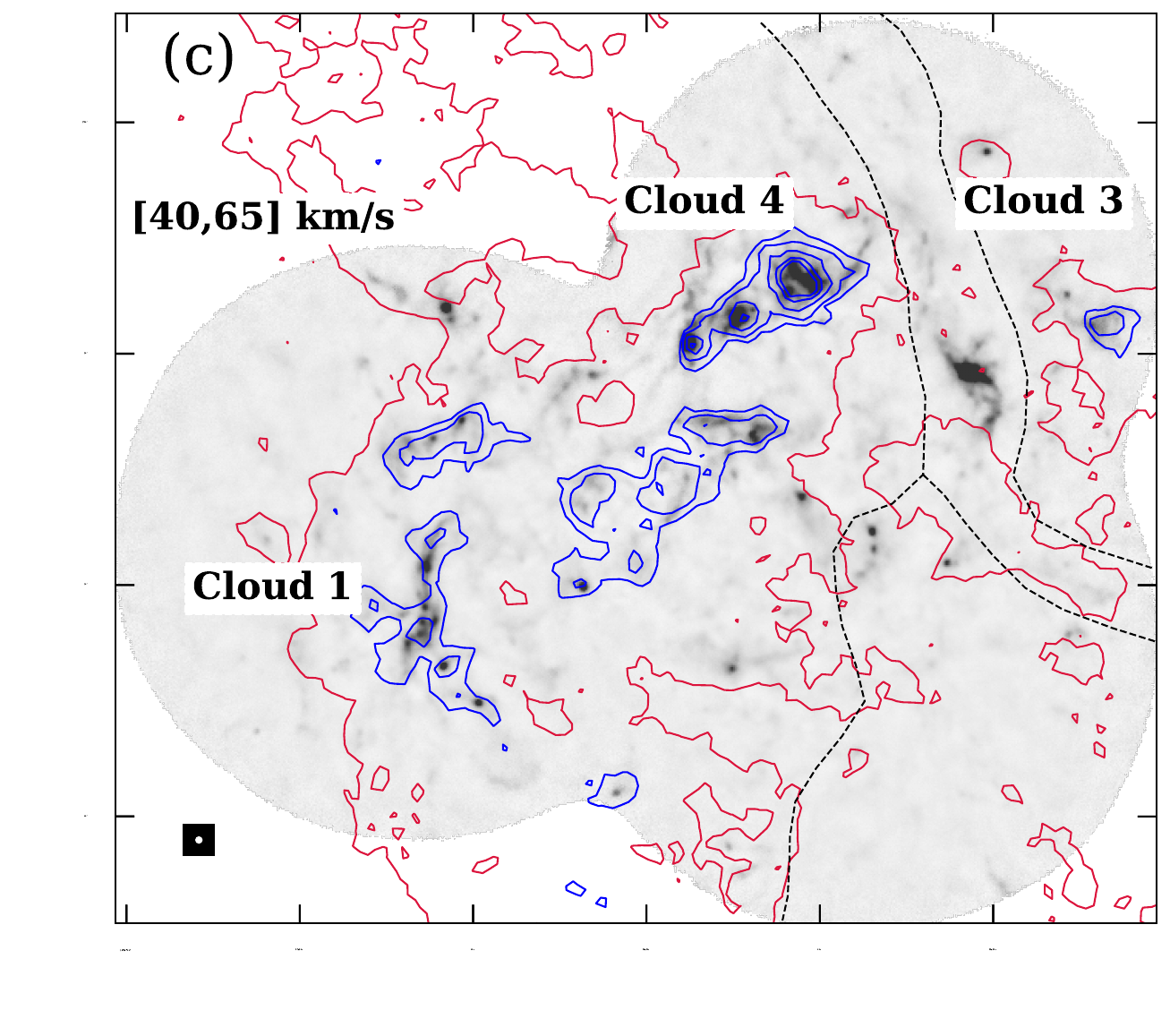}
    \includegraphics[scale=0.35]{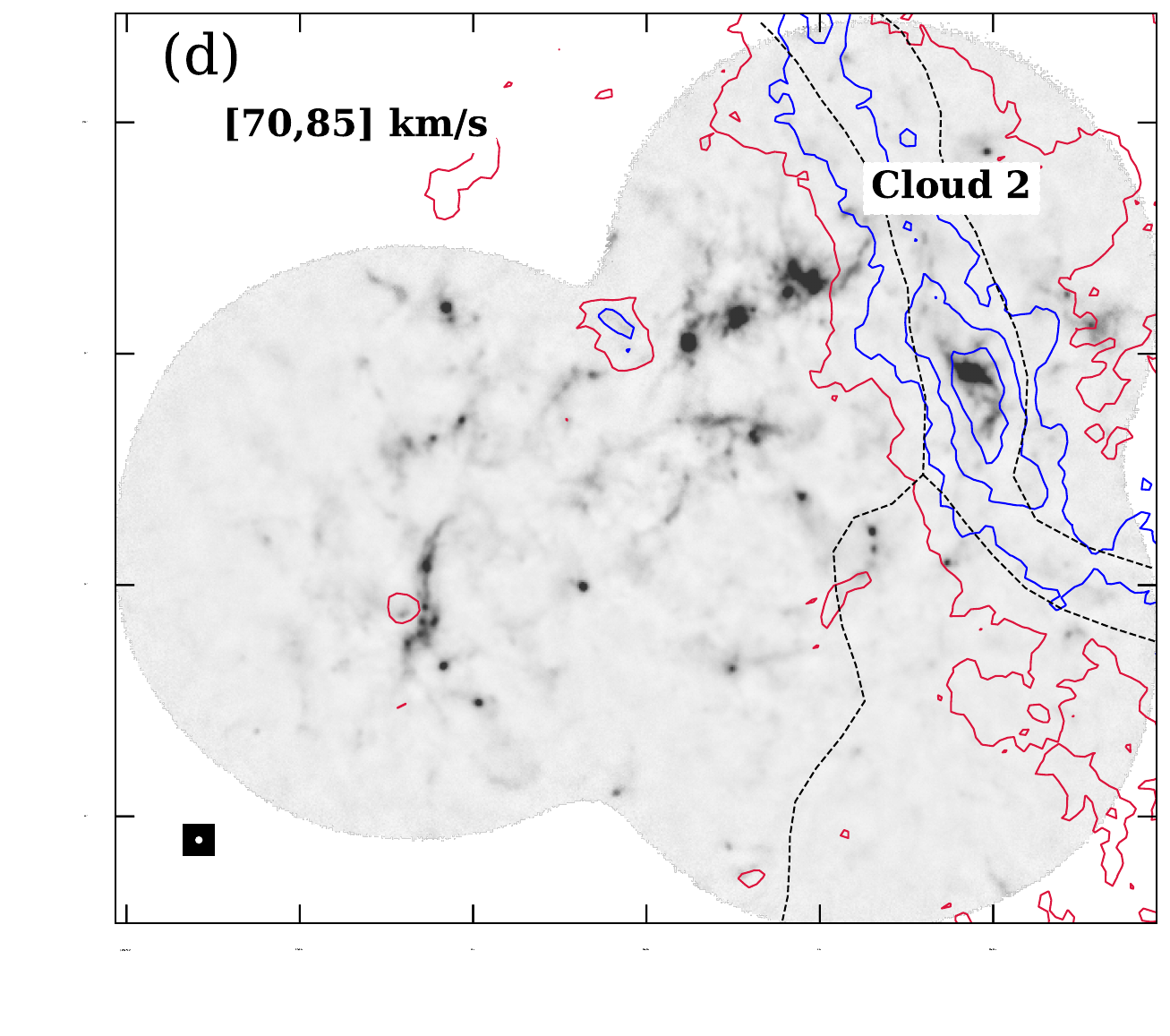}
    \caption{ JCMT 850\,\micron\ continuum image overlaid with $^{13}$CO integrated intensity contours of different velocity components. They correspond to [5, 20], [40, 65], and [70, 85] km/s. The {\it rms} levels ($\sigma$) of the four integrated intensities from (a) to (b) are $\sigma_1$=0.73\,K~km~s${^{-1}}$, $\sigma_2$=0.72\,K~km~s${^{-1}}$, $\sigma_3$=2.82\,K~km~s${^{-1}}$ and $\sigma_4$=1.28\,K~km~s${^{-1}}$, respectively. The red contour in each panel indicates the  the 5$\sigma$ level. 
    The blue contour levels from panel (a) to (b) are [8$\sigma_1$, 10$\sigma_1$] K~km~s${^{-1}}$, [8$\sigma_2$, 10$\sigma_2$, 15$\sigma_2$] K~km~s${^{-1}}$, [10$\sigma_3$,
    12$\sigma_3$, 15$\sigma_3$, 18$\sigma_3$, 20$\sigma_3$] K~km~s${^{-1}}$, and [10$\sigma_4$, 15$\sigma_4$, 20$\sigma_4$] K~km~s${^{-1}}$, respectively. According to the spatial association between gas and continuum emission,  boundaries of different clouds are delineated by the black dashed lines. 
    The beam size of 850 \micron\ observations is also displayed at the bottom left corner of each panel. }
    \label{fig:int}
\end{figure*}

\subsection{Spatial distribution of dust filamentary structures in the G35 complex}
Fig.\,\ref{fig:Tg} shows the  filamentary structures identified in Sect.\,4.3.1 overlaid on the dust temperature map. Fig.\,\ref{fig:hfs} presents the spatial distribution of 10 HFSs overlaid on $Spitzer$ 8 $\micron$ infrared image.
\setcounter{figure}{3}
\renewcommand{\thefigure}{A\arabic{figure}}
\begin{figure}[htbp]
\centering
\includegraphics[scale=0.4]{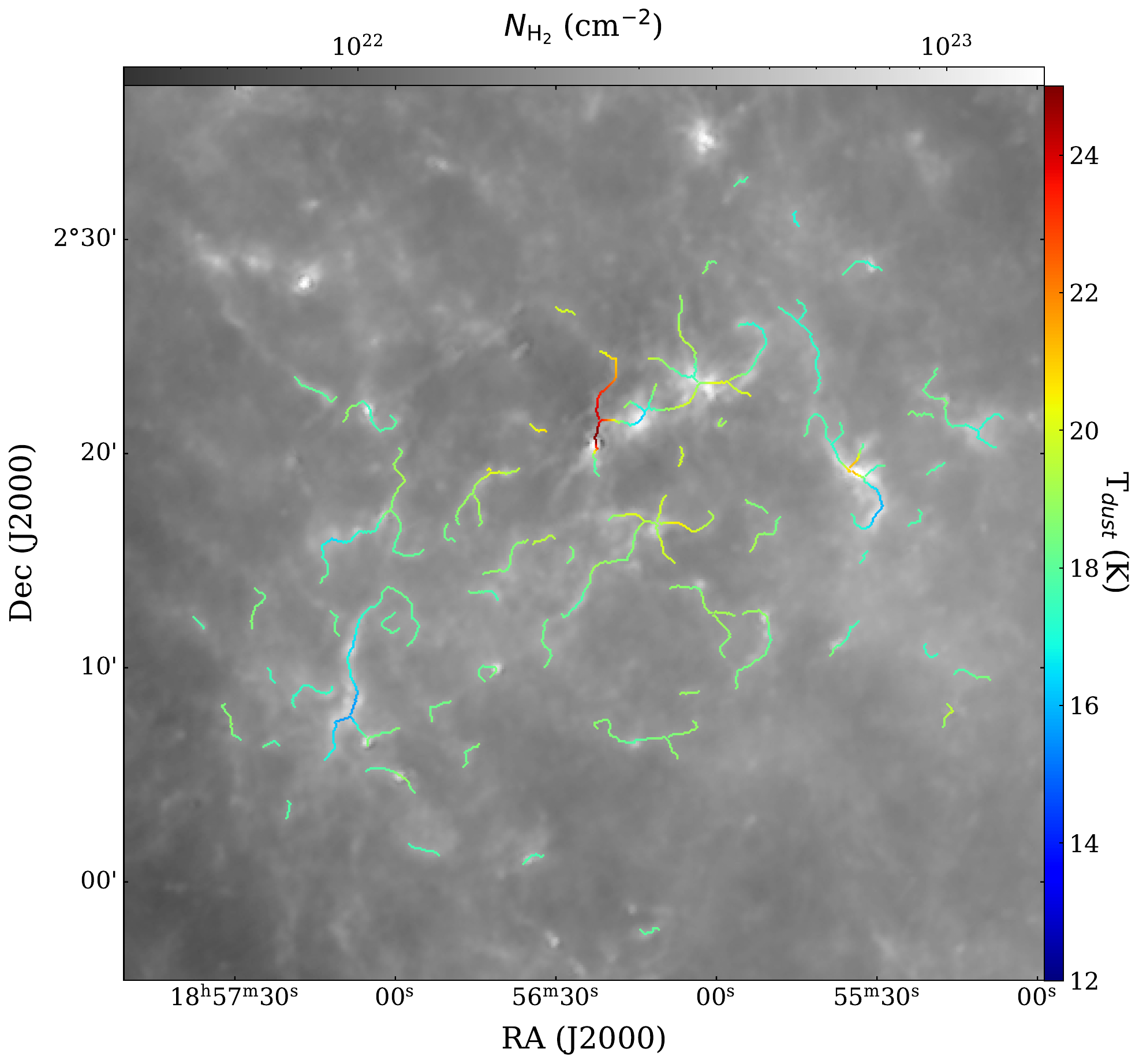}
\caption{Filament skeletons overlaid on the column density map. The color of each structure represents its average dust temperature derived from $Herschel$ observations. }
\label{fig:Tg}
\end{figure}

\renewcommand{\thefigure}{A\arabic{figure}}
\begin{figure*}[htbp]
    \centering
    \includegraphics[scale=0.4]{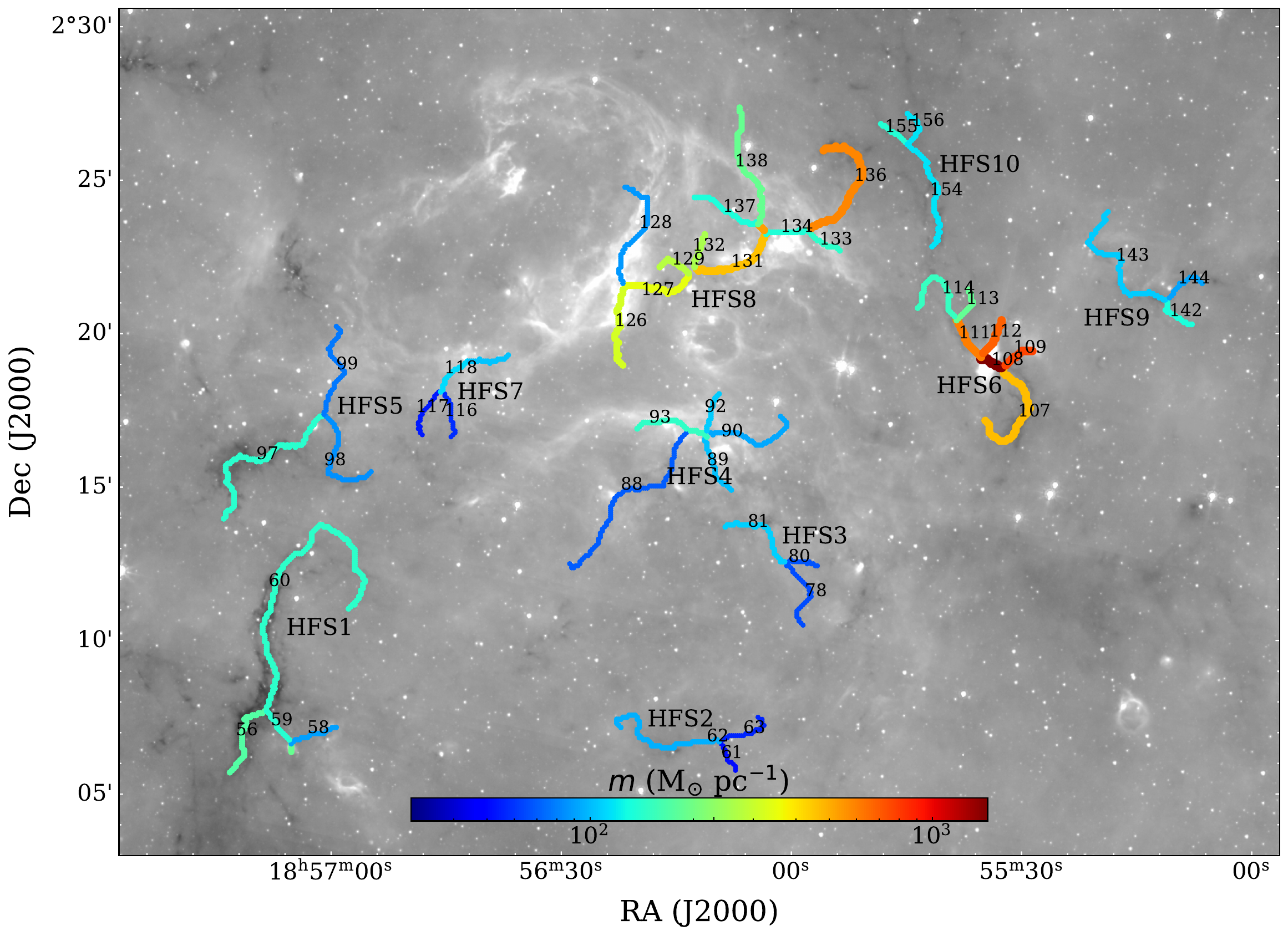}
    \caption{Spatial distribution of HFSs overlaid on the $Spitzer$ 8\,$\micron$ image. The different colored curves sketch the composing filaments of HFSs with each color indicating its respective line mass. }
    \label{fig:hfs}
\end{figure*}

\subsection{Flux comparison of the clumps observed both in previous 870\,\micron\ and our 850\,\micron\ observations}

We examined the variance in peak flux density of the 37 matched clumps between the two catalogues from the ATLASGAL survey \citep{Urquhart2018} at 870\,$\mu$m and our own at 850\,$\mu$m. Owing to the higher angular resolution (14$\arcsec$) of our observations, we recalibrated the 850\,$\mu$m peak flux density to match the angular resolution (19$\arcsec$) at 870\,$\mu$m. This recalibrated value of the 850\,$\mu$m peak flux density was then utilized to predict the 870\,$\mu$m peak flux density using a power-law relation ($S_{\nu}\propto\nu^{\alpha}$) and a spectral index $\alpha$\,=\,-2. Fig.\,\ref{fig:fp} illustrates the comparison between the 870\,$\mu$m peak flux density observed from the ATLASGAL survey and the result predicted from our 850\,$\mu$m observations. The peak flux densities display a strong linear correlation with a Pearson correlation coefficient of 0.99, with the predicted flux densities exceeding the observed ones by approximately 21\%. This implies that the flux measurements for clump sources are consistent between the previous 870\,$\mu$m and our 850\,$\mu$m observations, considering the systematic flux uncertainties of about 15\% in the former \citep[e.g.,][]{ATLASGAL2009} and about 6\% in the latter (see Sect.\,3.1).

\subsection{NH$_3$~(1-1) data seclected for filament kinematic analysis}\label{sec:NH3}
Higher-J CO line data are assumed to be superior for tracing filament kinematics compared to the CO J=1--0 line observed for example in the FUGIN \citep{FUGIN2017} and GRS \citep{GRS2006} surveys. This arises from their enhanced accessibility to the dense filament's physical properties, due to their higher critical excitation density. Furthermore, the G35 complex has access to NH$_3$ line data from the RAMPS \citep{RAMPS2018ApJS} survey, which has an angular resolution of 32$\arcsec$. Despite the low detection rate of the NH$_3$  line data in the G35 complex, similar to higher-J CO line data, the NH$_3$ line is assumed to be more sensitive to the kinematics of dense gas than other lines available here. Consequently, we utilize the RAMPS NH$_3$ line data to approximately quantify the turbulence contribution. Fig.\,\ref{fig:lw} displays the linewidth distribution of this species toward the G35 molecular complex.

\renewcommand{\thefigure}{A\arabic{figure}}
\begin{figure}
    \centering
    \includegraphics[scale=0.5]{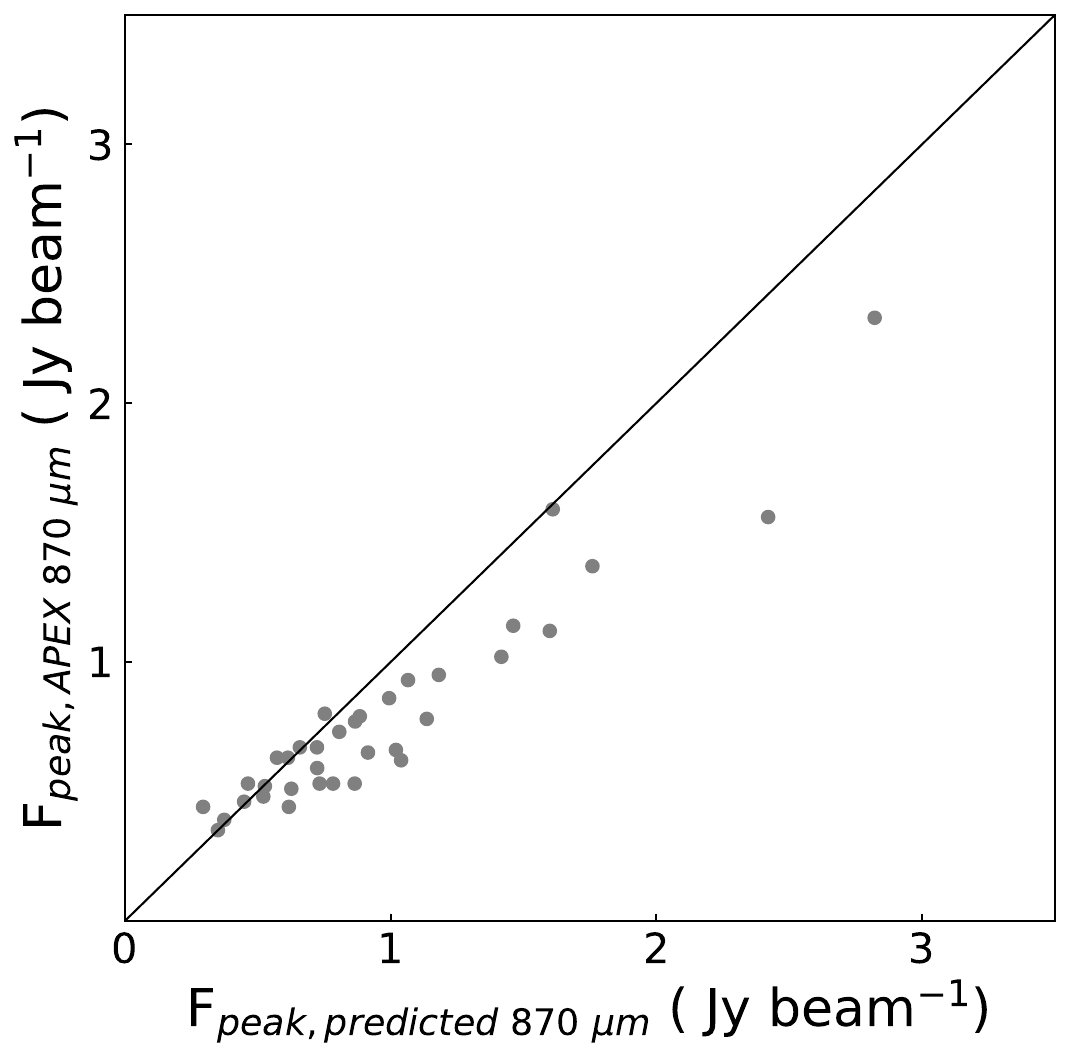}
    \caption{Peak flux density comparison of matched ATLASGAL clumps at 870\,$\micron$ and predicted results from our observed 850\,$\micron$ dust emission. The black line indicates the equality.}
    \label{fig:fp}
\end{figure}
\renewcommand{\thefigure}{A\arabic{figure}}
\begin{figure}
    \centering
    \includegraphics[scale=0.5]{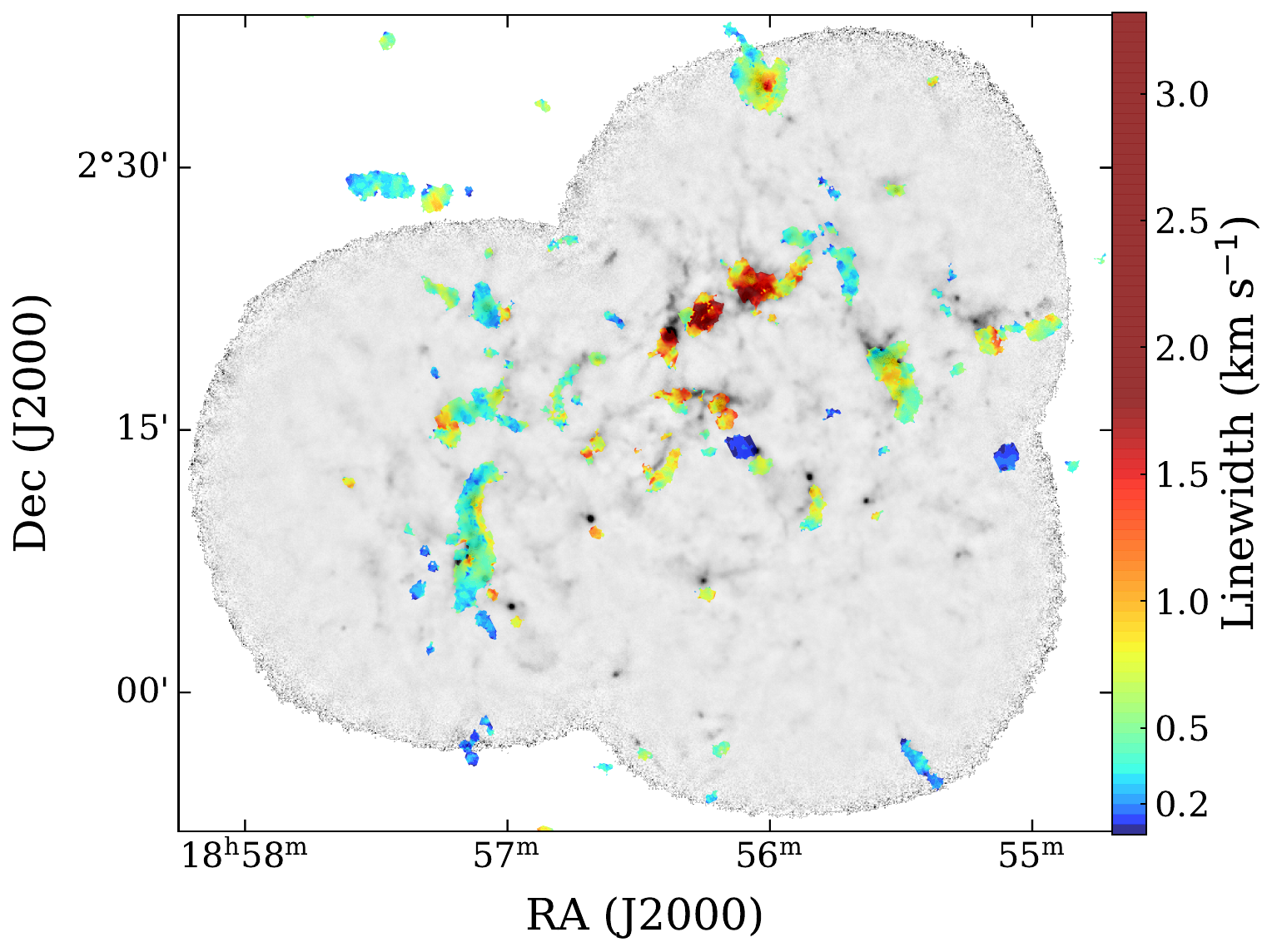}
    \caption{NH$_3$~(1-1) linewidth (rainbow color) overlaid on 850\,\micron\ continuum map (gray color). }
    \label{fig:lw}
\end{figure}

\clearpage

\section{Complementary Tables}
We list the tables of detailed parameters for the clumps and filaments. Table\,\ref{table:filament properties1} gives basic information of each filament,  the resultant physical properties for statistical analysis, their association (Assoc.) with both HMSF and HFSs. Table\,\ref{table:clump properties} presents the basic parameters of 350 clumps, including their position, position angle (PA), flux, distance, as well as the derived physical parameters such as mass and surface density.

\setcounter{table}{0}
\renewcommand{\thetable}{B\arabic{table}}
\startlongtable
\begin{deluxetable*}{lllllllllllll}\label{table:filament properties1}
\tabletypesize{\footnotesize}
\tablecaption{Parameters of filamentary structures} 
\tablehead{
\colhead{ID} & \colhead{Name}  & \colhead{RA} & \colhead{Dec} & \colhead{Flux} & \colhead{D} & \colhead{Mass} & \colhead{Length} & \colhead{Width} & \colhead{T$_{\mathrm{dust}}$} & \colhead{N$_{\mathrm{H_2}}$} & \colhead{Assoc.} & \colhead{Group}\\
& & \colhead{($\degr$)}  & \colhead{($\degr$)} & \colhead{(mJy)} & \colhead{(kpc)} & \colhead{(M$_{\sun}$)} & \colhead{(pc)} &\colhead{(pc)} & \colhead{(K)} &\colhead{($\times 10^{21}$ cm$^{-2}$)} & &\\ }
\colnumbers 
\startdata 
44 & G35.220-0.164 & 284.049 & 1.962 & 868$\pm$347 & 2.9 & 50$\pm$29 & 1.0 & 0.3 & 17.3 &  $8.6\pm4.7$ & - & - \\ 
46 & G35.312-0.216 & 284.137 & 2.020 & 1032$\pm671$ & 2.9 & 59$\pm46$ & 0.9 & 0.3 & 17.4 & $8.7\pm6.5$ & - &  -\\ 
47 & G35.358-0.294 & 284.228 & 2.025 & 1352$\pm$641 & 2.9 & 77$\pm$49 & 1.4 & 0.3 & 17.4 & $7.3\pm4.4$ & - & -\\ 
51 & G35.433-0.373 & 284.333 & 2.055 & 465$\pm$206 & 2.9 & 26$\pm$16 & 0.8 & 0.3 & 17.6 & $4.9\pm2.8$ & - & -\\ 
54 & G35.433-0.306 & 284.273 & 2.086 & 4325$\pm$2393 & 2.9 & 232$\pm$161 & 2.5 & 0.5 & 18.1 & $8.5\pm5.7$ & - & -\\ 
55 & G35.411-0.229 & 284.195 & 2.102 & 1269$\pm$916 & 2.9 & 67$\pm$56 & 1.3 & 0.4 & 18.2 & $5.3\pm4.3$ & - & - \\ 
56 & G35.470-0.315 & 284.298 & 2.115 & 6412$\pm$4680 & 2.9 & 408$\pm$350 & 2.4 & 0.4 & 15.9 & $19.0\pm16.0$ & D & HFS1\\  
58 & G35.454-0.280 & 284.259 & 2.116 & 2300$\pm$1679 & 2.9 & 121$\pm$106 & 1.3 & 0.4 & 18.0 & $10.1\pm8.4$ & Y & HFS1\\
\enddata
\tablecomments{ Column 12 show the filaments that are linked to massive star forming sources summarized in Fig.\ref{fig:cross}. The letter 'D' denotes the IRDCs, 'Y' represents the MYSOs, 'H' marks the H II regions, and 'M' stands for class II methanol masers. 'Group' indicates the structures to which the filaments belong. '-' indicates individual filaments, and the designation of the HFS to which the hub-composing filaments belong is also marked. The structure with ID 108 (belongs to HFS6) shows its width slightly greater than its length due to its location in a very bright HII region (see Fig.\,\ref{fig:hfs}). The relative uncertainty of parameters of both length and width is 10\%.}

\end{deluxetable*}
\setcounter{table}{1}
\renewcommand{\thetable}{B\arabic{table}}

\startlongtable
\begin{deluxetable*}{llllllllllllll}
\tabletypesize{\footnotesize}
\setlength{\tabcolsep}{3pt} 
\tablecaption{Phyical parameters of dense clumps}
\label{table:clump properties}
\tablehead{
    \colhead{ID} & \colhead{RA} & \colhead{Dec} & \colhead{Size} & \colhead{PA} & \colhead{R} & \colhead{Flux} & \colhead{$F_{\mathrm{peak}}$} & \colhead{SNR} & \colhead{D} & \colhead{T$_{\mathrm{dust}}$} & \colhead{Mass} & \colhead{$\Sigma$} & \colhead{N$_{\mathrm{H_2}}$} \\
    & \colhead{($\degr$)} & \colhead{($\degr$)} & \colhead{($\arcsec\times\arcsec$)} & \colhead{($\degr$)} & \colhead{(pc)} & \colhead{(mJy)} & \colhead{(mJy~beam$^{-1}$)} & & \colhead{(kpc)} & \colhead{(K)} & \colhead{(M$_{\sun}$)} & \colhead{(g~cm$^{-2}$)} & \colhead{($\times$10$^{21}$cm$^{-2}$)}\\
}
\colnumbers
\startdata
1 & 283.869 & 1.951 & 18.4 $\times$ 14.3 & -50.4 & 0.1 & 300$\pm$46 & 163 & 11 & 2.1 & 17.2 & 9$\pm$4 & 0.09$\pm$0.036 & 19.0$\pm$7.7 \\ 
2 & 284.056 & 1.959 & 21.7 $\times$ 15.9 & -3.8 & 0.1 & 320$\pm$47 & 148 & 13 & 2.9 & 16.9 & 19$\pm$9 & 0.07$\pm$0.03 & 15.8$\pm$6.3 \\ 
3 & 284.042 & 1.978 & 27.5 $\times$ 13.7 & -11.7 & 0.1 & 151$\pm$26 & 95 & 11 & 2.9 & 18.2 & 8$\pm$4 & 0.03$\pm$0.012 & 6.1$\pm$2.5 \\ 
4 & 283.968 & 1.975 & 21.4 $\times$ 14.4 & 47.9 & 0.1 & 88$\pm$16 & 51 & 8 & 2.1 & 18.6 & 2$\pm$1 & 0.02$\pm$0.008 & 4.2$\pm$1.8 \\ 
5 & 284.066 & 1.979 & 19.9 $\times$ 14.1 & -44.7 & 0.1 & 328$\pm$39 & 196 & 21 & 2.9 & 18.3 & 17$\pm$8 & 0.08$\pm$0.033 & 17.8$\pm$7.0 \\ 
6 & 283.931 & 2.002 & 25.2 $\times$ 14.3 & -75.3 & 0.1 & 109$\pm$19 & 48 & 9 & 2.1 & 18.8 & 3$\pm$1 & 0.02$\pm$0.009 & 4.4$\pm$1.8 \\ 
7 & 284.145 & 2.019 & 30.3 $\times$ 17.0 & 20.5 & 0.2 & 834$\pm$104 & 367 & 25 & 2.9 & 17.2 & 48$\pm$22 & 0.13$\pm$0.05 & 26.9$\pm$10.6 \\ 
8 & 283.962 & 2.025 & 28.9 $\times$ 15.7 & 58.2 & 0.1 & 103$\pm$18 & 45 & 10 & 2.1 & 19.3 & 3$\pm$1 & 0.01$\pm$0.006 & 3.1$\pm$1.3 \\
\enddata
\tablecomments{ The relative uncertainty of parameters of both length and width is 10\%.}
\end{deluxetable*}
\bibliography{sample631}{}

\begin{thebibliography}{}
\expandafter\ifx\csname natexlab\endcsname\relax\def\natexlab#1{#1}\fi
\providecommand{\url}[1]{\href{#1}{#1}}
\providecommand{\dodoi}[1]{doi:~\href{http://doi.org/#1}{\nolinkurl{#1}}}
\providecommand{\doeprint}[1]{\href{http://ascl.net/#1}{\nolinkurl{http://ascl.net/#1}}}
\providecommand{\doarXiv}[1]{\href{https://arxiv.org/abs/#1}{\nolinkurl{https://arxiv.org/abs/#1}}}

\bibitem[{{Anderson} {et~al.}(2014){Anderson}, {Bania}, {Balser}, {Cunningham}, {Wenger}, {Johnstone}, \& {Armentrout}}]{HIIcat2014}
{Anderson}, L.~D., {Bania}, T.~M., {Balser}, D.~S., {et~al.} 2014, \apjs, 212, 1, \dodoi{10.1088/0067-0049/212/1/1}

\bibitem[{{Andr{\'e}} {et~al.}(2014){Andr{\'e}}, {Di Francesco}, {Ward-Thompson}, {Inutsuka}, {Pudritz}, \& {Pineda}}]{Andro2014}
{Andr{\'e}}, P., {Di Francesco}, J., {Ward-Thompson}, D., {et~al.} 2014, in Protostars and Planets VI, ed. H.~{Beuther}, R.~S. {Klessen}, C.~P. {Dullemond}, \& T.~{Henning}, 27--51, \dodoi{10.2458/azu_uapress_9780816531240-ch002}

\bibitem[{{Andr{\'e}} {et~al.}(2010){Andr{\'e}}, {Men'shchikov}, {Bontemps}, {K{\"o}nyves}, {Motte}, {Schneider}, {Didelon}, {Minier}, {Saraceno}, {Ward-Thompson}, {di Francesco}, {White}, {Molinari}, {Testi}, {Abergel}, {Griffin}, {Henning}, {Royer}, {Mer{\'\i}n}, {Vavrek}, {Attard}, {Arzoumanian}, {Wilson}, {Ade}, {Aussel}, {Baluteau}, {Benedettini}, {Bernard}, {Blommaert}, {Cambr{\'e}sy}, {Cox}, {di Giorgio}, {Hargrave}, {Hennemann}, {Huang}, {Kirk}, {Krause}, {Launhardt}, {Leeks}, {Le Pennec}, {Li}, {Martin}, {Maury}, {Olofsson}, {Omont}, {Peretto}, {Pezzuto}, {Prusti}, {Roussel}, {Russeil}, {Sauvage}, {Sibthorpe}, {Sicilia-Aguilar}, {Spinoglio}, {Waelkens}, {Woodcraft}, \& {Zavagno}}]{Andre2010}
{Andr{\'e}}, P., {Men'shchikov}, A., {Bontemps}, S., {et~al.} 2010, \aap, 518, L102, \dodoi{10.1051/0004-6361/201014666}

\bibitem[{{Arzoumanian} {et~al.}(2023){Arzoumanian}, {Arakawa}, {Kobayashi}, {Iwasaki}, {Fukuda}, {Mori}, {Hirai}, {Kunitomo}, {Kumar}, \& {Kokubo}}]{Arzoumanian2023}
{Arzoumanian}, D., {Arakawa}, S., {Kobayashi}, M. I.~N., {et~al.} 2023, \apjl, 947, L29, \dodoi{10.3847/2041-8213/acc849}

\bibitem[{{Astropy Collaboration} {et~al.}(2013){Astropy Collaboration}, {Robitaille}, {Tollerud}, {Greenfield}, {Droettboom}, {Bray}, {Aldcroft}, {Davis}, {Ginsburg}, {Price-Whelan}, {Kerzendorf}, {Conley}, {Crighton}, {Barbary}, {Muna}, {Ferguson}, {Grollier}, {Parikh}, {Nair}, {Unther}, {Deil}, {Woillez}, {Conseil}, {Kramer}, {Turner}, {Singer}, {Fox}, {Weaver}, {Zabalza}, {Edwards}, {Azalee Bostroem}, {Burke}, {Casey}, {Crawford}, {Dencheva}, {Ely}, {Jenness}, {Labrie}, {Lim}, {Pierfederici}, {Pontzen}, {Ptak}, {Refsdal}, {Servillat}, \& {Streicher}}]{astropy2013}
{Astropy Collaboration}, {Robitaille}, T.~P., {Tollerud}, E.~J., {et~al.} 2013, \aap, 558, A33, \dodoi{10.1051/0004-6361/201322068}

\bibitem[{{Barbary}(2016)}]{Barbary2016}
{Barbary}, K. 2016, The Journal of Open Source Software, 1, 58, \dodoi{10.21105/joss.00058}

\bibitem[{{Barnes} {et~al.}(2016){Barnes}, {Kong}, {Tan}, {Henshaw}, {Caselli}, {Jim{\'e}nez-Serra}, \& {Fontani}}]{Barnes2016}
{Barnes}, A.~T., {Kong}, S., {Tan}, J.~C., {et~al.} 2016, \mnras, 458, 1990, \dodoi{10.1093/mnras/stw403}

\bibitem[{{Barnes} {et~al.}(2021){Barnes}, {Henshaw}, {Fontani}, {Pineda}, {Cosentino}, {Tan}, {Caselli}, {Jim{\'e}nez-Serra}, {Law}, {Avison}, {Bigiel}, {Feng}, {Kong}, {Longmore}, {Moser}, {Parker}, {S{\'a}nchez-Monge}, \& {Wang}}]{Barnes2021}
{Barnes}, A.~T., {Henshaw}, J.~D., {Fontani}, F., {et~al.} 2021, \mnras, 503, 4601, \dodoi{10.1093/mnras/stab803}

\bibitem[{{Benjamin} {et~al.}(2003){Benjamin}, {Churchwell}, {Babler}, {Bania}, {Clemens}, {Cohen}, {Dickey}, {Indebetouw}, {Jackson}, {Kobulnicky}, {Lazarian}, {Marston}, {Mathis}, {Meade}, {Seager}, {Stolovy}, {Watson}, {Whitney}, {Wolff}, \& {Wolfire}}]{GLIMPSE2003}
{Benjamin}, R.~A., {Churchwell}, E., {Babler}, B.~L., {et~al.} 2003, \pasp, 115, 953, \dodoi{10.1086/376696}

\bibitem[{{Bertin} \& {Arnouts}(1996)}]{SExtractor1996}
{Bertin}, E., \& {Arnouts}, S. 1996, \aaps, 117, 393, \dodoi{10.1051/aas:1996164}

\bibitem[{{Bronfman} {et~al.}(1996){Bronfman}, {Nyman}, \& {May}}]{Bronfman1996}
{Bronfman}, L., {Nyman}, L.~A., \& {May}, J. 1996, \aaps, 115, 81

\bibitem[{{Brunthaler} {et~al.}(2021){Brunthaler}, {Menten}, {Dzib}, {Cotton}, {Wyrowski}, {Dokara}, {Gong}, {Medina}, {M{\"u}ller}, {Nguyen}, {Ortiz-Le{\'o}n}, {Reich}, {Rugel}, {Urquhart}, {Winkel}, {Yang}, {Beuther}, {Billington}, {Carrasco-Gonzalez}, {Csengeri}, {Murugeshan}, {Pandian}, \& {Roy}}]{glostar2021A&A}
{Brunthaler}, A., {Menten}, K.~M., {Dzib}, S.~A., {et~al.} 2021, \aap, 651, A85, \dodoi{10.1051/0004-6361/202039856}

\bibitem[{{Chen} \& {Ostriker}(2015)}]{Chen2015}
{Chen}, C.-Y., \& {Ostriker}, E.~C. 2015, \apj, 810, 126, \dodoi{10.1088/0004-637X/810/2/126}

\bibitem[{{Contreras} {et~al.}(2013){Contreras}, {Schuller}, {Urquhart}, {Csengeri}, {Wyrowski}, {Beuther}, {Bontemps}, {Bronfman}, {Henning}, {Menten}, {Schilke}, {Walmsley}, {Wienen}, {Tackenberg}, \& {Linz}}]{csc2013}
{Contreras}, Y., {Schuller}, F., {Urquhart}, J.~S., {et~al.} 2013, \aap, 549, A45, \dodoi{10.1051/0004-6361/201220155}

\bibitem[{{Currie} {et~al.}(2014){Currie}, {Berry}, {Jenness}, {Gibb}, {Bell}, \& {Draper}}]{starlink2014}
{Currie}, M.~J., {Berry}, D.~S., {Jenness}, T., {et~al.} 2014, in Astronomical Society of the Pacific Conference Series, Vol. 485, Astronomical Data Analysis Software and Systems XXIII, ed. N.~{Manset} \& P.~{Forshay}, 391

\bibitem[{{Dunham} {et~al.}(2011){Dunham}, {Robitaille}, {Evans}, {Schlingman}, {Cyganowski}, \& {Urquhart}}]{Dunham2011}
{Dunham}, M.~K., {Robitaille}, T.~P., {Evans}, Neal~J., I., {et~al.} 2011, \apj, 731, 90, \dodoi{10.1088/0004-637X/731/2/90}

\bibitem[{{Dzib} {et~al.}(2023){Dzib}, {Yang}, {Urquhart}, {Medina}, {Brunthaler}, {Menten}, {Wyrowski}, {Cotton}, {Dokara}, {Ortiz-Le{\'o}n}, {Rugel}, {Nguyen}, {Gong}, {Chakraborty}, {Beuther}, {Billington}, {Carrasco-Gonzalez}, {Csengeri}, {Hofner}, {Ott}, {Pandian}, {Roy}, \& {Yanza}}]{Dzib2023}
{Dzib}, S.~A., {Yang}, A.~Y., {Urquhart}, J.~S., {et~al.} 2023, \aap, 670, A9, \dodoi{10.1051/0004-6361/202143019}

\bibitem[{{Feng} {et~al.}(2024){Feng}, {Smith}, {Hacar}, {Clark}, \& {Seifried}}]{Feng2024}
{Feng}, J., {Smith}, R.~J., {Hacar}, A., {Clark}, S.~E., \& {Seifried}, D. 2024, \mnras, 528, 6370, \dodoi{10.1093/mnras/stae407}

\bibitem[{{Fiege} \& {Pudritz}(2000)}]{Fiege2000}
{Fiege}, J.~D., \& {Pudritz}, R.~E. 2000, \mnras, 311, 85, \dodoi{10.1046/j.1365-8711.2000.03066.x}

\bibitem[{{Giannetti} {et~al.}(2013){Giannetti}, {Brand}, {S{\'a}nchez-Monge}, {Fontani}, {Cesaroni}, {Beltr{\'a}n}, {Molinari}, {Dodson}, \& {Rioja}}]{Giannetti2013}
{Giannetti}, A., {Brand}, J., {S{\'a}nchez-Monge}, {\'A}., {et~al.} 2013, \aap, 556, A16, \dodoi{10.1051/0004-6361/201321456}

\bibitem[{{G{\'o}mez} \& {V{\'a}zquez-Semadeni}(2014)}]{Gilberto2014}
{G{\'o}mez}, G.~C., \& {V{\'a}zquez-Semadeni}, E. 2014, \apj, 791, 124, \dodoi{10.1088/0004-637X/791/2/124}

\bibitem[{{Gong} \& {Ostriker}(2015)}]{Gong2015}
{Gong}, M., \& {Ostriker}, E.~C. 2015, \apj, 806, 31, \dodoi{10.1088/0004-637X/806/1/31}

\bibitem[{{Guzm{\'a}n} {et~al.}(2015){Guzm{\'a}n}, {Sanhueza}, {Contreras}, {Smith}, {Jackson}, {Hoq}, \& {Rathborne}}]{Guzman2015}
{Guzm{\'a}n}, A.~E., {Sanhueza}, P., {Contreras}, Y., {et~al.} 2015, \apj, 815, 130, \dodoi{10.1088/0004-637X/815/2/130}

\bibitem[{{Hacar} {et~al.}(2023){Hacar}, {Clark}, {Heitsch}, {Kainulainen}, {Panopoulou}, {Seifried}, \& {Smith}}]{Hacar2023}
{Hacar}, A., {Clark}, S.~E., {Heitsch}, F., {et~al.} 2023, in Astronomical Society of the Pacific Conference Series, Vol. 534, Protostars and Planets VII, ed. S.~{Inutsuka}, Y.~{Aikawa}, T.~{Muto}, K.~{Tomida}, \& M.~{Tamura}, 153, \dodoi{10.48550/arXiv.2203.09562}

\bibitem[{{He} {et~al.}(2015){He}, {Zhou}, {Esimbek}, {Ji}, {Wu}, {Tang}, {Yuan}, {Li}, \& {Baan}}]{He2015}
{He}, Y.-X., {Zhou}, J.-J., {Esimbek}, J., {et~al.} 2015, \mnras, 450, 1926, \dodoi{10.1093/mnras/stv732}

\bibitem[{{He} {et~al.}(2023){He}, {Liu}, {Tang}, {Qin}, {Zhou}, {Esimbek}, {Pan}, {Li}, {Zhao}, {Ji}, \& {Komesh}}]{He23}
{He}, Y.-X., {Liu}, H.-L., {Tang}, X.-D., {et~al.} 2023, \apj, 957, 61, \dodoi{10.3847/1538-4357/acf766}

\bibitem[{{Heyer} {et~al.}(2009){Heyer}, {Krawczyk}, {Duval}, \& {Jackson}}]{Heyer2009}
{Heyer}, M., {Krawczyk}, C., {Duval}, J., \& {Jackson}, J.~M. 2009, \apj, 699, 1092, \dodoi{10.1088/0004-637X/699/2/1092}

\bibitem[{{Hildebrand}(1983)}]{mass1983}
{Hildebrand}, R.~H. 1983, \qjras, 24, 267

\bibitem[{{Hogge} {et~al.}(2018){Hogge}, {Jackson}, {Stephens}, {Whitaker}, {Foster}, {Camarata}, {Anish Roshi}, {Di Francesco}, {Longmore}, {Loughnane}, {Moore}, {Rathborne}, {Sanhueza}, \& {Walsh}}]{RAMPS2018ApJS}
{Hogge}, T., {Jackson}, J., {Stephens}, I., {et~al.} 2018, \apjs, 237, 27, \dodoi{10.3847/1538-4365/aacf94}

\bibitem[{{Hou} \& {Han}(2014)}]{Hou2014}
{Hou}, L.~G., \& {Han}, J.~L. 2014, \aap, 569, A125, \dodoi{10.1051/0004-6361/201424039}

\bibitem[{{Hu} {et~al.}(2016){Hu}, {Menten}, {Wu}, {Bartkiewicz}, {Rygl}, {Reid}, {Urquhart}, \& {Zheng}}]{Hu2016}
{Hu}, B., {Menten}, K.~M., {Wu}, Y., {et~al.} 2016, \apj, 833, 18, \dodoi{10.3847/0004-637X/833/1/18}

\bibitem[{{Inoue} \& {Fukui}(2013)}]{Inoue2013}
{Inoue}, T., \& {Fukui}, Y. 2013, \apjl, 774, L31, \dodoi{10.1088/2041-8205/774/2/L31}

\bibitem[{{Inoue} {et~al.}(2018){Inoue}, {Hennebelle}, {Fukui}, {Matsumoto}, {Iwasaki}, \& {Inutsuka}}]{Inoue2018}
{Inoue}, T., {Hennebelle}, P., {Fukui}, Y., {et~al.} 2018, \pasj, 70, S53, \dodoi{10.1093/pasj/psx089}

\bibitem[{{Jackson} {et~al.}(2006){Jackson}, {Rathborne}, {Shah}, {Simon}, {Bania}, {Clemens}, {Chambers}, {Johnson}, {Dormody}, {Lavoie}, \& {Heyer}}]{GRS2006}
{Jackson}, J.~M., {Rathborne}, J.~M., {Shah}, R.~Y., {et~al.} 2006, \apjs, 163, 145, \dodoi{10.1086/500091}

\bibitem[{{Jayasinghe} {et~al.}(2019){Jayasinghe}, {Dixon}, {Povich}, {Binder}, {Velasco}, {Lepore}, {Xu}, {Offner}, {Kobulnicky}, {Anderson}, {Kendrew}, \& {Simpson}}]{MWP2019}
{Jayasinghe}, T., {Dixon}, D., {Povich}, M.~S., {et~al.} 2019, \mnras, 488, 1141, \dodoi{10.1093/mnras/stz1738}

\bibitem[{{Jiao} {et~al.}(2022){Jiao}, {Lin}, {Shui}, {Wu}, {Ren}, \& {Li}}]{Jiao2022}
{Jiao}, S., {Lin}, Y., {Shui}, X., {et~al.} 2022, Science China Physics, Mechanics, and Astronomy, 65, 299511, \dodoi{10.1007/s11433-021-1902-3}

\bibitem[{{Jim{\'e}nez-Serra} {et~al.}(2014){Jim{\'e}nez-Serra}, {Caselli}, {Fontani}, {Tan}, {Henshaw}, {Kainulainen}, \& {Hernandez}}]{Jim2014}
{Jim{\'e}nez-Serra}, I., {Caselli}, P., {Fontani}, F., {et~al.} 2014, \mnras, 439, 1996, \dodoi{10.1093/mnras/stu078}

\bibitem[{{Kalcheva} {et~al.}(2018){Kalcheva}, {Hoare}, {Urquhart}, {Kurtz}, {Lumsden}, {Purcell}, \& {Zijlstra}}]{Kalcheva2018}
{Kalcheva}, I.~E., {Hoare}, M.~G., {Urquhart}, J.~S., {et~al.} 2018, \aap, 615, A103, \dodoi{10.1051/0004-6361/201832734}

\bibitem[{{Kauffmann} {et~al.}(2008){Kauffmann}, {Bertoldi}, {Bourke}, {Evans}, \& {Lee}}]{mug2008}
{Kauffmann}, J., {Bertoldi}, F., {Bourke}, T.~L., {Evans}, N.~J., I., \& {Lee}, C.~W. 2008, \aap, 487, 993, \dodoi{10.1051/0004-6361:200809481}

\bibitem[{{Koch} \& {Rosolowsky}(2015)}]{FilFinder2015}
{Koch}, E.~W., \& {Rosolowsky}, E.~W. 2015, \mnras, 452, 3435, \dodoi{10.1093/mnras/stv1521}

\bibitem[{{Krumholz} \& {McKee}(2008)}]{Nature2008}
{Krumholz}, M.~R., \& {McKee}, C.~F. 2008, \nat, 451, 1082, \dodoi{10.1038/nature06620}

\bibitem[{{Kuhn} {et~al.}(2021){Kuhn}, {de Souza}, {Krone-Martins}, {Castro-Ginard}, {Ishida}, {Povich}, {Hillenbrand}, \& {COIN Collaboration}}]{SPICY2021}
{Kuhn}, M.~A., {de Souza}, R.~S., {Krone-Martins}, A., {et~al.} 2021, \apjs, 254, 33, \dodoi{10.3847/1538-4365/abe465}

\bibitem[{{Kumar} {et~al.}(2022){Kumar}, {Arzoumanian}, {Men'shchikov}, {Palmeirim}, {Matsumura}, \& {Inutsuka}}]{Kumar2022}
{Kumar}, M.~S.~N., {Arzoumanian}, D., {Men'shchikov}, A., {et~al.} 2022, \aap, 658, A114, \dodoi{10.1051/0004-6361/202140363}

\bibitem[{{Kumar} {et~al.}(2020){Kumar}, {Palmeirim}, {Arzoumanian}, \& {Inutsuka}}]{Kumar2020}
{Kumar}, M.~S.~N., {Palmeirim}, P., {Arzoumanian}, D., \& {Inutsuka}, S.~I. 2020, \aap, 642, A87, \dodoi{10.1051/0004-6361/202038232}

\bibitem[{{Ladeyschikov} {et~al.}(2019){Ladeyschikov}, {Bayandina}, \& {Sobolev}}]{Ladeyschikov2019}
{Ladeyschikov}, D.~A., {Bayandina}, O.~S., \& {Sobolev}, A.~M. 2019, \aj, 158, 233, \dodoi{10.3847/1538-3881/ab4b4c}

\bibitem[{{Larson}(1981)}]{Larson1981}
{Larson}, R.~B. 1981, \mnras, 194, 809, \dodoi{10.1093/mnras/194.4.809}

\bibitem[{{Li} {et~al.}(2016){Li}, {Urquhart}, {Leurini}, {Csengeri}, {Wyrowski}, {Menten}, \& {Schuller}}]{Li2016}
{Li}, G.-X., {Urquhart}, J.~S., {Leurini}, S., {et~al.} 2016, \aap, 591, A5, \dodoi{10.1051/0004-6361/201527468}

\bibitem[{{Li} {et~al.}(2018){Li}, {Zhou}, {Esimbek}, {He}, {Baan}, {Li}, {Wu}, {Tang}, {Ji}, \& {Zhexeray}}]{LiQ2018}
{Li}, Q., {Zhou}, J., {Esimbek}, J., {et~al.} 2018, \apj, 867, 167, \dodoi{10.3847/1538-4357/aae2b8}

\bibitem[{Lisa \& Bot(2017)}]{aplpy2017}
Lisa, M., \& Bot, H. 2017, {My Research Software}, 2.0.4, \dodoi{10.5281/zenodo.1234}

\bibitem[{{Liu} {et~al.}(2012{\natexlab{a}}){Liu}, {Jim{\'e}nez-Serra}, {Ho}, {Chen}, {Zhang}, \& {Li}}]{LiuH2012}
{Liu}, H.~B., {Jim{\'e}nez-Serra}, I., {Ho}, P. T.~P., {et~al.} 2012{\natexlab{a}}, \apj, 756, 10, \dodoi{10.1088/0004-637X/756/1/10}

\bibitem[{{Liu} {et~al.}(2012{\natexlab{b}}){Liu}, {Quintana-Lacaci}, {Wang}, {Ho}, {Li}, {Zhang}, \& {Zhang}}]{Liu2012}
{Liu}, H.~B., {Quintana-Lacaci}, G., {Wang}, K., {et~al.} 2012{\natexlab{b}}, \apj, 745, 61, \dodoi{10.1088/0004-637X/745/1/61}

\bibitem[{{Liu} {et~al.}(2020){Liu}, {Sanhueza}, {Liu}, {Zavagno}, {Tang}, {Wu}, \& {Zhang}}]{LHL2020}
{Liu}, H.-L., {Sanhueza}, P., {Liu}, T., {et~al.} 2020, \apj, 901, 31, \dodoi{10.3847/1538-4357/abadfe}

\bibitem[{{Liu} {et~al.}(2019){Liu}, {Stutz}, \& {Yuan}}]{LiuHL2019}
{Liu}, H.-L., {Stutz}, A., \& {Yuan}, J.-H. 2019, \mnras, 487, 1259, \dodoi{10.1093/mnras/stz1340}

\bibitem[{{Liu} {et~al.}(2021){Liu}, {Liu}, {Evans}, {Wang}, {Garay}, {Qin}, {Li}, {Stutz}, {Goldsmith}, {Liu}, {Tej}, {Zhang}, {Juvela}, {Li}, {Wang}, {Bronfman}, {Ren}, {Wu}, {Kim}, {Lee}, {Tatematsu}, {Cunningham}, {Liu}, {Wu}, {Hirota}, {Lee}, {Li}, {Kang}, {Mardones}, {Ristorcelli}, {Zhang}, {Luo}, {Toth}, {Yi}, {Yun}, {Peng}, {Li}, {Zhu}, {Shen}, {Baug}, {Dewangan}, {Chakali}, {Liu}, {Xu}, {Wang}, {Zhang}, {Li}, {Zhang}, {Zhou}, {Tang}, {Xue}, {Issac}, {Soam}, \& {{\'A}lvarez-Guti{\'e}rrez}}]{LiuH21}
{Liu}, H.-L., {Liu}, T., {Evans}, N.J., I., {et~al.} 2021, \mnras, 505, 2801, \dodoi{10.1093/mnras/stab1352}

\bibitem[{{Liu} {et~al.}(2022){Liu}, {Tej}, {Liu}, {Issac}, {Saha}, {Goldsmith}, {Wang}, {Zhang}, {Qin}, {Wang}, {Li}, {Soam}, {Dewangan}, {Lee}, {Li}, {Liu}, {Zhang}, {Ren}, {Juvela}, {Bronfman}, {Wu}, {Tatematsu}, {Chen}, {Li}, {Stutz}, {Zhang}, {Viktor Toth}, {Luo}, {Xu}, {Li}, {Liu}, {Zhou}, {Zhang}, {Tang}, {Zhang}, {Baug}, {Mannfors}, {Chakali}, \& {Dutta}}]{LHL2022}
{Liu}, H.-L., {Tej}, A., {Liu}, T., {et~al.} 2022, \mnras, 510, 5009, \dodoi{10.1093/mnras/stab2757}

\bibitem[{{Liu} {et~al.}(2023){Liu}, {Tej}, {Liu}, {Sanhueza}, {Qin}, {He}, {Goldsmith}, {Garay}, {Pan}, {Morii}, {Li}, {Stutz}, {Tatematsu}, {Xu}, {Bronfman}, {Saha}, {Issac}, {Baug}, {Toth}, {Dewangan}, {Wang}, {Zhou}, {Lee}, {Yang}, {Luo}, {Shen}, {Zhang}, {Wu}, {Ren}, {Liu}, {Soam}, {Zhang}, \& {Luo}}]{LiuHL2023}
---. 2023, \mnras, 522, 3719, \dodoi{10.1093/mnras/stad047}

\bibitem[{{Liu} {et~al.}(2018){Liu}, {Li}, {Juvela}, {Kim}, {Evans}, {Di Francesco}, {Liu}, {Yuan}, {Tatematsu}, {Zhang}, {Ward-Thompson}, {Fuller}, {Goldsmith}, {Koch}, {Sanhueza}, {Ristorcelli}, {Kang}, {Chen}, {Hirano}, {Wu}, {Sokolov}, {Lee}, {White}, {Wang}, {Eden}, {Li}, {Thompson}, {Pattle}, {Soam}, {Nasedkin}, {Kim}, {Kim}, {Lai}, {Park}, {Qiu}, {Zhang}, {Alina}, {Eswaraiah}, {Falgarone}, {Fich}, {Greaves}, {Gu}, {Kwon}, {Li}, {Malinen}, {Montier}, {Parsons}, {Qin}, {Rawlings}, {Ren}, {Tang}, {Tang}, {Toth}, {Wang}, {Wouterloot}, {Yi}, \& {Zhang}}]{Liu2018}
{Liu}, T., {Li}, P.~S., {Juvela}, M., {et~al.} 2018, \apj, 859, 151, \dodoi{10.3847/1538-4357/aac025}

\bibitem[{{L{\'o}pez-Sepulcre} {et~al.}(2010){L{\'o}pez-Sepulcre}, {Cesaroni}, \& {Walmsley}}]{Lopez2010}
{L{\'o}pez-Sepulcre}, A., {Cesaroni}, R., \& {Walmsley}, C.~M. 2010, \aap, 517, A66, \dodoi{10.1051/0004-6361/201014252}

\bibitem[{{Lu} {et~al.}(2014){Lu}, {Zhang}, {Liu}, {Wang}, \& {Gu}}]{Lu2014}
{Lu}, X., {Zhang}, Q., {Liu}, H.~B., {Wang}, J., \& {Gu}, Q. 2014, \apj, 790, 84, \dodoi{10.1088/0004-637X/790/2/84}

\bibitem[{{Lu} {et~al.}(2018){Lu}, {Zhang}, {Liu}, {Sanhueza}, {Tatematsu}, {Feng}, {Smith}, {Myers}, {Sridharan}, \& {Gu}}]{Lu2018}
{Lu}, X., {Zhang}, Q., {Liu}, H.~B., {et~al.} 2018, \apj, 855, 9, \dodoi{10.3847/1538-4357/aaad11}

\bibitem[{{Mairs} {et~al.}(2021){Mairs}, {Dempsey}, {Bell}, {Parsons}, {Currie}, {Friberg}, {Jiang}, {Tetarenko}, {Bintley}, {Cookson}, {Li}, {Rawlings}, {Wouterloot}, {Berry}, {Graves}, {Mizuno}, {Acohido}, {Clark}, {Cox}, {Fuchs}, {Hoge}, {Kemp}, {Lee}, {Matulonis}, {Montgomerie}, {Silva}, \& {Smith}}]{jcmt2021}
{Mairs}, S., {Dempsey}, J.~T., {Bell}, G.~S., {et~al.} 2021, \aj, 162, 191, \dodoi{10.3847/1538-3881/ac18bf}

\bibitem[{{Molinari} {et~al.}(2010){Molinari}, {Swinyard}, {Bally}, {Barlow}, {Bernard}, {Martin}, {Moore}, {Noriega-Crespo}, {Plume}, {Testi}, {Zavagno}, {Abergel}, {Ali}, {Andr{\'e}}, {Baluteau}, {Benedettini}, {Bern{\'e}}, {Billot}, {Blommaert}, {Bontemps}, {Boulanger}, {Brand}, {Brunt}, {Burton}, {Campeggio}, {Carey}, {Caselli}, {Cesaroni}, {Cernicharo}, {Chakrabarti}, {Chrysostomou}, {Codella}, {Cohen}, {Compiegne}, {Davis}, {de Bernardis}, {de Gasperis}, {Di Francesco}, {di Giorgio}, {Elia}, {Faustini}, {Fischera}, {Fukui}, {Fuller}, {Ganga}, {Garcia-Lario}, {Giard}, {Giardino}, {Glenn}, {Goldsmith}, {Griffin}, {Hoare}, {Huang}, {Jiang}, {Joblin}, {Joncas}, {Juvela}, {Kirk}, {Lagache}, {Li}, {Lim}, {Lord}, {Lucas}, {Maiolo}, {Marengo}, {Marshall}, {Masi}, {Massi}, {Matsuura}, {Meny}, {Minier}, {Miville-Desch{\^e}nes}, {Montier}, {Motte}, {M{\"u}ller}, {Natoli}, {Neves}, {Olmi}, {Paladini}, {Paradis}, {Pestalozzi}, {Pezzuto}, {Piacentini}, {Pomar{\`e}s}, {Popescu}, {Reach}, {Richer}, {Ristorcelli},
  {Roy}, {Royer}, {Russeil}, {Saraceno}, {Sauvage}, {Schilke}, {Schneider-Bontemps}, {Schuller}, {Schultz}, {Shepherd}, {Sibthorpe}, {Smith}, {Smith}, {Spinoglio}, {Stamatellos}, {Strafella}, {Stringfellow}, {Sturm}, {Taylor}, {Thompson}, {Tuffs}, {Umana}, {Valenziano}, {Vavrek}, {Viti}, {Waelkens}, {Ward-Thompson}, {White}, {Wyrowski}, {Yorke}, \& {Zhang}}]{HiGAL2010}
{Molinari}, S., {Swinyard}, B., {Bally}, J., {et~al.} 2010, \pasp, 122, 314, \dodoi{10.1086/651314}

\bibitem[{{Motte} {et~al.}(2018){Motte}, {Bontemps}, \& {Louvet}}]{Motte2018}
{Motte}, F., {Bontemps}, S., \& {Louvet}, F. 2018, \araa, 56, 41, \dodoi{10.1146/annurev-astro-091916-055235}

\bibitem[{{Myers}(2009)}]{Myers2009}
{Myers}, P.~C. 2009, \apj, 700, 1609, \dodoi{10.1088/0004-637X/700/2/1609}

\bibitem[{{Nguyen} {et~al.}(2022){Nguyen}, {Rugel}, {Murugeshan}, {Menten}, {Brunthaler}, {Urquhart}, {Dokara}, {Dzib}, {Gong}, {Khan}, {Medina}, {Ortiz-Le{\'o}n}, {Reich}, {Wyrowski}, {Yang}, {Beuther}, {Cotton}, \& {Pandian}}]{Nguyen2022}
{Nguyen}, H., {Rugel}, M.~R., {Murugeshan}, C., {et~al.} 2022, \aap, 666, A59, \dodoi{10.1051/0004-6361/202244115}

\bibitem[{{Nguyen Luong} {et~al.}(2011){Nguyen Luong}, {Motte}, {Hennemann}, {Hill}, {Rygl}, {Schneider}, {Bontemps}, {Men'shchikov}, {Andr{\'e}}, {Peretto}, {Anderson}, {Arzoumanian}, {Deharveng}, {Didelon}, {di Francesco}, {Griffin}, {Kirk}, {K{\"o}nyves}, {Martin}, {Maury}, {Minier}, {Molinari}, {Pestalozzi}, {Pezzuto}, {Reid}, {Roussel}, {Sauvage}, {Schuller}, {Testi}, {Ward-Thompson}, {White}, \& {Zavagno}}]{ministarburst2011}
{Nguyen Luong}, Q., {Motte}, F., {Hennemann}, M., {et~al.} 2011, \aap, 535, A76, \dodoi{10.1051/0004-6361/201117831}

\bibitem[{{Ossenkopf} \& {Henning}(1994)}]{kv1994}
{Ossenkopf}, V., \& {Henning}, T. 1994, \aap, 291, 943

\bibitem[{{Ostriker}(1964)}]{Ostriker1964}
{Ostriker}, J. 1964, \apj, 140, 1056, \dodoi{10.1086/148005}

\bibitem[{{Ouyang} {et~al.}(2019){Ouyang}, {Chen}, {Shen}, {Yang}, {Li}, {Chen}, {Zhao}, \& {Sobolev}}]{Ouyang2019}
{Ouyang}, X.-J., {Chen}, X., {Shen}, Z.-Q., {et~al.} 2019, \apjs, 245, 12, \dodoi{10.3847/1538-4365/ab4db2}

\bibitem[{{Padoan} {et~al.}(2020){Padoan}, {Pan}, {Juvela}, {Haugb{\o}lle}, \& {Nordlund}}]{Pad20}
{Padoan}, P., {Pan}, L., {Juvela}, M., {Haugb{\o}lle}, T., \& {Nordlund}, {\r{A}}. 2020, \apj, 900, 82, \dodoi{10.3847/1538-4357/abaa47}

\bibitem[{{Pan} {et~al.}(2023){Pan}, {Liu}, \& {Qin}}]{Pan23}
{Pan}, S., {Liu}, H.-L., \& {Qin}, S.-L. 2023, \mnras, 519, 3851, \dodoi{10.1093/mnras/stac3658}

\bibitem[{{Pan} {et~al.}(2024){Pan}, {Liu}, \& {Qin}}]{Pan24}
---. 2024, \apj, 960, 76, \dodoi{10.3847/1538-4357/ad10ac}

\bibitem[{{Peretto} \& {Fuller}(2009)}]{IRDC2009}
{Peretto}, N., \& {Fuller}, G.~A. 2009, \aap, 505, 405, \dodoi{10.1051/0004-6361/200912127}

\bibitem[{{Peretto} {et~al.}(2016){Peretto}, {Lenfestey}, {Fuller}, {Traficante}, {Molinari}, {Thompson}, \& {Ward-Thompson}}]{Tdust2016}
{Peretto}, N., {Lenfestey}, C., {Fuller}, G.~A., {et~al.} 2016, \aap, 590, A72, \dodoi{10.1051/0004-6361/201527064}

\bibitem[{{Peretto} {et~al.}(2013){Peretto}, {Fuller}, {Duarte-Cabral}, {Avison}, {Hennebelle}, {Pineda}, {Andr{\'e}}, {Bontemps}, {Motte}, {Schneider}, \& {Molinari}}]{Peretto2013}
{Peretto}, N., {Fuller}, G.~A., {Duarte-Cabral}, A., {et~al.} 2013, \aap, 555, A112, \dodoi{10.1051/0004-6361/201321318}

\bibitem[{{Peretto} {et~al.}(2014){Peretto}, {Fuller}, {Andr{\'e}}, {Arzoumanian}, {Rivilla}, {Bardeau}, {Duarte Puertas}, {Guzman Fernandez}, {Lenfestey}, {Li}, {Olguin}, {R{\"o}ck}, {de Villiers}, \& {Williams}}]{Peretto2014}
{Peretto}, N., {Fuller}, G.~A., {Andr{\'e}}, P., {et~al.} 2014, \aap, 561, A83, \dodoi{10.1051/0004-6361/201322172}

\bibitem[{{Pineda} {et~al.}(2023){Pineda}, {Arzoumanian}, {Andre}, {Friesen}, {Zavagno}, {Clarke}, {Inoue}, {Chen}, {Lee}, {Soler}, \& {Kuffmeier}}]{Pineda2023}
{Pineda}, J.~E., {Arzoumanian}, D., {Andre}, P., {et~al.} 2023, in Astronomical Society of the Pacific Conference Series, Vol. 534, Protostars and Planets VII, ed. S.~{Inutsuka}, Y.~{Aikawa}, T.~{Muto}, K.~{Tomida}, \& M.~{Tamura}, 233, \dodoi{10.48550/arXiv.2205.03935}

\bibitem[{{Polychroni} {et~al.}(2013){Polychroni}, {Schisano}, {Elia}, {Roy}, {Molinari}, {Martin}, {Andr{\'e}}, {Turrini}, {Rygl}, {Di Francesco}, {Benedettini}, {Busquet}, {di Giorgio}, {Pestalozzi}, {Pezzuto}, {Arzoumanian}, {Bontemps}, {Hennemann}, {Hill}, {K{\"o}nyves}, {Men'shchikov}, {Motte}, {Nguyen-Luong}, {Peretto}, {Schneider}, \& {White}}]{Polychroni2013}
{Polychroni}, D., {Schisano}, E., {Elia}, D., {et~al.} 2013, \apjl, 777, L33, \dodoi{10.1088/2041-8205/777/2/L33}

\bibitem[{{Rathborne} {et~al.}(2010){Rathborne}, {Jackson}, {Chambers}, {Stojimirovic}, {Simon}, {Shipman}, \& {Frieswijk}}]{Rathborne2010}
{Rathborne}, J.~M., {Jackson}, J.~M., {Chambers}, E.~T., {et~al.} 2010, \apj, 715, 310, \dodoi{10.1088/0004-637X/715/1/310}

\bibitem[{{Reid} {et~al.}(2016){Reid}, {Dame}, {Menten}, \& {Brunthaler}}]{bessel2016}
{Reid}, M.~J., {Dame}, T.~M., {Menten}, K.~M., \& {Brunthaler}, A. 2016, \apj, 823, 77, \dodoi{10.3847/0004-637X/823/2/77}

\bibitem[{{Reid} {et~al.}(2019){Reid}, {Menten}, {Brunthaler}, {Zheng}, {Dame}, {Xu}, {Li}, {Sakai}, {Wu}, {Immer}, {Zhang}, {Sanna}, {Moscadelli}, {Rygl}, {Bartkiewicz}, {Hu}, {Quiroga-Nu{\~n}ez}, \& {van Langevelde}}]{bessel2019}
{Reid}, M.~J., {Menten}, K.~M., {Brunthaler}, A., {et~al.} 2019, \apj, 885, 131, \dodoi{10.3847/1538-4357/ab4a11}

\bibitem[{{Ren} {et~al.}(2021){Ren}, {Zhu}, {Shi}, {Yue}, {Li}, {Zhang}, {Mardones}, {Wu}, {Jiao}, {Liu}, {Luo}, {Xie}, {Zhang}, \& {Xu}}]{Ren2021}
{Ren}, Z., {Zhu}, L., {Shi}, H., {et~al.} 2021, \mnras, 505, 5183, \dodoi{10.1093/mnras/stab1509}

\bibitem[{{Rigby} {et~al.}(2021){Rigby}, {Peretto}, {Adam}, {Ade}, {Anderson}, {Andr{\'e}}, {Andrianasolo}, {Aussel}, {Bacmann}, {Beelen}, {Beno{\^\i}t}, {Berta}, {Bourrion}, {Bracco}, {Calvo}, {Catalano}, {De Petris}, {D{\'e}sert}, {Doyle}, {Driessen}, {Garc{\'\i}a}, {Gomez}, {Goupy}, {K{\'e}ruzor{\'e}}, {Kramer}, {Ladjelate}, {Lagache}, {Leclercq}, {Lestrade}, {Mac{\'\i}as-P{\'e}rez}, {Mauskopf}, {Mayet}, {Monfardini}, {Perotto}, {Pisano}, {Ponthieu}, {Rev{\'e}ret}, {Ristorcelli}, {Ritacco}, {Romero}, {Roussel}, {Ruppin}, {Schuster}, {Shu}, {Sievers}, {Tucker}, \& {Watkins}}]{Rigby2021GASTON}
{Rigby}, A.~J., {Peretto}, N., {Adam}, R., {et~al.} 2021, \mnras, 502, 4576, \dodoi{10.1093/mnras/stab200}

\bibitem[{{Rosolowsky} {et~al.}(2008){Rosolowsky}, {Pineda}, {Kauffmann}, \& {Goodman}}]{Rosolowsky2008}
{Rosolowsky}, E.~W., {Pineda}, J.~E., {Kauffmann}, J., \& {Goodman}, A.~A. 2008, \apj, 679, 1338, \dodoi{10.1086/587685}

\bibitem[{{Saha} {et~al.}(2022){Saha}, {Tej}, {Liu}, {Liu}, {Issac}, {Lee}, {Garay}, {Goldsmith}, {Juvela}, {Qin}, {Stutz}, {Li}, {Wang}, {Baug}, {Bronfman}, {Xu}, {Zhang}, \& {Eswaraiah}}]{Saha22}
{Saha}, A., {Tej}, A., {Liu}, H.-L., {et~al.} 2022, \mnras, 516, 1983, \dodoi{10.1093/mnras/stac2353}

\bibitem[{{S{\'a}nchez-Monge} {et~al.}(2013){S{\'a}nchez-Monge}, {Palau}, {Fontani}, {Busquet}, {Ju{\'a}rez}, {Estalella}, {Tan}, {Sep{\'u}lveda}, {Ho}, {Zhang}, \& {Kurtz}}]{Sanchez2013}
{S{\'a}nchez-Monge}, {\'A}., {Palau}, A., {Fontani}, F., {et~al.} 2013, \mnras, 432, 3288, \dodoi{10.1093/mnras/stt679}

\bibitem[{{Sanhueza} {et~al.}(2012){Sanhueza}, {Jackson}, {Foster}, {Garay}, {Silva}, \& {Finn}}]{Sanhueza2012}
{Sanhueza}, P., {Jackson}, J.~M., {Foster}, J.~B., {et~al.} 2012, \apj, 756, 60, \dodoi{10.1088/0004-637X/756/1/60}

\bibitem[{{Sanhueza} {et~al.}(2017){Sanhueza}, {Jackson}, {Zhang}, {Guzm{\'a}n}, {Lu}, {Stephens}, {Wang}, \& {Tatematsu}}]{Sanhueza2017G28}
{Sanhueza}, P., {Jackson}, J.~M., {Zhang}, Q., {et~al.} 2017, \apj, 841, 97, \dodoi{10.3847/1538-4357/aa6ff8}

\bibitem[{{Sanhueza} {et~al.}(2021){Sanhueza}, {Girart}, {Padovani}, {Galli}, {Hull}, {Zhang}, {Cortes}, {Stephens}, {Fern{\'a}ndez-L{\'o}pez}, {Jackson}, {Frau}, {Kock}, {Wu}, {Zapata}, {Olguin}, {Lu}, {Silva}, {Tang}, {Sakai}, {Guzm{\'a}n}, {Tatematsu}, {Nakamura}, \& {Chen}}]{Sanhueza2021}
{Sanhueza}, P., {Girart}, J.~M., {Padovani}, M., {et~al.} 2021, \apjl, 915, L10, \dodoi{10.3847/2041-8213/ac081c}

\bibitem[{{Schisano} {et~al.}(2020){Schisano}, {Molinari}, {Elia}, {Benedettini}, {Olmi}, {Pezzuto}, {Traficante}, {Brescia}, {Cavuoti}, {di Giorgio}, {Liu}, {Moore}, {Noriega-Crespo}, {Riccio}, {Baldeschi}, {Becciani}, {Peretto}, {Merello}, {Vitello}, {Zavagno}, {Beltr{\'a}n}, {Cambr{\'e}sy}, {Eden}, {Li Causi}, {Molinaro}, {Palmeirim}, {Sciacca}, {Testi}, {Umana}, \& {Whitworth}}]{Schisano2020}
{Schisano}, E., {Molinari}, S., {Elia}, D., {et~al.} 2020, \mnras, 492, 5420, \dodoi{10.1093/mnras/stz3466}

\bibitem[{{Schneider} {et~al.}(2012){Schneider}, {Csengeri}, {Hennemann}, {Motte}, {Didelon}, {Federrath}, {Bontemps}, {Di Francesco}, {Arzoumanian}, {Minier}, {Andr{\'e}}, {Hill}, {Zavagno}, {Nguyen-Luong}, {Attard}, {Bernard}, {Elia}, {Fallscheer}, {Griffin}, {Kirk}, {Klessen}, {K{\"o}nyves}, {Martin}, {Men'shchikov}, {Palmeirim}, {Peretto}, {Pestalozzi}, {Russeil}, {Sadavoy}, {Sousbie}, {Testi}, {Tremblin}, {Ward-Thompson}, \& {White}}]{Schneider2012}
{Schneider}, N., {Csengeri}, T., {Hennemann}, M., {et~al.} 2012, \aap, 540, L11, \dodoi{10.1051/0004-6361/201118566}

\bibitem[{{Schuller} {et~al.}(2009){Schuller}, {Menten}, {Contreras}, {Wyrowski}, {Schilke}, {Bronfman}, {Henning}, {Walmsley}, {Beuther}, {Bontemps}, {Cesaroni}, {Deharveng}, {Garay}, {Herpin}, {Lefloch}, {Linz}, {Mardones}, {Minier}, {Molinari}, {Motte}, {Nyman}, {Reveret}, {Risacher}, {Russeil}, {Schneider}, {Testi}, {Troost}, {Vasyunina}, {Wienen}, {Zavagno}, {Kovacs}, {Kreysa}, {Siringo}, \& {Wei{\ss}}}]{ATLASGAL2009}
{Schuller}, F., {Menten}, K.~M., {Contreras}, Y., {et~al.} 2009, \aap, 504, 415, \dodoi{10.1051/0004-6361/200811568}

\bibitem[{{Shimajiri} {et~al.}(2019){Shimajiri}, {Andr{\'e}}, {Ntormousi}, {Men'shchikov}, {Arzoumanian}, \& {Palmeirim}}]{Shimajiri2019}
{Shimajiri}, Y., {Andr{\'e}}, P., {Ntormousi}, E., {et~al.} 2019, \aap, 632, A83, \dodoi{10.1051/0004-6361/201935689}

\bibitem[{{Shirley} {et~al.}(2011){Shirley}, {Huard}, {Pontoppidan}, {Wilner}, {Stutz}, {Bieging}, \& {Evans}}]{kv2011}
{Shirley}, Y.~L., {Huard}, T.~L., {Pontoppidan}, K.~M., {et~al.} 2011, \apj, 728, 143, \dodoi{10.1088/0004-637X/728/2/143}

\bibitem[{{Simon} {et~al.}(2006){Simon}, {Rathborne}, {Shah}, {Jackson}, \& {Chambers}}]{Simon2006}
{Simon}, R., {Rathborne}, J.~M., {Shah}, R.~Y., {Jackson}, J.~M., \& {Chambers}, E.~T. 2006, \apj, 653, 1325, \dodoi{10.1086/508915}

\bibitem[{{Smith} {et~al.}(2016){Smith}, {Glover}, {Klessen}, \& {Fuller}}]{Smith2016}
{Smith}, R.~J., {Glover}, S. C.~O., {Klessen}, R.~S., \& {Fuller}, G.~A. 2016, \mnras, 455, 3640, \dodoi{10.1093/mnras/stv2559}

\bibitem[{{Smith} {et~al.}(2009){Smith}, {Longmore}, \& {Bonnell}}]{clump-fed2009}
{Smith}, R.~J., {Longmore}, S., \& {Bonnell}, I. 2009, \mnras, 400, 1775, \dodoi{10.1111/j.1365-2966.2009.15621.x}

\bibitem[{{Sokolov} {et~al.}(2017){Sokolov}, {Wang}, {Pineda}, {Caselli}, {Henshaw}, {Tan}, {Fontani}, {Jim{\'e}nez-Serra}, \& {Lim}}]{G35Td2017}
{Sokolov}, V., {Wang}, K., {Pineda}, J.~E., {et~al.} 2017, \aap, 606, A133, \dodoi{10.1051/0004-6361/201630350}

\bibitem[{{Solomon} {et~al.}(1987){Solomon}, {Rivolo}, {Barrett}, \& {Yahil}}]{Solomon1987}
{Solomon}, P.~M., {Rivolo}, A.~R., {Barrett}, J., \& {Yahil}, A. 1987, \apj, 319, 730, \dodoi{10.1086/165493}

\bibitem[{{Suri} {et~al.}(2019){Suri}, {S{\'a}nchez-Monge}, {Schilke}, {Clarke}, {Smith}, {Ossenkopf-Okada}, {Klessen}, {Padoan}, {Goldsmith}, {Arce}, {Bally}, {Carpenter}, {Ginsburg}, {Johnstone}, {Kauffmann}, {Kong}, {Lis}, {Mairs}, {Pillai}, {Pineda}, \& {Duarte-Cabral}}]{Suri2019}
{Suri}, S., {S{\'a}nchez-Monge}, {\'A}., {Schilke}, P., {et~al.} 2019, \aap, 623, A142, \dodoi{10.1051/0004-6361/201834049}

\bibitem[{{Tafalla} \& {Hacar}(2015)}]{Tafalla2015}
{Tafalla}, M., \& {Hacar}, A. 2015, \aap, 574, A104, \dodoi{10.1051/0004-6361/201424576}

\bibitem[{{Taniguchi} {et~al.}(2023){Taniguchi}, {Sanhueza}, {Olguin}, {Gorai}, {Das}, {Nakamura}, {Saito}, {Zhang}, {Lu}, {Li}, \& {Chen}}]{Taniguchi2023}
{Taniguchi}, K., {Sanhueza}, P., {Olguin}, F.~A., {et~al.} 2023, \apj, 950, 57, \dodoi{10.3847/1538-4357/acca1d}

\bibitem[{{Thompson} {et~al.}(2006){Thompson}, {Hatchell}, {Walsh}, {MacDonald}, \& {Millar}}]{Thompson2006}
{Thompson}, M.~A., {Hatchell}, J., {Walsh}, A.~J., {MacDonald}, G.~H., \& {Millar}, T.~J. 2006, \aap, 453, 1003, \dodoi{10.1051/0004-6361:20054383}

\bibitem[{{Umemoto} {et~al.}(2017){Umemoto}, {Minamidani}, {Kuno}, {Fujita}, {Matsuo}, {Nishimura}, {Torii}, {Tosaki}, {Kohno}, {Kuriki}, {Tsuda}, {Hirota}, {Ohashi}, {Yamagishi}, {Handa}, {Nakanishi}, {Omodaka}, {Koide}, {Matsumoto}, {Onishi}, {Tokuda}, {Seta}, {Kobayashi}, {Tachihara}, {Sano}, {Hattori}, {Onodera}, {Oasa}, {Kamegai}, {Tsuboi}, {Sofue}, {Higuchi}, {Chibueze}, {Mizuno}, {Honma}, {Muller}, {Inoue}, {Morokuma-Matsui}, {Shinnaga}, {Ozawa}, {Takahashi}, {Yoshiike}, {Costes}, \& {Kuwahara}}]{FUGIN2017}
{Umemoto}, T., {Minamidani}, T., {Kuno}, N., {et~al.} 2017, \pasj, 69, 78, \dodoi{10.1093/pasj/psx061}

\bibitem[{{Urquhart} {et~al.}(2008){Urquhart}, {Hoare}, {Lumsden}, {Oudmaijer}, \& {Moore}}]{MYSO2008}
{Urquhart}, J.~S., {Hoare}, M.~G., {Lumsden}, S.~L., {Oudmaijer}, R.~D., \& {Moore}, T.~J.~T. 2008, in Astronomical Society of the Pacific Conference Series, Vol. 387, Massive Star Formation: Observations Confront Theory, ed. H.~{Beuther}, H.~{Linz}, \& T.~{Henning}, 381, \dodoi{10.48550/arXiv.0711.4715}

\bibitem[{{Urquhart} {et~al.}(2011){Urquhart}, {Morgan}, {Figura}, {Moore}, {Lumsden}, {Hoare}, {Oudmaijer}, {Mottram}, {Davies}, \& {Dunham}}]{Urquhart2011}
{Urquhart}, J.~S., {Morgan}, L.~K., {Figura}, C.~C., {et~al.} 2011, \mnras, 418, 1689, \dodoi{10.1111/j.1365-2966.2011.19594.x}

\bibitem[{{Urquhart} {et~al.}(2014{\natexlab{a}}){Urquhart}, {Csengeri}, {Wyrowski}, {Schuller}, {Bontemps}, {Bronfman}, {Menten}, {Walmsley}, {Contreras}, {Beuther}, {Wienen}, \& {Linz}}]{csc2014}
{Urquhart}, J.~S., {Csengeri}, T., {Wyrowski}, F., {et~al.} 2014{\natexlab{a}}, \aap, 568, A41, \dodoi{10.1051/0004-6361/201424126}

\bibitem[{{Urquhart} {et~al.}(2014{\natexlab{b}}){Urquhart}, {Moore}, {Csengeri}, {Wyrowski}, {Schuller}, {Hoare}, {Lumsden}, {Mottram}, {Thompson}, {Menten}, {Walmsley}, {Bronfman}, {Pfalzner}, {K{\"o}nig}, \& {Wienen}}]{Urquhart2014}
{Urquhart}, J.~S., {Moore}, T.~J.~T., {Csengeri}, T., {et~al.} 2014{\natexlab{b}}, \mnras, 443, 1555, \dodoi{10.1093/mnras/stu1207}

\bibitem[{{Urquhart} {et~al.}(2018){Urquhart}, {K{\"o}nig}, {Giannetti}, {Leurini}, {Moore}, {Eden}, {Pillai}, {Thompson}, {Braiding}, {Burton}, {Csengeri}, {Dempsey}, {Figura}, {Froebrich}, {Menten}, {Schuller}, {Smith}, \& {Wyrowski}}]{Urquhart2018}
{Urquhart}, J.~S., {K{\"o}nig}, C., {Giannetti}, A., {et~al.} 2018, \mnras, 473, 1059, \dodoi{10.1093/mnras/stx2258}

\bibitem[{{V{\'a}zquez-Semadeni} {et~al.}(2019){V{\'a}zquez-Semadeni}, {Palau}, {Ballesteros-Paredes}, {G{\'o}mez}, \& {Zamora-Avil{\'e}s}}]{GHC2019}
{V{\'a}zquez-Semadeni}, E., {Palau}, A., {Ballesteros-Paredes}, J., {G{\'o}mez}, G.~C., \& {Zamora-Avil{\'e}s}, M. 2019, \mnras, 490, 3061, \dodoi{10.1093/mnras/stz2736}

\bibitem[{{Wang} {et~al.}(2016){Wang}, {Testi}, {Burkert}, {Walmsley}, {Beuther}, \& {Henning}}]{Wang2016}
{Wang}, K., {Testi}, L., {Burkert}, A., {et~al.} 2016, \apjs, 226, 9, \dodoi{10.3847/0067-0049/226/1/9}

\bibitem[{{Wienen} {et~al.}(2012){Wienen}, {Wyrowski}, {Schuller}, {Menten}, {Walmsley}, {Bronfman}, \& {Motte}}]{Wienen2012}
{Wienen}, M., {Wyrowski}, F., {Schuller}, F., {et~al.} 2012, \aap, 544, A146, \dodoi{10.1051/0004-6361/201118107}

\bibitem[{{Xie} {et~al.}(2021){Xie}, {Fuller}, {Li}, {Chen}, {Ren}, {Wu}, {Duan}, {Wang}, {Li}, {Peretto}, {Liu}, \& {Shen}}]{Xie2021}
{Xie}, J., {Fuller}, G.~A., {Li}, D., {et~al.} 2021, Science China Physics, Mechanics, and Astronomy, 64, 279511, \dodoi{10.1007/s11433-021-1695-0}

\bibitem[{{Xu} {et~al.}(2024){Xu}, {Wang}, {Liu}, {Tang}, {Evans}, {Palau}, {Morii}, {He}, {Sanhueza}, {Liu}, {Stutz}, {Zhang}, {Chen}, {Li}, {G{\'o}mez}, {V{\'a}zquez-Semadeni}, {Li}, {Mai}, {Lu}, {Liu}, {Chen}, {Li}, {Shi}, {Ren}, {Li}, {Garay}, {Bronfman}, {Dewangan}, {Juvela}, {Lee}, {Zhang}, {Yue}, {Wang}, {Ge}, {Jiao}, {Luo}, {Zhou}, {Tatematsu}, {Chibueze}, {Su}, {Sun}, {Ristorcelli}, \& {Toth}}]{Xu2024ASSEMBLE}
{Xu}, F., {Wang}, K., {Liu}, T., {et~al.} 2024, \apjs, 270, 9, \dodoi{10.3847/1538-4365/acfee5}

\bibitem[{{Xu} {et~al.}(2023){Xu}, {Wang}, {Liu}, {Goldsmith}, {Zhang}, {Juvela}, {Liu}, {Qin}, {Li}, {Tej}, {Garay}, {Bronfman}, {Li}, {Wu}, {G{\'o}mez}, {V{\'a}zquez-Semadeni}, {Tatematsu}, {Ren}, {Zhang}, {Toth}, {Liu}, {Yue}, {Zhang}, {Baug}, {Issac}, {Stutz}, {Liu}, {Fuller}, {Tang}, {Zhang}, {Dewangan}, {Lee}, {Zhou}, {Xie}, {Jiao}, {Wang}, {Liu}, {Luo}, {Soam}, \& {Eswaraiah}}]{Xu2023}
{Xu}, F.-W., {Wang}, K., {Liu}, T., {et~al.} 2023, \mnras, 520, 3259, \dodoi{10.1093/mnras/stad012}

\bibitem[{{Yang} {et~al.}(2018){Yang}, {Thompson}, {Urquhart}, \& {Tian}}]{YangAY2018}
{Yang}, A.~Y., {Thompson}, M.~A., {Urquhart}, J.~S., \& {Tian}, W.~W. 2018, \apjs, 235, 3, \dodoi{10.3847/1538-4365/aaa297}

\bibitem[{{Yang} {et~al.}(2023{\natexlab{a}}){Yang}, {Dzib}, {Urquhart}, {Brunthaler}, {Medina}, {Menten}, {Wyrowski}, {Ortiz-Le{\'o}n}, {Cotton}, {Gong}, {Dokara}, {Rugel}, {Beuther}, {Pandian}, {Csengeri}, {Veena}, {Roy}, {Nguyen}, {Winkel}, {Ott}, {Carrasco-Gonzalez}, {Khan}, \& {Cheema}}]{YangAY2023}
{Yang}, A.~Y., {Dzib}, S.~A., {Urquhart}, J.~S., {et~al.} 2023{\natexlab{a}}, \aap, 680, A92, \dodoi{10.1051/0004-6361/202347563}

\bibitem[{{Yang} {et~al.}(2023{\natexlab{b}}){Yang}, {Liu}, {Tej}, {Liu}, {Sanhueza}, {Qin}, {Lu}, {Wang}, {Pan}, {Xu}, {V{\'a}zquez-Semadeni​}, {Li}, {G{\'o}mez}, {Palau}, {Garay}, {Goldsmith}, {Juvela}, {Saha}, {Bronfman}, {Lee}, {Tatematsu}, {Dewangan}, {Zhou}, {Zhang}, {Stutz}, {Eswaraiah}, {Toth}, {Ristorcelli}, {Shen}, {Luo}, \& {Chibueze}}]{Yang23}
{Yang}, D., {Liu}, H.-L., {Tej}, A., {et~al.} 2023{\natexlab{b}}, \apj, 953, 40, \dodoi{10.3847/1538-4357/acdf42}

\bibitem[{{Yuan} {et~al.}(2018){Yuan}, {Li}, {Wu}, {Ellingsen}, {Henkel}, {Wang}, {Liu}, {Liu}, {Zavagno}, {Ren}, \& {Huang}}]{Yuan2018}
{Yuan}, J., {Li}, J.-Z., {Wu}, Y., {et~al.} 2018, \apj, 852, 12, \dodoi{10.3847/1538-4357/aa9d40}

\bibitem[{{Zavagno} {et~al.}(2023){Zavagno}, {Dup{\'e}}, {Bensaid}, {Schisano}, {Li Causi}, {Gray}, {Molinari}, {Elia}, {Lambert}, {Brescia}, {Arzoumanian}, {Russeil}, {Riccio}, \& {Cavuoti}}]{Annie2023}
{Zavagno}, A., {Dup{\'e}}, F.~X., {Bensaid}, S., {et~al.} 2023, \aap, 669, A120, \dodoi{10.1051/0004-6361/202244103}

\bibitem[{{Zhang} {et~al.}(2023){Zhang}, {Zhang}, {Li}, \& {Yuan}}]{WEGO2023}
{Zhang}, C., {Zhang}, G.-Y., {Li}, J.-Z., \& {Yuan}, J.-H. 2023, \apjs, 264, 24, \dodoi{10.3847/1538-4365/aca325}

\bibitem[{{Zhou} {et~al.}(2022){Zhou}, {Liu}, {Evans}, {Garay}, {Goldsmith}, {G{\'o}mez}, {V{\'a}zquez-Semadeni}, {Liu}, {Stutz}, {Wang}, {Juvela}, {He}, {Li}, {Bronfman}, {Liu}, {Xu}, {Tej}, {Dewangan}, {Li}, {Zhang}, {Zhang}, {Ren}, {Tatematsu}, {Shing Li}, {Won Lee}, {Baug}, {Qin}, {Wu}, {Peng}, {Zhang}, {Liu}, {Luo}, {Ge}, {Saha}, {Chakali}, {Zhang}, {Kim}, {Ristorcelli}, {Shen}, \& {Li}}]{Zhou2022}
{Zhou}, J.-W., {Liu}, T., {Evans}, N.~J., {et~al.} 2022, \mnras, 514, 6038, \dodoi{10.1093/mnras/stac1735}

\bibitem[{{Zinnecker} \& {Yorke}(2007)}]{Zinnecker&Yorke2007}
{Zinnecker}, H., \& {Yorke}, H.~W. 2007, \araa, 45, 481, \dodoi{10.1146/annurev.astro.44.051905.092549}

\end{thebibliography}
\bibliographystyle{aasjournal}



\end{document}